\documentclass[11pt]{article}
\usepackage{verbatim}
\usepackage[T1]{fontenc}
\DeclareTextSymbolDefault{\dh}{T1}
\usepackage{hyperref}
\hypersetup{
	colorlinks=true,
	linkcolor=red,
	filecolor=magenta,      
	urlcolor=cyan,
}
\usepackage{graphicx}    
\usepackage{overpic}
\usepackage{subcaption}

\usepackage{amsmath,amssymb,stmaryrd}
\usepackage{color}
\usepackage{xcolor}
\usepackage{wrapfig}
\usepackage[numbers,sort&compress]{natbib}
\usepackage{setspace}
\usepackage[subtle]{savetrees}
\usepackage{times}
\usepackage{xurl}
\newtheorem{theorem}{Theorem}
\usepackage{bbm}
\usepackage{mathtools}
\usepackage{multirow}
\usepackage[normalem]{ulem}
\usepackage{cleveref}
\usepackage{epstopdf}

\newcommand{\D}{\mathrm{d}}

\def\div{\nabla\cdot}

\newcommand{\R}{\mathbb R}

\newcommand{\bV}{\mathbf V}

\newcommand{\bk}{\mathbf k}

\newcommand{\bn}{\mathbf n}
\newcommand{\be}{\mathbf e}

\newcommand{\br}{\mathbf r}

\newcommand{\bs}{\mathbf s}

\newcommand{\bu}{\mathbf u}
\newcommand{\bU}{\mathbf U}
\newcommand{\bv}{\mathbf v}
\newcommand{\bw}{\mathbf w}

\newcommand{\bx}{\mathbf x}
\newcommand{\by}{\mathbf y}

\newcommand{\bbf}{\mathbf f}
\newcommand{\A}{\mathcal A}
\newcommand{\F}{\mathcal F}
\newcommand{\G}{\mathcal G}
\newcommand{\I}{\mathcal I}
\newcommand{\K}{\mathcal K}
\newcommand{\M}{\mathcal M}
\newcommand{\N}{\mathcal N}
\newcommand{\Q}{\mathcal Q}
\newcommand{\J}{\mathcal J}
\newcommand{\T}{\mathcal T}
\newcommand{\V}{\mathcal V}
\newcommand{\Div}{\mathop{\rm div}}

\newcommand{\cD}{\mathcal D}
\newcommand{\cF}{\mathcal F}
\newcommand{\cI}{\mathcal I}

\newcommand{\cL}{\mathcal L}
\newcommand{\cM}{\mathcal M}

\def\div{\nabla\cdot}

\newcommand{\anna}[2][cyan]{\textcolor{#1}{#2}}
\newcommand{\ti}[1]{{\color{blue}TI: #1}}
\newcommand{\alex}[2][magenta]{\textcolor{#1}{#2}}

\newcommand{\ou}{\overline{u}}
\newcommand{\uu}{\underline{u}}
\newcommand{\obu}{\overline{\boldsymbol u}}

\newcommand{\bphi}{\boldsymbol{\varphi}}
\newcommand{\bX}{\mathbf X}

\newtheorem{remark}{Remark}[section]
\newcommand{\Deriv}[3]{\dfrac{\partial^{#1} #2}{\partial #3^{#1}}}

\def\el {\nonumber }

\begin{document}

\title{Bridging Large Eddy Simulation and \\ Reduced Order Modeling 
of Convection-Dominated Flows through 
Spatial Filtering: Review and Perspectives}
\author{Annalisa Quaini$^1$, Omer San$^2$, Alessandro Veneziani$^3$, Traian Iliescu$^4$}

\maketitle

\begin{center}
\footnotesize
$^1$ Department of Mathematics, University of Houston, 3551 Cullen Blvd, Houston TX 77204, USA \\ 
$^2$ Department of Mechanical, Aerospace and Biomedical Engineering, University of Tennessee, \\ 1512 Middle Drive, Knoxville TN 37996, USA \\
$^3$ Department of Mathematics, Department of Computer Science, Emory University, \\ 
400 Dowman Drive, Atlanta GA 30322, USA \\
$^4$ Department of Mathematics, Virginia Tech, 225 Stanger Street,
Blacksburg VA 24061, USA 
\end{center}

\begin{abstract}
Reduced order models (ROMs) have achieved a lot of success in reducing the computational cost of traditional numerical methods across many disciplines.
In fluid dynamics, ROMs have been successful in providing efficient and relatively accurate solutions for the numerical simulation of laminar flows.
For convection-dominated (e.g., turbulent) flows, however, standard ROMs generally yield inaccurate results, usually affected by spurious oscillations.
Thus, ROMs are usually equipped with numerical stabilization or closure models in order to account for the effect of the discarded modes. 
The literature on ROM closures and stabilizations is large and growing fast.
In this paper, instead of reviewing all the ROM closures and stabilizations, we took a more modest step and focused on one particular type of ROM closures and stabilizations that are inspired by Large Eddy Simulation (LES), a classical strategy in Computational Fluid Dynamics (CFD). 

These ROMs, which we call LES-ROMs, are extremely easy to implement, very efficient, and accurate. 
Indeed, the LES-ROMs are modular and generally require minimal modifications to standard (``legacy'') ROM formulations.
Furthermore, the computational overhead of these modifications is minimal. 
Finally, carefully tuned LES-ROMs can accurately capture the average physical quantities of interest in challenging convection-dominated flows in science and
engineering applications.
LES-ROM are constructed by leveraging spatial filtering, which is the same principle used to build classical LES models.
This ensures a modeling consistency between LES-ROMs and the approaches that generated the data used to train them.
It also ``bridges'' two distinct research fields (LES and ROMs), that have been disconnected until now.

This paper is a review of LES-ROMs. It starts with a description
of a versatile LES strategy called evolve-filter-relax (EFR)
that has been successfully used as a full order method
for both incompressible and compressible convection-dominated flows. We present evidence of this success. 
We then show how the EFR strategy, and spatial filtering in general, can be leveraged to construct LES-ROMs (e.g., EFR-ROM).
Several applications of LES-ROMs to the numerical simulation of incompressible and compressible convection-dominated flows
are presented. 
Finally, we draw conclusions and outline several research directions and open questions in the LES-ROM development.
While we do not claim this review to be comprehensive, 
we certainly hope it serves as a brief and friendly introduction to this exciting research area, which we believe has a lot of potential in practical numerical simulation of convection-dominated flows in science, engineering, and medicine.



\end{abstract}


\section{Introduction}\label{sec:intro}

In a couple of seminal papers appeared in 1941 \cite{Kolmogorov41-1,Kolmogorov41-2}, Kolmogorov showed
that the vortices and eddies in a fluid flow span an increasingly large range of scales as the inertial forces become dominant over the viscous forces. The aim of a  Direct Numerical Simulation (DNS) 
is to capture all these flow structures with a computational mesh
sufficiently refined to resolve
even the smallest scales. 
For a large number of important applications in science, engineering, and medicine, 
a DNS requires extremely fine meshes, leading to exorbitant computational costs for current computing technology. Examples of such applications include atmospheric flow for weather and climate predictions, flow around wind turbines for efficient energy production, and blood flow in larger arteries to predict the progression of a disease or plan for surgery.


The goal to reduce the computational cost of DNS while maintaining a reasonable level of accuracy
has motivated a vast body of literature.
Among the many proposed approaches, Reynolds Averaged Navier-Stokes (RANS) and Large Eddy Simulation (LES) have become popular and widespread across different sciences and industries. 
In broad terms, RANS models average the 
Navier-Stokes equations in different ways (quite often in time), while LES techniques
filter them (usually in space). This considerably lowers the computational cost but requires closures to represent certain terms emerging from the averaging/filtering process. For closure, RANS
models typically rely on empirical
descriptions calibrated with available data. 
While one important line of research seeks to improve 
the accuracy of RANS models by systematically informing them with larger datasets \cite{duraisamy2019turbulence},
this review focuses on LES, which is a
mathematically and physically justified approach \cite{sagaut2006large,BIL05,layton2012approximate}.


LES techniques directly resolve a portion of the flow scales and require a model to account for the remaining (small) unresolved scales that are not directly captured due to mesh under-refinement.
Much literature has been dedicated to
the derivation and performance analysis of 
LES models, which inevitably feature parameters. 
The combination of properly tuned LES models and high mesh resolution has been shown to yield 
accurate 
computational results in 
fields ranging from blood flow studies \cite{delorme2013large,xu2018coupled,manchester2021analysis} 
to weather and climate forecasts \cite{Caldwelletal2019,Teraietal2018,Wehneretal2014,Bacmeisteretal2014,DelworthetAl2012,Loveetal2011,Atlasetal2005,Iorioetal2004,DuffyetAl2003,PopeStratton2002}.
However, while feasible (unlike DNS), these highly accurate LES simulations 
still carry a high computational cost. 
These hefty computational costs could be acceptable in exploratory studies, but they are incompatible with pressing deadlines and/or multi-query contexts. 


Multi-query contexts arise from needing to, e.g., assess the uncertainty in the computed solution, optimally control a system, or solve an inverse problem for parameter identification or optimization. 
They are as ubiquitous as tight deadlines in science, engineering, and medicine. 
Unable to be patient for the long computational times of accurate simulations, LES practitioners typically trade accuracy (high-resolution meshes and costly parameter tuning) for computational efficiency (coarse meshes and parameters set empirically). This trade-off limits the full potential of computational simulations and has possible detrimental consequences on the reliability of a prediction (e.g., weather and climate), the structural integrity of a physical asset (e.g., a wind turbine), or the life of an individual (e.g., surgical planning for cardiovascular disease). 
Drastically reducing the computational cost of highly accurate LES simulations is a significant research goal that would  enable 
tremendous advancements across sciences and industries.

A viable way to  rectify the imbalance between computational efficiency and accuracy 
is by bridging two fields that have so far remained disconnected: 
LES and advanced Reduced Order Modeling (ROM) techniques.
ROMs are 
methods specifically designed to speed up computations in multi-query settings using state-of-the-art linear algebra techniques and error estimation. 
The natural connections between these two fields emerge around the mathematical operations of \textit{spatial (low-pass) filtering} and \textit{approximate deconvolution}, both in the classical sense \cite{LEONARD1975237,germano_1992,sagaut2006large,Moser2021} and reinterpreted through a more modern data-driven lens \cite{BECK2019108910,SIRIGNANO2020109811,PhysRevFluids.5.054606,duraisamy2021perspectives,raissi2020hidden,di2021two}. {Filtering can be redefined through the integration of physics-based models and 
data-driven methodologies, fostering the evolution of hybrid solvers that
revere the underlying physics while harnessing the robust structure and pattern recognition capabilities of machine learning. 
}
 With a research effort on the complex computational mathematics behind
spatial filtering and approximate deconvolution for LES and ROM, 
fast, accurate, and robust simulations of convection-dominated flows will be possible with minimal user set-up. 
{
The modular bidirectional interaction between ROM and LES will yield a converged computational twin usable for detailed flow analysis (LES counterpart) and swift decision-making (ROM counterpart), dynamically updating itself with incoming measurements and synthetic data.}

The review is structured as follows. Sec.~\ref{sec:les-fom} introduces general concepts in LES modeling and then, for clarity of presentation, focuses on one of the most popular and successful models, the EFR strategy, with details on the application of the EFR algorithm to 
the incompressible Navier-Stokes equations and the weakly compressible Euler equations. 
Sec.~\ref{sec:app_FOM} presents results obtained with the EFR method as Full Order Method for complex applications arising in hemodynamics and physics of the atmosphere and discusses related open problems. 
In Sec.~\ref{sec:LES_ROM}, we show how key concepts in the construction of LES models, i.e., spatial filters and approximate deconvolution, can be leveraged to construct LES-ROMs. An important feature in  Sec.~\ref{sec:LES_ROM} is the discussion of preliminary numerical analysis results obtained for LES-ROMs. Selected achievements 
and prospected applications of LES-ROMs for incompressible and compressible fluids are reported in Sec.~\ref{sec:app_ROM}.
Sec.~\ref{sec:concl} draws conclusions from this review and summarizes the open perspectives. 


\section{LES as a Full Order Model}
    \label{sec:les-fom}

In order to explain the main ideas of LES, let us briefly summarize some key concepts in Kolmogorov's theory \cite{Kolmogorov41-1,Kolmogorov41-2}.

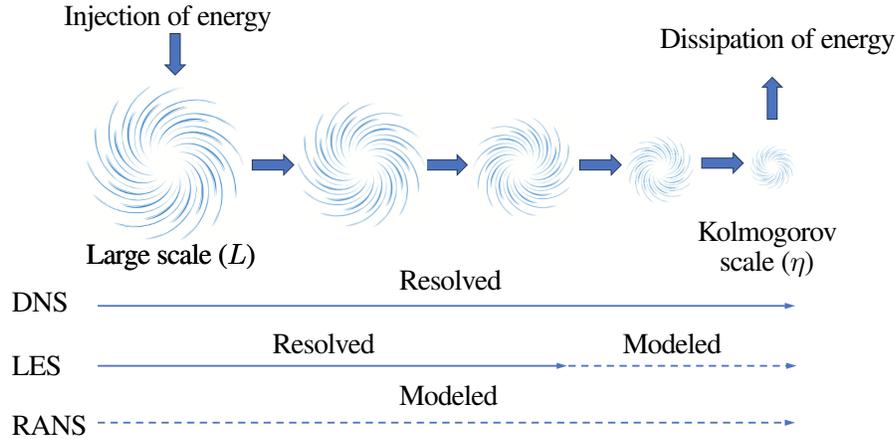
\begin{figure}[htb]
\centering
 \begin{overpic}[width=0.6\textwidth, grid=false]{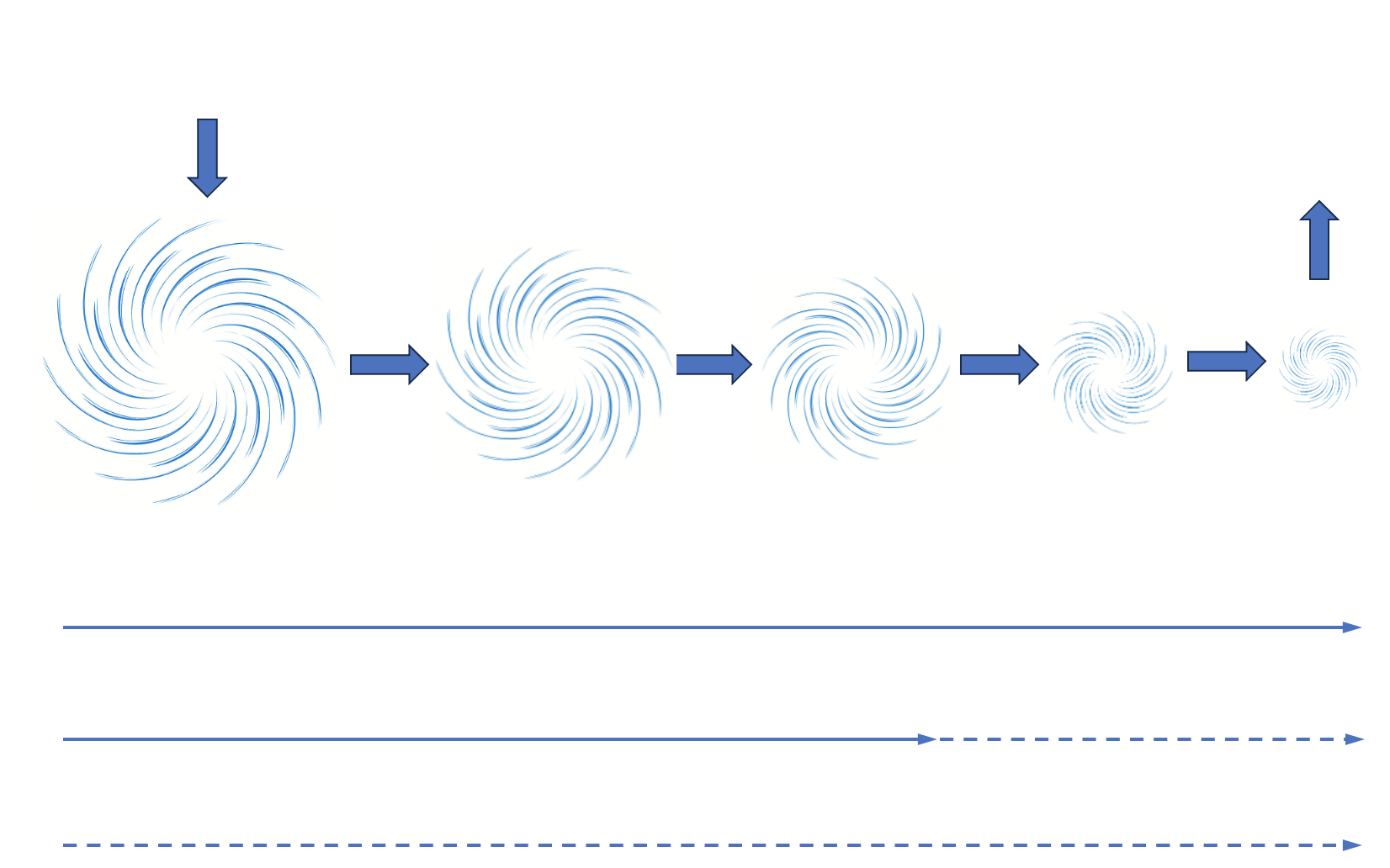}
        \put(0,55){Injection of energy}
        \put(3,23){Large scale ($L$)}
        \put(80,52){Dissipation of energy}
        \put(3,23){Large scale ($L$)}
        \put(85,26){Kolmogorov}
        \put(88,22){scale ($\eta$)}
        \put(-7,16.5){DNS}
        \put(45,19.5){Resolved}
        \put(-7,8){LES}
        \put(28,11){Resolved}
        \put(75,11){Modeled}
        \put(-7,0){RANS}
        \put(45,4){Modeled}
      \end{overpic}
\caption{Overall view of the energy cascade, from injection to dissipation of energy, and associated types of modeling.}
\label{fig:energy_cascade}
\end{figure}

The 
{kinetic energy (KE)} of a fluid, which is the kinetic energy associated with eddies in the flow,
defined as {$\frac{1}{2}$}$\rho \int_\Omega \|\bu\|^2 d\omega$, {where $\rho$ is the fluid density, $\bu$ the fluid velocity, 
$\Omega$ the spatial domain where the fluid is flowing, and $||\cdot||_{L^2}$ denotes the $L^2$ norm,} is injected in the system at the large scales (low wave numbers). Since the large scale eddies are unstable, they break down, transferring the energy to smaller {and smaller} eddies. Finally, the 
{KE} is dissipated by the viscous forces at the small scales (high wave numbers). This process is usually referred to as \emph{energy cascade}. See Fig.~\ref{fig:energy_cascade}. The scale at which the viscous forces dissipate energy takes the name of \emph{Kolmogorov scale}:
\begin{equation}
\eta = \left(\dfrac{\nu^3} {\varepsilon}\right)^{1/4}\label{eq:eta},
\end{equation}
where $\nu$ is the {kinematic} viscosity of the fluid, and
$\varepsilon$ is the time-average of the rate at which the energy is dissipated (see, e.g., \cite{frisch}). In order to 
formally define $\varepsilon$, let $[0, t_f]$ be the time interval of interest for the fluid flow. 
Then,
\begin{equation}
\varepsilon = \limsup_{t_f\to\infty} \dfrac{1}{t_f|\Omega|}\int_{0}^{t_f} \nu||\nabla\bu||^2_{L^2}~dt, \el
\end{equation}
where 
$|\Omega|$ denotes the measure of the domain.

For a flow in a developed turbulent regime, the dissipation rate has to be of the same magnitude as the production rate, which is the rate at which the 
{KE} is supplied to the small scales. A common way to express $\varepsilon$ is
in terms of the macro-scale variables associated with the flow. For this, let $U$ and $L$ be characteristic macroscopic speed and length, respectively. 
Then, we can write
$\varepsilon\sim U^3/L$ \cite{tennekes}. If we plug this into \eqref{eq:eta}, we obtain
\begin{equation}\label{eq:eta-re}
\eta \sim Re^{-3/4}L, 
\end{equation}
where $Re$ is the Reynolds number
\begin{equation}\label{eq:re}
Re = \frac{U L}{\nu}.
\end{equation}
The Reynolds number, which weights the inertial forces over the viscous forces, is often used as an indication of whether a flow is turbulent or not. Roughly speaking, at ``high'' Reynolds numbers inertial forces dominate and flows tend to be turbulent.


Equation~\eqref{eq:eta-re} explains the difficulty 
associated with the numerical solution of flows at high Reynolds numbers: 
in order to correctly capture the dissipated energy, 
one needs a grid with spacing $h\sim\eta$. This approach, called DNS,
leads in general to a large number of unknowns and prohibitive computational costs as the Reynolds number increases. If one tries to reduce the computational cost simply by making the mesh coarser (i.e., using a mesh with size $h$, which fails to resolve the Kolmogorov scale), {the energy physically dissipated at the small scales is not properly dissipated at the numerical level.}
The under-diffusion in the simulation leads to nonphysical computed velocities. In the best-case scenario,  nonphysical oscillations in the velocity field eventually make the simulation crash, which points to an obvious problem in the simulation set-up. In the worst-case scenario, the simulation does not crash and one accepts a nonphysical solution as good.

A possible way to reduce the computational cost without sacrificing accuracy is to
solve for the flow average using a mesh coarser than that required by DNS and introduce a model for the effects not captured by the flow average alone. Among such alternatives, 
RANS and LES are the most widely used for practical applications in science and engineering. In the RANS approach, none of the eddies in the flow 
are directly resolved from the flow equations, meaning that the effect of all eddies, from largest to smallest, is modeled. See Fig.~\ref{fig:energy_cascade}. 
In a sense, RANS could be considered the opposite extreme to DNS. RANS models are formally obtained from decomposing instantaneous quantities (velocity and pressure) into a time-averaged component and a fluctuating (over the average) component, and applying an averaging procedure to the governing equations. Extra  stresses and scalar fluxes emerging from this averaging procedure require closure, i.e., additional equations or assumptions to close the new system of equations.
In the LES approach, one resolves only the larger scales by using a ``coarse'' mesh (again with respect to DNS) and models the effect of the smaller scales that are not directly solved.
See Fig.~\ref{fig:energy_cascade}. 
LES models, which are a versatile compromise between DNS and RANS, are devised from applying a low-pass filter to the governing equations. 
Such a low-pass filtering can be viewed as a 
space-averaging. 

Traditionally, LES models introduce dissipation via momentum fluxes that are 
linearly dependent upon the rate of strain of the resolved scales. To clarify this, let us consider the incompressible Navier-Stokes equations (NSE): Denoting the velocity by $\bu$ and the pressure by $p$, the NSE read
\begin{align}
\rho\, \frac{\partial \bu}{\partial t} + \rho\,\nabla \cdot \left(\bu \otimes \bu\right)  - \nabla \cdot ( -p \mathbf{I} +2 \mu  \boldsymbol{\epsilon} (\bu) ) & = \bbf \quad \text{in } \Omega \times (0,t_f),\label{eq:NS1}\\
\nabla \cdot \bu & = 0\quad\, \text{in }\Omega \times(0,t_f),\label{eq:NS2}
\end{align}
where 
$\boldsymbol{\epsilon} (\bu) = (\nabla\bu + \nabla\bu^T)/2$ is the strain-rate tensor, $\mu$ is the dynamic viscosity, and $\bbf$ is a forcing term. 
{In real applications, appropriate initial and boundary conditions define the problem to solve, as we will see below.}
With a traditional LES model, one ultimately solves the following problem: find $\bu$ and $p$ such that
\begin{align}
\rho\, \frac{\partial \bu}{\partial t} + \rho\,\nabla \cdot \left(\bu \otimes \bu\right)  - \nabla \cdot ( -p \mathbf{I} +2 (\mu + \mu_a) \boldsymbol{\epsilon} (\bu) ) & = \bbf \quad \text{in } \Omega \times (0,t_f),\label{eq:LES1}\\
\nabla \cdot \bu & = 0\quad\, \text{in }\Omega \times(0,t_f),\label{eq:LES2}
\end{align}
where $\mu_a$ is an artificial viscosity introduced by the LES model. This is known as 
\textit{eddy viscosity closure}~\cite{sagaut2006large,BIL05}. {Different LES models arise for different definitions of $\mu_a$.}

One of the most widely used eddy viscosity models is the Smagorinsky model \cite{smagorinsky1963}. Introduced in the early 60s, it defines the artificial viscosity as:
\begin{align}
\mu_a = \rho (C_s \alpha)^2 \sqrt{ 2 \boldsymbol{\epsilon} (\bu) : \boldsymbol{\epsilon} (\bu)} , \quad C_s^2 = C_k \sqrt{\dfrac{C_k}{C_{\epsilon}}}, \label{eq:smago}
\end{align}
where $\alpha$ is a filter width (typically comparable with the mesh size), and $C_k$ and $C_{\epsilon}$ are model parameters. 
The values of such parameters are not universal and need to be tuned for each specific application.
The Smagorinsky model has enjoyed much success
for several reasons. It is relatively simple and computationally inexpensive compared to other LES models. It is easy to implement and can be used with a wide range of numerical methods, also thanks to the tunable parameters. However, it has one important limitation:
the assumption of local balance between 
the subgrid scale energy production and dissipation. Since such equilibrium conditions do 
not hold in many practical applications, the Smagorinsky model often results in over-diffusive simulations. 
A large body of research has been motivated by improving upon the Smagorinsky model.


Some alternative methods introduce artificial diffusion that can be solution-dependent (see, e.g., \cite{ABGRALL2001277,,klockner_warburton_hesthaven_2011,rispoliSaavedra2006,persson2006sub}) or residual-based (see, e.g., \cite{guermond2011entropy,guermondPasquetti2008,Guermond_Popov_2014,KURGANOV20128114,marrasNazarovGiraldo2015,wang2019entropy}), so that the artificial viscosity vanishes where the solution is smooth 
and/or decreases as the mesh is refined. 
Other methods add a set of equations to the discrete governing equations formulated on a coarse mesh.
This extra problem can be devised in different ways, for example, by a functional
splitting of the solved and unresolved scales
as in variational multiscale methods (see, e.g., \cite{bazilevsCaloCottrellHughesRealiScovazzi2007, codina2002,Codinaetal2017,hughesFeijoo1998}) or by 
resorting to the concept of ``suitability'' of weak solutions \cite{guermond2011suitable}. 
All these methods entail adding one or more extra terms to the governing equations, in the spirit of \eqref{eq:LES1}, and have had great accomplishments. 
However, one practical downside, in this case, is the need to implement ad hoc extensions that are not part of standard solvers, requiring access to legacy source codes (if not rewriting 
the solver). 
An attractive option is, therefore,  to add the extra problem sequentially to the fluid dynamics model, 
since in this case the implementation of the LES model would not require any major modification of a legacy solver. 
This can be achieved by applying a differential nonlinear low-pass filter to the (non-physical) solution computed with the fluid dynamics model and a ``coarse'' mesh.
Since this particular LES strategy, i.e., nonlinear spatial filtering, 
has already made an impact in ROM, below we focus on this approach. 
This will improve the clarity of our presentation on how to establish a two-way connection  between (i.e, ``bridging'') LES and ROM. 

\subsection{Nonlinear Spatial Filtering for LES}\label{sec:EFR}

Let us consider a generic fluid model 
{(e.g., the momentum equation of the NSE~\eqref{eq:NS1})}
\begin{equation}\label{eq:model}
    \frac{\partial \bU}{\partial t} + \cM(\bU) = \bf{0}
\end{equation}
and discretize it in time with, e.g., the Backward Euler scheme for simplicity:
\begin{equation}\label{eq:model_dt}
    \frac{\bU^{n+1} - \bU^{n}}{\Delta t} + \cM(\bU^{n+1}) = \bf{0}.
\end{equation}



If we further apply space discretization to \eqref{eq:model_dt} and use a mesh 
of size
$h \gg \eta$, the computed solution 
typically features nonphysical oscillations. Such oscillations can be attenuated or removed through a filtering operation. This is the idea at the core of a sequential algorithm called Evolve-Filter-Relax (EFR).
The EFR algorithm applied to \eqref{eq:model} reads:
\begin{itemize}
\item[-] \emph{Evolve}: find $\bV^{n+1}$ such that
\begin{align}
    \frac{\bV^{n+1} - \bU^{n}}{\Delta t} + \cM(\bV^{n+1}) = \bf{0}. \label{eq:EFR1}
\end{align}
For this step, one could adopt the same space discretiziation technique used for \eqref{eq:model_dt} and, hence, the same solver.
\item[-] \emph{Filter}:  find filtered variable $\overline{\bV}^{n+1}$ such that 
\begin{align}
\overline{\bV}^{n+1} = \cF \bV^{n+1},    \label{eq:EFR2}
\end{align}
where $\cF$ is a generic spatial filter. The solver for this step can be added as a module to the solver for \eqref{eq:model_dt}.
\item[-] \emph{Relax}: set 
\begin{align}
\bU^{n+1} = (1-\chi){\bV}^{n+1} + \chi \overline{\bV}^{n+1},\label{eq:EFR3}
\end{align}
with relaxation parameter $\chi \in [0, 1]$.
\end{itemize}
The appeal of the EFR algorithm lies not only in \emph{modularity}, but also in \emph{flexibility}, since there exists a large variety of spatial filters that one could use (e.g., Gaussian, box, sharp cutoff)~\cite{BIL05,layton2012approximate}.

\begin{figure}[h]
    \centering
    \includegraphics[width=7.0cm]{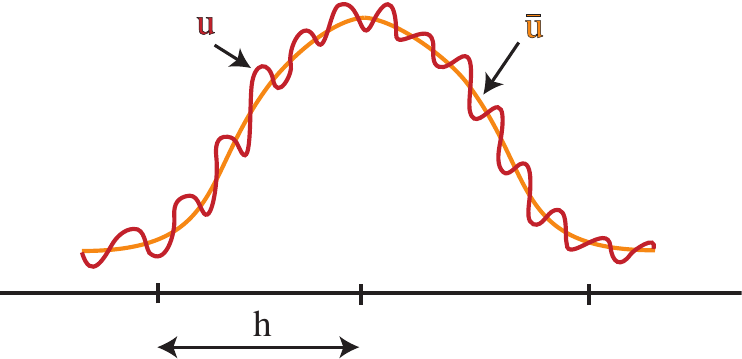} 
    \caption{LES schematic showing
            the input flow variable, $\bu$, that cannot be represented on a given coarse mesh, and
            the filtered flow variable, $\obu$, 
            that can be accurately represented on the coarse mesh.
            \label{fig:les-schematic}
            }       
\end{figure}

{The main idea of using spatial filtering can be illustrated with a schematic like in Figure~\ref{fig:les-schematic}:
The challenge is to represent a flow variable, e.g., the velocity, $\bu$ (represented by the maroon curve in Figure~\ref{fig:les-schematic}), on a given coarse mesh of size $h$.
As mentioned in Sec.~\ref{sec:intro},
having to work with a coarse mesh is generally the case in realistic engineering, geophysical, and, often, medical applications of convection-dominated flows.
As illustrated in Figure~\ref{fig:les-schematic}, if $\bu$ has small scale components of size below the mesh size, $h$, then these components will not be accurately represented on the given mesh.
More importantly, the inaccuracies in representing $\bu$ on the given coarse mesh are generally propagated and amplified in convection-dominated, nonlinear problems (e.g., the NSE), yielding spurious numerical oscillations.
To eliminate/alleviate these spurious oscillations, the central idea in LES is extremely simple:
Instead of trying to approximate all the spatial scales in $\bu$ on the given coarse mesh, try to approximate only the {\it } large scales in $\bu$.
This can be achieved by using a low pass spatial filter, $\cF$, to filter $\bu$.
The filtered flow variable, $\obu := \cF \bu$, which is represented by the orange curve in Figure~\ref{fig:les-schematic}, retains only the large spatial structures in $\bu$.
Thus, $\obu$ can be represented accurately on the given coarse mesh.
Furthermore, the spatial filtering eliminates the small scale, oscillatory component of $\bu$.
This, in a nutshell, is the guiding principle of LES~\cite{BIL05,sagaut2006large}.
}

The EFR algorithms, which 
are also known 
as filter stabilizations {(or regularizations~\cite{layton2012approximate})} because they reduce or eliminate unphysical fluctuations in the computed solution, can be categorized as eddy viscosity models for certain discretization choices  \cite{Olshanskii2013,clinco2023filter}. 
The connection between the EFR algorithm and LES modeling is easily seen by shifting the index $n+1$ to $n$ in \eqref{eq:EFR2}-\eqref{eq:EFR3} and plugging them into \eqref{eq:EFR1} to obtain:
\begin{align}
    \frac{\bV^{n+1} - \bV^{n}}{\Delta t} + \cM(\bV^{n+1}) + \frac{\chi}{\Delta t} (\cI -  \cF) \bV^{n} = \bf{0}, \label{eq:EFR1bis}
\end{align}
where $\cI$ is the identity operator.  Eq.~\eqref{eq:EFR1bis} gives us an implicit discretization of problem \eqref{eq:model} with the Backward Euler scheme and an additional,
explicitly treated, dissipation term (compare \eqref{eq:EFR1bis} with \eqref{eq:model_dt}). 
Let us assume that $\chi = \chi_0 \Delta t$,
where $\chi_0$ is a time-independent constant.
Then, \eqref{eq:EFR1bis} can be seen as a time-stepping scheme for problem:
\begin{equation}\label{eq:model_LES}
    \frac{\partial \bV}{\partial t} + \cM(\bV) + {\chi_0}(\cI -  \cF) \bV = \bf{0}.
\end{equation}
Thus, the EFR algorithm \eqref{eq:EFR1}-\eqref{eq:EFR3} can be interpreted as a splitting scheme for problem \eqref{eq:model_LES}. In turn, this can be read as problem \eqref{eq:model} with an additional term designed to be dissipative.

Stabilization based on linear filters \cite{Boyd1998283,Fischer2001265,mullen1999filtering} has been widely studied (see, e.g., \cite{Mathew2003,Visbal2002,B-garnier}). However, the breaking down of eddies into smaller ones until they get damped 
is a highly nonlinear process. Hence, it seems more appropriate to use nonlinear filters to select 
the eddies to be damped. 
A well-established \cite{GIRFOGLIO201927,bertagna2016deconvolution,Bowers2012,abigail_CMAME,layton2012modular,LAYTON20113183,viguerie2019deconvolution} nonlinear filter for
\eqref{eq:EFR2} is:
\begin{equation}\label{eq:FH}
    \cF = (\cI + \cL)^{-1},\quad \cL=-\nabla \cdot (\delta \nabla),  \quad \text{with} \
\delta =\alpha^2 a(\bV^{n+1})    
\end{equation}
where $\alpha$ can be interpreted as the \textit{filtering radius}
and $a(\cdot) \in (0,1]$ is the so-called \textit{indicator function}. This is a critical function that determines where and
how much eddy viscosity is added and it is crucial to the success of the EFR method.
We will discuss possible definitions of $a(\cdot)$ in Sec.~\ref{sec:ind_f}.
Notice that $\cF$ can be interpreted in view of the theory of maximal monotone operators and their Yosida regularized operators (see \cite{B-brezis}, Ch. 7). This provides a solid mathematical foundation to the methodology.

The rest of this section focuses on the application of the EFR algorithm to the incompressible NSE (in Sec.~\ref{sec:EFR_NS}) and the weakly compressible Euler equations (in Sec.~\ref{sec:EFR_Euler}). However, it could be easily applied to other fluid dynamics model, like, e.g, the one-layer
\cite{Girfoglio_JCP2023} and two-layer \cite{Besabe2024}
quasi-geostrophic equations.





\subsubsection{The EFR method for the incompressible Navier-Stokes equations}\label{sec:EFR_NS}



For the sake of completeness, let us consider the equations   \eqref{eq:NS1}-\eqref{eq:NS2} supplemented with initial conditions
and boundary conditions. Precisely, we are given initial data $\bu_0(\bx)$ such that
$$
\bu(\bx,0) = \bu_0(\bx), \quad \bx \in \Omega,
$$
and 
boundary data $\bu_D$ and ${\bf g}$ such that
\begin{align}
\bu(\bx,t) &= \bu_D (\bx, t), \quad \bx \in \Gamma_D, \label{eq:BC_D}\\
-p \bn + 2 \mu \epsilon(\bu) \cdot \bn &= {\bf g} (\bx,t), \quad ~~~ \bx \in \Gamma_N, \label{eq:BC_N}
\end{align}
where $\Gamma_D \cup \Gamma_N = \partial \Omega$, $\Gamma_D \cap \Gamma_N = \emptyset$, and $\bn$ is the outward unit vector normal to $\Gamma_N$.
A mix of Dirichlet and Neumann boundary conditions as in \eqref{eq:BC_D}-\eqref{eq:BC_N} occurs in many practical problems. 
Other possible boundary conditions (Robin type) are addressed in Remark \ref{eq:BC_R}.

Algorithm \eqref{eq:EFR1}-\eqref{eq:EFR3}, with filter as in \eqref{eq:FH}, applied to problem \eqref{eq:NS1}-\eqref{eq:NS2} reads:
\begin{itemize}
\item[-] \emph{Evolve}: find intermediate velocity $\bv^{n+1}$ and pressure $q^{n+1}$ such that
\begin{align}
\frac{\rho}{\Delta t}\, \bv^{n+1} + \rho\, \div \left(\bu^n \otimes \bv^{n+1}\right) - 2\mu\div \boldsymbol{\epsilon}(\bv^{n+1}) +\nabla q^{n+1}  
&= \bbf^{n+1} + \frac{\rho}{\Delta t}\, \bv^{n},
\label{eq:evolve-1.1}\\
\div \bv^{n+1} & = 0\label{eq:evolve-1.2},
\end{align}
with boundary conditions
\begin{align}
\bv^{n+1}(\bx) &= \bu_D (\bx, t^{n+1}), \quad \bx \in \Gamma_D, \\
-q^{n+1} \bn + 2 \mu \epsilon(\bv^{n+1}) \cdot \bn &= {\bf g} (\bx,t^{n+1}), \quad ~~~ \bx \in \Gamma_N.
\end{align}
Again, for simplicity, we have used the Backward Euler scheme for time discretization. Of course, other time discretization schemes are possible. Notice that, consistently with the time-discretization order, we have also linearized the problem with a first-order time extrapolation of the convective velocity. We also set $\bv^0 = \bu_0$.

\item \textit{Filter}:  find filtered velocity $\overline{\bv}^{n+1}$ and Lagrange multiplier $\lambda^{n+1}$ such that:
\begin{align}
 -\div \left( \delta \nabla\overline{\bv}^{n+1}\right) +\overline{\bv}^{n+1} +\nabla \lambda^{n+1} & = \bv^{n+1}, \label{eq:evolve-2.1}\\
\div \overline{\bv}^{n+1} & = 0 \label{eq:filter-2.2},
\end{align}
with $\delta = \alpha^2 a(\bv^{n+1})$ and $a(\cdot)$ is the indicator function.
In this case, we apply boundary conditions
\begin{align}
\overline{\bv}^{n+1}(\bx) &= \bu_D (\bx, t^{n+1}), ~ \bx \in \Gamma_D, \\
-\lambda^{n+1} \bn + 2 \alpha^2 a(\bv^{n+1}) \epsilon(\overline{\bv}^{n+1}) \cdot \bn &= {\bf 0},\quad \quad \quad \quad ~ \bx \in \Gamma_N. \label{eq:filter_BC_N}
\end{align}

The filtered velocity is introduced only in the time discrete problem, so we do not need initial conditions for it.
The choice of the boundary conditions for the filtered velocity 
is driven by the need for consistency with the original conditions. 

\item \textit{Relax}: set
\begin{align}
\bu^{n+1}&=(1-\chi)\bv^{n+1} + \chi\overline{\bv}^{n+1}, \label{eq:relax-1} \\
p^{n+1}&= q^{n+1}. \label{eq:relax-2} 
\end{align}
\end{itemize}
This particular version of the EFR method for incompressible
flows was first studied in \cite{layton2012modular}.

Note that filter problem \eqref{eq:evolve-2.1}-\eqref{eq:filter-2.2} can be conveniently rewritten as
a generalized Stokes problem. In fact, by multiplying eq.~\eqref{eq:evolve-2.1}
by $\rho/\Delta t$ and rearranging the terms, we obtain:
\begin{align}
\frac{\rho}{\Delta t} \overline{\bv}^{n+1}  - \div \left( \mu_a \nabla \overline{\bv}^{n+1}\right) + \nabla \overline{q}^{n+1} & = \frac{\rho}{\Delta t} \bv^{n+1}, \quad \mu_a = \rho \frac{\alpha^2}{\Delta t} a(\bv^{n+1}), \label{eq:filter-1.1}
\end{align}
where $\overline{q}^{n+1} = \rho \lambda^{n+1}/\Delta t$. 
Problem \eqref{eq:filter-1.1},\eqref{eq:filter-2.2} can be
seen as a time-dependent Stokes problem with a non-constant artificial viscosity $\mu_a$, discretized
by the Backward Euler scheme. Unlike $\delta$, which has the dimension of a length squared, $\mu_a$ is dimensionally a dynamic viscosity. The big advantage of this rewriting is that a solver for problem \eqref{eq:filter-1.1},\eqref{eq:filter-2.2} can then be obtained 
by adopting a standard linearized Navier-Stokes solver. 
Another advantage is that, unlike $\lambda^{n+1}$, $\overline{q}^{n+1}$ has the same dimensional 
units as $q^{n+1}$.
Hence, if one replaces \eqref{eq:evolve-2.1}-\eqref{eq:filter-2.2} with \eqref{eq:filter-1.1},\eqref{eq:filter-2.2}, it is possible to relax the pressure too. In \cite{bertagna2016deconvolution}, it was proposed to replace 
\eqref{eq:relax-2} with $p^{n+1} =(1-\chi)q^{n+1} + \chi\overline{q}^{n+1}$. 

\begin{remark}\label{eq:BC_R}
It can happen that a Robin boundary condition is enforced on part of the domain:
\begin{align}
-p \bn + 2 \mu \epsilon(\bu) \cdot \bn + \gamma \bu(\bx,t) = {\bf g}_R (\bx,t), \quad \bx \in \Gamma_R, \el
\end{align}
where ${\bf g}_R$ and $\gamma$ are given. 
This kind of boundary condition can arise in certain cases, such as domain decomposition algorithms (see, e.g., \cite{B-quarteroniv1}), or the use of ``transpiration'' techniques for fluid-structure interaction problems (see, e.g., \cite{transpiration}). The corresponding boundary condition for the evolve step is 
\begin{align}
-q^{n+1} \bn + 2 \mu \epsilon(\bv^{n+1}) \cdot \bn + \gamma \bv^{n+1} = {\bf g}_R (\bx,t^{n+1}), \quad \bx \in \Gamma_R, \el
\end{align}
while in the filter step - still driven by consistency arguments - one could impose
\begin{align}
-\overline{q}^{n+1} \bn + 2 \mu \epsilon(\overline{\bv}^{n+1}) \cdot \bn + \gamma \overline{\bv}^{n+1} = \gamma \bv^{n+1}, \quad \bx \in \Gamma_R. \el
\end{align}
\end{remark}

\begin{remark}
While eq.~\eqref{eq:filter-2.2} is required to prove that the solution of the EFR algorithm 
\eqref{eq:evolve-1.1}-\eqref{eq:relax-2} converges to the solution of the NSE
\eqref{eq:NS1}-\eqref{eq:NS2} \cite{layton2012approximate}, it can be disregarded in the implementation of the algorithm. Releasing constraint \eqref{eq:filter-2.2} leads to 
substantial simplification and computational time savings since there is one less variable
(i.e., the Lagrange multiplier to enforce the incompressibility constraint). In fact, one would solve a simplified filter problem: find filtered velocity $\overline{\bv}^{n+1}$ such that
\begin{align}
 -\div \left( \delta \nabla\overline{\bv}^{n+1}\right) +\overline{\bv}^{n+1} & = \bv^{n+1}, \label{eq:sim_filter}
\end{align}
As noted in \cite{Ervin2012}, the incompressibility is exactly  preserved by the simplified differential filter 
\eqref{eq:sim_filter} only for periodic conditions. 
In all the other cases, the end-of-step velocity $\bu^{n+1}$ does not strictly satisfy mass conservation.
However, it can be shown numerically that at the discrete level, the mass conservation error is very low (see, e.g., \cite{girfoglio2021pod}).
\end{remark}

EFR algorithms
have been well studied for the incompressible Navier-Stokes equations with a finite element  \cite{bertagna2016deconvolution,Bowers2012,abigail_CMAME,Ervin2012,layton2012modular,LAYTON20113183,Olshanskii2013}, spectral element \cite{Boyd1998283,Fischer2001265,mullen1999filtering}, 
or finite volume method \cite{GIRFOGLIO201927}.
They can be seen as 
splitting schemes for a nonlinear
version of the 
Leray model.
The original Leray model was introduced in 1934 as a theoretical tool to prove existence of weak solutions of the NSE~\cite{leray1934sur}.
As a computational tool, Leray regularization was first used in~\cite{geurts2003regularization} as a stabilization strategy for under-resolved simulations of turbulent flows with classical numerical discretizations~\cite{layton2012approximate}. 
It was also shown \cite{guermond2004mathematical,guermond2011entropy} that when a differential filter is used, the Leray model is similar to the NS-$\alpha$ model of Foias, Holm, and Titi~\cite{FHT2001}.
The EFR algorithm with nonlinear differential filters
was shown to provide numerical results that are more precise in localizing 
where the eddy viscosity is needed and are overall more accurate than results obtained with Smagorinsky-type models while featuring comparable computational efficiency \cite{Bowers2012,abigail_CMAME,Ervin2012,layton2012modular,LAYTON20113183}.
In \cite{bertagna2016deconvolution,GIRFOGLIO201927},
the results given by the EFR algorithm for certain filter choices were shown to be very accurate when compared against measurements 
for complex 3D incompressible flow problems.

\subsubsection{The EFR method for the weakly compressible Euler equations}\label{sec:EFR_Euler}

Despite the encouraging results for incompressible flows, 
EFR 
algorithms have been used to simulate
compressible flows \cite{clinco2023filter} only recently. 
While ideas similar to filter stabilization have been applied to the Euler equations \cite{Chehab2021,Hesthaven_Warburton, Euler_comp_orig, Euler_comp}, its full potential for compressible flows remains to be tapped into.

Below we apply the EFR algorithm \eqref{eq:EFR1}-\eqref{eq:EFR3} to the weakly compressible
Euler equations in the formulation most used in the literature of atmospheric simulations (see, e.g, \cite{giraldo_2008,marrasEtAl2013a,Marras2016} and references therein). Let us state such formulation first.

Let $\rho$ be the air density, $\bu$ the wind velocity, $p$ the atmospheric pressure, and $\theta$ the potential temperature. Temperature 
$\theta$ is a quantity of interest in meteorology and is
defined as $\theta = {T}/{\pi}$, where 
 $T$ is the absolute temperature and $\pi$ is the Exner pressure.
Since one is typically interested in fluctuations of $\rho$ and $p$ with respect to a
constant (in time) and hydrostatically balanced
background state $\rho_0$ and $p_0$, we write $p = p_0 + p'$ and
$\rho = \rho_0 + \rho'$, with $\nabla p_0 + \rho_0 g \widehat{\bk} = \boldsymbol{0}$, where $g$ is the gravitational constant
and $\widehat{\bk}$ is the unit vector aligned with the vertical axis $z$. See, e.g., \cite{marchukBook1974} for more details on this. 
Then, the conservation of mass, momentum, and potential temperature in spatial domain $\Omega$ and
over a time interval $(0,t_f]$ can be written as:
\begin{align}
&\frac{\partial \rho}{\partial t} + \nabla \cdot (\rho \bu) = 0 &&\text{in } \Omega \times (0,t_f], \label{eq:mass}  \\
&\frac{\partial (\rho \bu)}{\partial t} +  \nabla \cdot (\rho \bu \otimes \bu) + \nabla p' + \rho' g \widehat{\bk} = \boldsymbol{f_u}  &&\text{in } \Omega \times (0,t_f]. \label{eq:momeCE2} \\
&\frac{\partial (\rho \theta)}{\partial t} +  \nabla \cdot (\rho \bu \theta) = f_\theta &&\text{in } \Omega \times (0,t_f], \label{eq:energyCE3}
\end{align}
where $\boldsymbol{f_u}$ and $f_\theta$ are possible forcing terms. We close system \eqref{eq:mass}-\eqref{eq:energyCE3} 
with a thermodynamics equation of state for $p$:
\begin{align}
p = p_g\left({\rho R \theta}/{p_g}\right)^{{c_p}/{c_v}},
\label{eq:p2}
\end{align}
where $p_g = 10^5$ Pa is a standard pressure, $R$ is the specific gas constant of dry air, and $c_{p}$ (respectively, $c_{v}$) is the specific heat capacity at constant 
pressure (respectively, volume).
Obviously, system \eqref{eq:mass}-\eqref{eq:p2} has to be supplemented with proper initial and boundary conditions. 
However, we do not specify them as they would be treated as explained in Sec.~\ref{sec:EFR_NS}.
Other perspectives on this are included in Sec.~\ref{sec:com_op}.

EFR algorithm \eqref{eq:EFR1}-\eqref{eq:EFR3}, with filter as in \eqref{eq:FH}, applied to problem \eqref{eq:mass}-\eqref{eq:p2} after time discretization by the Backward Euler scheme reads:
\begin{itemize}
\item[-] \emph{Evolve}: find density $\rho^{n+1}$, density fluctuation $\rho'^{,n+1}$ and intermediate variables  $\bv^{n+1}$, $q^{n+1}$, $q'^{,n+1}$, and $\beta^{n+1}$ 
such that:
\begin{align}
& \frac{\rho^{n+1} - \rho^n}{\Delta t} + \nabla \cdot (\rho^{n+1} \bv^{n+1}) = 0, \label{eq:mass_td}  \\
&\frac{\rho^{n+1} \bv^{n+1} - \rho^{n}\bu^n}{\Delta t} +  \nabla \cdot (\rho^{n+1} \bv^{n+1} \otimes \bv^{n+1}) + \nabla q'^{,n+1} + \rho'^{,n+1} g \nabla \rho^{n+1} = \boldsymbol{f}_{ \boldsymbol{u}}^{n+1},  \label{eq:mom_td} \\
&\frac{\rho^{n+1} \beta^{n+1} - \rho^{n}\theta^n}{\Delta t} +  \nabla \cdot (\rho^{n+1} \bv^{n+1} \beta^{n+1})  = \boldsymbol{f}_{\theta}^{n+1},  \label{eq:theta_td} \\
&q^{n+1}  = p_g\left({\rho^{n+1} R \beta^{n+1} }/{p_g}\right)^{{c_p}/{c_v}}, \label{eq:q_td} \\
&q^{n+1} = p_0 + q'^{,n+1}, \label{eq:qprime_td} \\
&\rho^{n+1} = \rho_0 + \rho'^{,n+1}. \label{eq:rho_td}
\end{align}
\item[-] \emph{Filter}: find filtered variables $\overline{\bv}^{n+1}$ and $\overline{\beta}^{n+1}$  such that 
\begin{align}
-\nabla \cdot (\delta_v \nabla(\overline{\bv}^{n+1})) + \overline{\bv}^{n+1} = \bv^{n+1},  \quad \delta_v = \alpha^2 a(\bv^{n+1}),
\label{eq:fv} \\
-\nabla \cdot (\delta_\beta \nabla(\overline{\beta}^{n+1})) + \overline{\beta}^{n+1} = \beta^{n+1},  \quad \delta_\beta = \alpha^2 a(\beta^{n+1}). \label{eq:fbeta}
\end{align}
\item[-] \emph{Step 3 - Relax}: find end-of-step ${\bu}^{n+1}$, $p^{n+1}$, $p'^{,n+1}$, and $\theta^{n+1}$  
such that
\begin{align}
&\bu^{n+1} = (1-\chi){\bv}^{n+1} + \chi \overline{\bv}^{n+1}, \label{eq:R1}\\ &\theta^{n+1} = (1 -\xi){\beta}^{n+1} + \xi \overline{\beta}^{n+1}, \label{eq:R2}\\
&p^{n+1}  = p_g\left({\rho^{n+1} R \theta^{n+1} }/{p_g}\right)^{{c_p}/{c_v}}, \label{eq:R3}\\
&p^{n+1} = p_0 + p'^{,n+1}, \label{eq:end-of-s-p}
\end{align}
with relaxation parameters $\chi, \  \xi \in [0, 1]$.
\end{itemize}

This particular version of the EFR method for compressible
flows was introduced in \cite{clinco2023filter}. Therein, the same filter was applied to the intermediate velocity and intermediate potential temperature. However, different variables have different physical properties and, thus, different types of filtering may be a better choice.

Similarly to what done for \eqref{eq:evolve-2.1}, one can multiply \eqref{eq:fv} by $\rho^{n+1}/\Delta t$ to obtain a Stokes-like problem:
\begin{equation}
    \frac{\rho^{n+1}}{\Delta t} (\overline{\bv}^{n+1}-\bv^{n+1}) -\nabla \cdot (\mu_a\nabla \overline{\bv}^{n+1}) = \boldsymbol{0},\quad \mu_a = \rho^{n+1}\frac{\alpha^2}{\Delta t}a(\bv^{n+1}), \label{eq:ufil}
\end{equation}
where $\mu_a$ is dimensionally a dynamic viscosity, i.e., a proper eddy viscosity.

\subsubsection{Indicator function}\label{sec:ind_f}

Once the filter is selected, 
the success of the EFR algorithm depends on the reliability of the indicator function. 
This function should take values close to zero where its argument (i.e., velocity 
or potential temperature) does not need regularization, and close to 1 where the argument 
needs to be regularized.
Linear filters, which are equivalent to setting
\begin{equation}\label{eq:a_lin}
a(\bv)= 1, 
\end{equation}
have limited efficacy, since they introduce the same amount of regularization everywhere in the domain and likely lead to overdiffusion. Much attention has been devoted to devising accurate indicator functions. 

Before presenting a brief survey of the indicator functions available in the literature, we remark that one can recover a Smagorinsky-like model from the EFR algorithm by taking the following indicator function \cite{borggaard2009bounded}:
\begin{equation}
a_S(\bv) = \frac{| \nabla \bv |}{\|\nabla \bv\|_{\infty}}. \label{eq:a_smago}
\end{equation}
This indicator function is convenient because of its strong monotonicity  properties. However, it is known to introduce excessive diffusion, just like \eqref{eq:a_lin}. In fact, notice that $a_S$ takes large values in the regions of the domain with shear flow, which is not turbulent.

Below, we group indicator functions into two categories (physical phenomenology-based and mathematics-based) and present their strengths and limitations.

\vskip 2mm
\noindent \underline{Physical phenomenology-based indicator functions}. 
In \cite{layton2012modular}, the authors propose indicator functions based on physical quantities that
are known to vanish for coherent flow structures, i.e., structures that persist in their form for ``long'' periods of time. 
One of the most popular methods to identify coherent vortices is the $Q$ criterion \cite{O-hunt1988},
which classifies as persistent and coherent vorteces those regions where:
\begin{equation}
Q(\bu, \bu) = \frac{1}{2} (\nabla^{s} \bu : \nabla^{s} \bu - \boldsymbol{\epsilon} (\bu) : \boldsymbol{\epsilon} (\bu)) > 0, \el
\end{equation}
where $\nabla^{s} \bu$ is the spin tensor (skew-symmetric part of $\nabla \bu$):
\begin{equation}\label{eq:gradss}
\nabla^{s} \bu = \nabla \bu - \boldsymbol{\epsilon} (\bu) =  \frac{1}{2} (\nabla\bu - \nabla\bu^T). 
\end{equation}
It is easy to see that $Q > 0$ in the regions of the domain where spin dominates deformation.
An indicator is obtained by rescaling $Q(\bu, \bu)$ so that the condition $Q(\bu, \bu) > 0$ implies $a(\bu) \simeq 0$, which means that regularization is not needed. A possibility is the following:
 \begin{align}
 a_Q (\bu) = \frac{1}{2}- \frac{1}{\pi} \arctan \left( \epsilon^{-1} \frac{Q(\bu, \bu)}{|Q(\bu, \bu)|+ \epsilon^2} \right), \el
 \end{align}
where $\epsilon$ is a parameter introduced to avoid the denominator vanishes.
 
A second indicator uses an eddy viscosity coefficient formula proposed by Vreman \cite{Vreman2004} that vanishes for 320 types of flow structures known to be coherent. The Vreman-based indicator function reads
\begin{align}
a_V(\bu) = \sqrt{\frac{B(\bu)}{| \nabla \bu|_F^4 }}, \el
\end{align}
where the subindex $F$ refers to the Frobenius norm and $B(\bu)$ is defined as 
\begin{align}
B = \beta_{11} \beta_{22} - \beta_{12}^2 + \beta_{11} \beta_{33} - \beta_{13}^2 + \beta_{22} \beta_{33} - \beta_{23}^2, \quad \beta_{ij} (\bu) 
= \sum_{m = 1,2,3} \frac{\partial u_i}{\partial x_m}\frac{\partial u_j}{\partial x_m}. \el
\end{align}
Since $0 \leq B(\bu)/| \nabla \bu|_F^4 \leq 1$, $a_V(\bu) \in [0, 1]$. The Vreman-based indicator function was shown to be successful in \cite{Bowers2012}.

Another physics-based indicator function uses the relative helicity density $RH$, which is a local quantity related to the macroscopic helicity $H$:
\begin{align}
RH = \frac{\bu \cdot \bw}{|\bu| |\bw|}, \quad
 H = \frac{1}{| \Omega |} \int_\Omega \bu \cdot \bw ~d \Omega, \el
\end{align}
where $\bw = \nabla \times \bu$ denotes vorticity. From the NSE in rotational form, it is possible to see that local high helicity suppresses local turbulent dissipation caused by breakdown of eddies into smaller ones. 
A helicity-based indicator can be created by ensuring that values of $RH$ near one imply $a(\bu) \simeq 0$, i.e.:
\begin{align}
a_H(\bu) = 1 - \left| \frac{\bu \cdot \bw}{|\bu| |\bw| + \epsilon^2} \right|, \el
\end{align}
where again $\epsilon$ is a parameter tha prevents the denominator from vanishing.

Notice that other (more selective) indicator functions can also be obtained by taking the geometric average of two (or more) indicator functions.

All these physical phenomenology-based indicator functions have the advantage of requiring only algebraic operations on $\bu$ and its derivatives. Hence,  their implementation may be quite straightforward. 
However, the major drawback is that they do not allow for a rigorous convergence theory to verify the robustness of the associated filtering method. This limitation is overcome by the mathematics-based (rather than physics-based) indicators discussed next.

\vskip 2mm

\noindent \underline{Mathematics-based indicator functions}.
In order to define a class of mathematics-based indicator functions, let us recall a basic result about linear filters $\cF$, which are invertible, self-adjoint, compact operators from a Hilbert space $V$ to itself. From the spectral theorem (see, e.g., \cite{B-brezis}), we have:
\begin{align}
\cF\bx = \sum_{i=0}^\infty \lambda_i \langle \bx,\be_i \rangle \be_i,\qquad
\cF^{-1}\by = \sum_{i=0}^\infty \dfrac{1}{\lambda_i} \langle \by,\be_i \rangle \be_i, \el
\end{align}
where $\lambda_i$ are the eigenvalues of $\cF$, and $\be_i$ are the corresponding eigenfunctions, which form an orthonormal basis for $V$.
Since $\cF$ is compact,  its eigenvalues accumulate to $0$ and the inverse operator $\cF^{-1}$ is unbounded.  

Let $\cD$ be a bounded regularized approximation of $\cF^{-1}$, whose action on $\by$ is given by
 \begin{equation*}
 \cD\by = \sum_{i=0}^\infty\phi\left(\dfrac{1}{\lambda_i}\right)\langle \by,\be_i\rangle\be_i 
 \end{equation*}
 with 
\begin{equation*}
 \phi\left(\dfrac{1}{\lambda_i}\right)\simeq
 \begin{cases}
 \dfrac{1}{\lambda_i}\qquad &\mbox{ if } i \mbox{ is ``small''},\\
 0 \qquad & \mbox{ if } i \mbox{ is ``large''}.
 \end{cases}
 \end{equation*}
Then, by composing $\cF$ and $\cD$ we can obtain a low-pass filter, so that:
 \begin{equation*}
 \|\bx-\cD\cF\bx\| \mbox{ is }
 \begin{cases}
 \mbox{small} \qquad & \mbox{ if } \bx \mbox{ is ``smooth''},\\
 \mbox{large} \qquad & \mbox{ if } \bx \mbox{ is not ``smooth''},
 \end{cases}
 \end{equation*}
 where ``smooth'' is intended with respect to the eigenfunction of the operator $\cF$. This means that $\bx$ is smooth if $\langle \bx,\be_i\rangle$ is significantly different from zero only for small values of $i$. The composition of the two operators $\cF$ and $\cD$ can be interpreted as a low-pass filter. This motivates the definition of indicator function
\begin{align}
a_{\cD}(\bu) = \frac{\left|  \bu - \cD (\cF(\bu)) \right|}{\left| \left| \bu - \cD (\cF(\bu)) \right| \right|_\infty}. \label{eq:a_deconv_scaled}
\end{align}

A well-studied choice for the linear filter $\cF$ in \eqref{eq:a_deconv_scaled}
is the linear Helmholtz filter, i.e., \eqref{eq:FH} with $\delta$ constant in space and time. Let us denote it with $\cF_H$.
A popular choice for $\cD$ is the Van Cittert deconvolution operator $\cD_N$, defined as
\begin{equation}
\cD_N = \sum_{n = 0}^N (\cI - \cF)^n. \el
\end{equation}
Note that the evaluation of $a_{\cD}$ with $\cD=\cD_N$ requires to apply the linear filter $\cF$ a total of $N+1$ times. This is an argument in favor of limiting the value of $N$, since large $N$ would increase the computational cost. Another argument is provided by interpreting $\cD_N$ as the $N$-th iteration of a Richardson scheme to solve the filter problem. See \cite{bertagna2016deconvolution} for more details. In practice,  one takes $N = 0, 1$, corresponding to $\cD_0=\cI$ and $\cD_1 = 2\cI - \cF$, respectively. For these choices of $N$, the indicator function \eqref{eq:a_deconv_scaled} becomes
\begin{align}
a_{\cD_0}(\bu) = \frac{\left|  \bu - \cF(\bu) \right|}{\max(1,\left| \left| \bu - \cF(\bu) \right| \right|_\infty)}, \quad a_{\cD_1}(\bu) = \frac{\left|  \bu - 2 \cF(\bu) + \cF(\cF(\bu)) \right|}{\max(1,\left| \left| \bu - 2 \cF(\bu) + \cF(\cF(\bu)) \right| \right|_\infty)}. \label{eq:a_D0_a_D1}
\end{align}

Indicator function \eqref{eq:a_deconv_scaled} with $\cD=\cD_N$ and $\cF=\cF_H$, which we will refer to as \emph{approximate deconvolution} (AD) indicator function, 
has been proposed in \cite{abigail_CMAME}. However, the idea of using van Cittert approximate deconvolution in fluid models to increase accuracy is well established and has strong mathematical foundations \cite{Stolz1999,Stolz2001,Dunca2005}.
It is possible to prove \cite{Dunca2005} that
\begin{align}
\phi - D_N (F_H(\phi)) = (-1)^{N+1} \delta^{2N+2} \Delta^{N+1} \cF_H^{N+1} \phi. \label{eq:duncan_eps}
\end{align}
From this, it follows that, indeed, $a_{\cD_N}(\bu)$ takes values close to zero in the regions of the domain where $\bu$ is smooth. 

In addition to allowing for rigorous mathematical analysis, the AD indicator function has been shown to be more selective than indicator $a_S(\cdot)$ \eqref{eq:a_smago} in identifying the regions of the domain where diffusion is needed for both incompressible \cite{abigail_CMAME} and compressible flows \cite{clinco2023filter}. Hence, the EFR algorithm with indicator function $a_{\cD}(\cdot)$ corrects the overdiffusivity of $a_S(\cdot)$.
For incompressible flows, $a_{\cD}(\cdot)$ has also been shown to be particularly accurate for realistic 3D problems
\cite{bertagna2016deconvolution,viguerie2019deconvolution,GIRFOGLIO201927,xu2018coupled}.

\subsection{Machine learning for LES}


Recent years have seen an escalating interest in leveraging machine learning techniques for CFD, as elucidated in \cite{duraisamy2019turbulence,brunton2020machine}. 
This subsection is devoted to highlighting some opportunities offered 
by machine learning for LES modeling, starting with a literature review.
Noteworthy endeavors within various research groups have demonstrated the feasibility of training deep learning-based closure models \cite{ling2016reynolds,BECK2019108910,beck2021perspective,sirignano2023deep}, while others have focused on the acquisition of closure models through symbolic regression techniques \cite{schmelzer2020discovery,reissmann2021application,wu2023enhancing}. It is imperative to underscore that a relatively underexplored domain is the integration of blending and classification methodologies, particularly utilizing indicator functions as presented in the previous subsection.

There is an existing body of knowledge detailing the efficacy of various closure models for specific types of flows. Furthermore, current efforts are marked by a growing emphasis on generating extensive turbulent flow datasets \cite{ren2023superbench}. However, a critical gap remains in establishing comprehensive maps between underlying patterns and flow structures, along with the identification of the best-suited, verified, tested, and mathematically analyzed closure models.
On a more granular level, once trained, the same maps can also be used to find the most suitable filters or filtering parameters. This approach allows the utilization of indicator functions within certain latent spaces to accurately identify a low-pass filter or its associated shape and parameters, which can then be applied to mainstream closure approaches. In essence, machine learning tools can be leveraged not only to identify a filter but also to build a hybrid closure model atop it. Furthermore, it is important to note that references \cite{maulik2019sub,maulik2020spatiotemporally} demonstrate how a single hypothesis or map can be used to create a simple selection mechanism. This mechanism proves valuable for choosing between different structural and functional closures or even various numerical methods.

\begin{figure}
    \centering
    \includegraphics[width=0.95\textwidth]{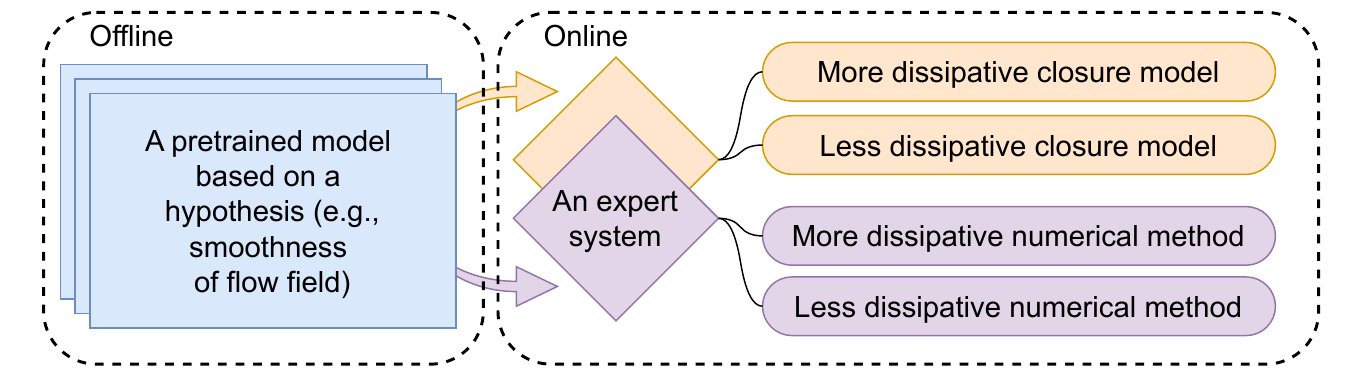} 
    \caption{Schematic of the concept proposed in \cite{maulik2020spatiotemporally}.
            \label{fig:expertsystem}
            }       
\end{figure}

A hypothesis segregation concept proposed in \cite{maulik2019sub} uses functional and structural closure models, and in \cite{maulik2020spatiotemporally} tests upwind and central numerical methods. The key idea is to train a modular neural network for structure identification offline and use that model with defining an expert systems to adjust underlying closure models and numerical approaches in a weighted manner temporary or spatially online as needed. A key aspect is the modular nature of offline-online split approaches, where the same trained model defines filter strengths, radius, and shapes for use in a structural model. See Fig.~\ref{fig:expertsystem} for a schematic representation. 

We also advocate for a judicious use of machine learning tools to facilitate the generation of diverse maps within the latent space \cite{kim2019deep,kontolati2023learning,du2024confild}. Such learned maps serve as foundational elements for constructing an expert system capable of discerning flow regimes and their associated closure models from an array of existing models. This approach offers a departure from the conventional approach of relying purely on
black-box closure models or resource-intensive symbolic regression practices.

By harnessing the pattern and structure recognition capabilities inherent in machine learning, one can effectively employ existing structural or functional closure models in a dynamic and adaptive manner. For example, an approach with two types of artificial neural networks for ML has been proposed \cite{lozano2023machine}: a classifier to identify the contribution of each building block in the flow and a predictor to estimate wall shear stress by combining these building-block flows. Such approaches with structure recognition capabilities are poised to enhance the efficiency and accuracy of flow regime identification, thereby presenting a compelling alternative to traditional methodologies. 
The use of flow physics-based indicator functions brings the additional
advantage of exploiting existing structural and functional knowledge in a dynamic and adaptive framework. We conclude this section by referring to recent survey articles that provide further insights into machine learning for turbulence modeling \cite{duraisamy2021perspectives,vadrot2023survey}.


\section{Applications of LES for FOM}\label{sec:app_FOM}

\subsection{Incompressible flows}\label{sec:inc_FOM}

The use of LES models for incompressible flows covers a wide range of applications. 
Among these, we want to mention here biomedical applications, with particular reference to computational hemodynamics.
This is a field of CFD that has progressively received more attention when the availability of patient-specific data
and reliable methods of image processing enabled numerical methods from the benchmark to the bedside/operating room stage \cite{veneziani2013inverse}.
Using CFD in the clinical routine for the assessment of cardiovascular diseases - and sometimes also for the planning of 
therapeutic options - is nowadays a consolidated practice, with some outstanding examples involving entrepreneurial activities \cite{taylor2013computational,taylor2023patient}.
This circumstance clearly raises some specific challenges that go beyond the mathematical modeling stage, involving the efficiency required by clinical timelines and quality certification/uncertainty quantification.

In general, human blood flow does not feature a turbulent regime in most of the vascular districts of interest. However, in some cases,
highly disturbed flow is possible either for physiological reasons (like in the ascending aorta during the systolic phase, when the aortic valve is open and the heart pumps blood into the system) or for pathological conditions, like in the presence of significant stenoses. In some cases, turbulence may be induced by devices required by therapeutical options. In this section, we focus on an application involving a pathological condition called Type B Aortic Dissection (AoD) because it highlights the practical usefulness of the EFR method in a complex real-life setting and the associated challenges.

\begin{figure}[h!]
    \centering
    \includegraphics[width=0.5\linewidth]{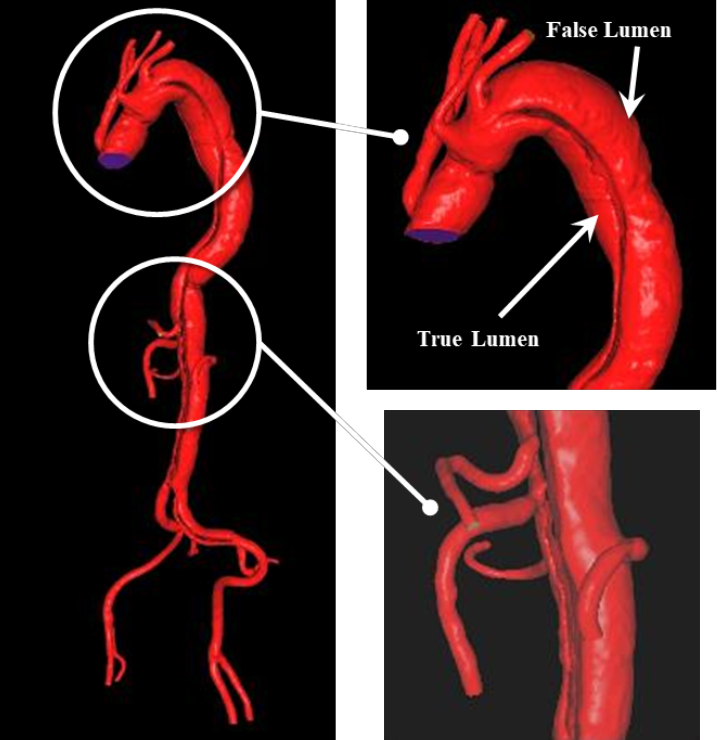}
    \caption{Images of a patient-specific AoD showing the true lumen and the false lumen.}
    \label{fig:AoD1}
\end{figure}

\subsubsection{Computational Hemodynamics in Type B Aortic Dissections}\label{sec:AoD}

AoD is a pathology consisting of 
a splitting of the natural lumen of the descending aorta into two separate channels: {true lumen} and false {lumen} - see Fig. \ref{fig:AoD1}. 
This splitting may have different reasons, including traumatic ones, and  represents a danger for the patient: 
the tissue of the two lumens is weaker than the physiological one, so it may break,
causing life-threatening hemorrhages. Clinically, there is a strong interest in predicting the evolution of the false lumen in time, in 
particular, whether it remains stable or grows, the latter condition being the most dangerous for the patient.
While drugs are an option for a stable false lumen, surgery is the usual choice for cases in which growth is expected. Since surgery is invasive, it should be considered only when necessary.

\begin{figure}
    \centering
    \includegraphics[width=0.75\linewidth]{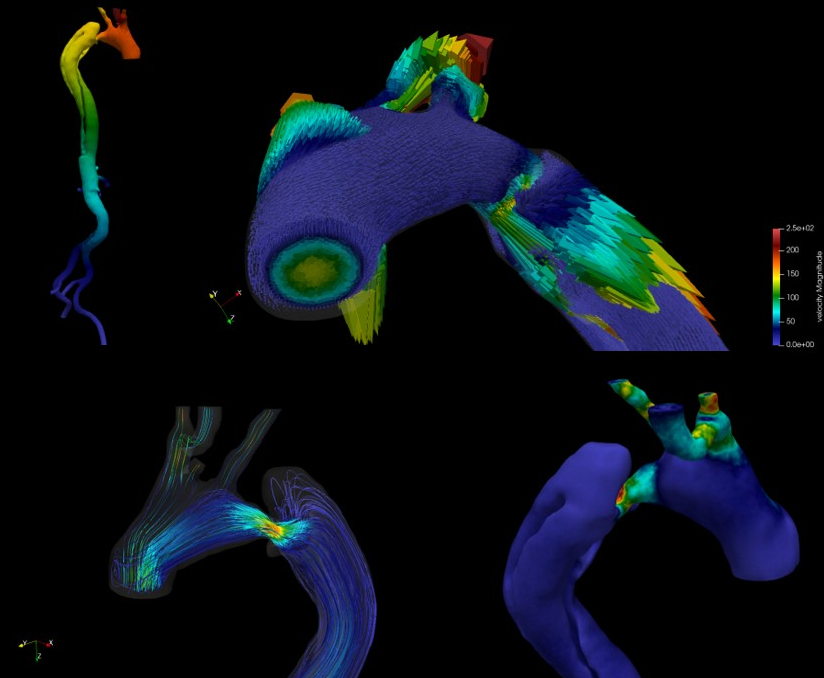}
    \caption{Simulation in a patient-specific AoD. Top left: pressure. Top right: velocity (in $cm/s$) in the descending 
    aorta and at the entrance of the false lumen. The two bottom panels outline the complexity of the flow induced 
    by the entry tear for the velocity (left) and the Wall Shear Stress (right).
}
    \label{fig:AD_JY}
\end{figure}

Currently, decisions are 
made only on the geometry of the dissection. 
The addition of hemodynamics information, in particular of quantities that are difficult or impossible to measure in vivo, is anticipated to improve significantly the reliability of the predictions 
\cite{xu2018coupled}. In particular, the Wall Shear Stress (WSS), i.e., the tangential component of the normal stress at the wall, may play a significant role for prognostic purposes. It is defined as
\begin{equation}\label{eq:wss}
\boldsymbol{\tau} = 2\mu \boldsymbol{\epsilon}(\bu) \cdot \bn -
2\mu \left(\bn \cdot \boldsymbol{\epsilon}(\bu) \cdot \bn\right) \bn \qquad {\rm on} \quad \Gamma, 
\end{equation}
where $\Gamma$ is a portion of $\partial \Omega$ of interest and $\bn$ is its outward normal unit vector. 
Hence, CFD of aortic dissections may become a critical component in the clinical routine. 
However, 
the complexity of patient-specific geometries and the high convective fields induced by the small entry tears between the true and the false lumen (see Fig. \ref{fig:AD_JY})
pose significant challenges. It is important to stress that, in clinical applications,
there is the need to run many patient-specific cases, as required by the concept of {\it Computer Aided Clinical Trials} (CACT), to get statistically significant data. This means that beyond accuracy, computational efficiency is a critical aspect. In the case of AoD, a one-shot simulation might be possible with DNS, however the computational cost is not compatible with the large number of patients required by CACT.
The trade-off between efficiency and accuracy - where accuracy has to be intended in the clinical sense, 
i.e., the quantification of indexes to enable correct clinical decisions - is of primary importance, and the EFR scheme, in this context, seems to be the ideal approach.

\begin{figure}[hbtp]
    \centering
    \includegraphics[width=\linewidth]{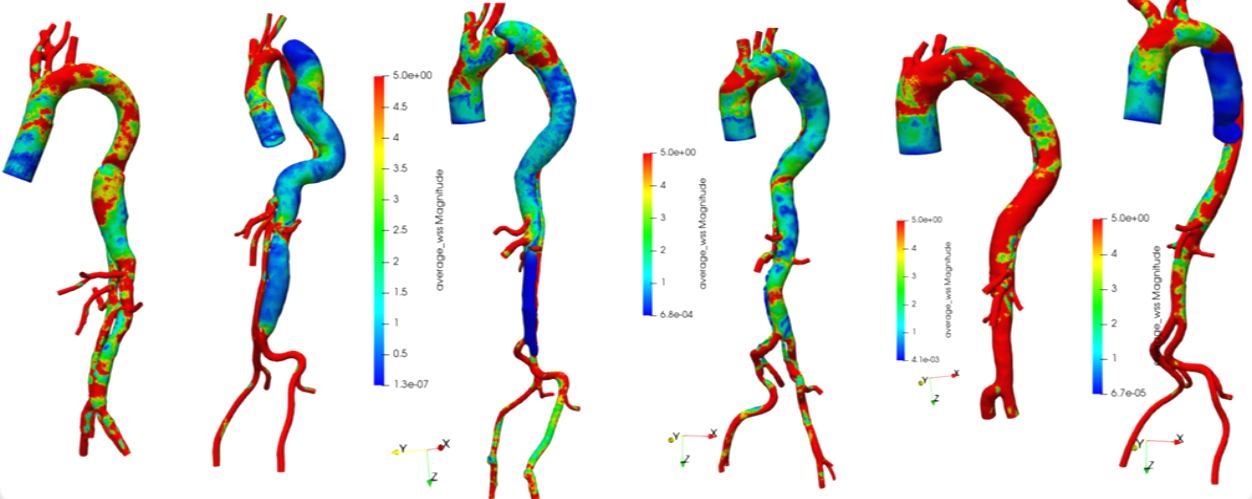}
    \caption{Anatomies of several AoDs, pinpointing the diversity of the possible morphologies. Geometries reconstructed at Emory University with Vascular Modeling ToolKit - \cite{xu2020thesis}.}
    \label{fig:AoD2}
\end{figure}

\paragraph{Geometry.} The geometries of AoD may be significantly diverse - see Fig. \ref{fig:AoD2}. The location of the first tear connecting the true and the false lumen (
also known as Primal Entry Tear - PET) has a critical importance. In fact, clinical guidelines suggest that the evolution of the dissection depends on the distance between the PET and the left subclavian artery. A short distance generally indicates the evolution into complicated clinical cases requiring surgery \cite{weiss2012location}.
However, there are cases with a distal PET evolving in cases to be surgically operated \cite{xu2018coupled}.
For certain sites of interest in the cardiovascular system (e.g., the coronary arteries), the geometry can be generally templatized, i.e., the patient-specific anatomy can be regarded as a map or a minor modification of a template geometry.
In AoDs, it is very hard to find a realistic template. While AI-based methodologies can be used for the segmentation and reconstruction of coronaries,
at this time, the classical level-set method is the method of choice for complex geometries like AoD. In fact, the lack of sufficiently large data sets prevents reliable training of data-based
methodologies, like neural networks. 
The geometries presented in this paper
were reconstructed with the Vascular Modeling ToolKit \cite{izzo2018vascular}, a free, open-source tool based on the level-set method, using images from Computer Tomography. The segmentation was carried out by an expert, trained 
for the identification of the tears.  The mesh was successively obtained by the NetGen mesh generator, with an in-home Python script to guide the refinement of the mesh close to the boundaries (the so-called {\it mesh boundary layers}). 
The introduction of a mesh more refined close to the boundary is a common practice in biomedical applications whenever one of the main quantities of interest is 
computed at the boundary, like the WSS. The meshes we deemed to be appropriate for simulations of AoD with EFR turbulence modeling have a number of elements in the range 
$6.0e5-1.0e6$

\paragraph{Boundary conditions and backflows.}
The patient-specific domain $\Omega$ generally features physical boundaries 
$\Gamma_w$ represented by the arterial walls and artificial boundaries represented by the regions where the blood enters ($\Gamma_{in}$) and leaves ($\Gamma_{out}$)  the region of interest $\Omega$.  The arterial walls include also the flow divider separating the true and the false lumen. On the wall, we postulate the non-slip condition prescribing the continuity of the velocity field of the wall and the blood.
Since we assume a rigid structure, this leads to the homogeneous Dirichlet condition 
$$
\bu = {\bf 0}, \qquad {\rm on} \quad \Gamma_w.
$$
The role of fluid-structure interaction, in particular at the flap separating the two lumens, has recently attracted the attention of the biomedical engineering community \cite{baumler2020fluid,zimmermann2023hemodynamic}.
We considered rigid structures for simplicity. However, nothing prevents the extension of the framework presented here to the case of compliant structures (see, e.g., \cite{Girfoglio_CAIM2022}).

When considering the artificial inflow/outflow portions of the boundaries, 
computational hemodynamics applied to clinical problems generally suffers from a long-term condition, sometimes called the ``defective boundary conditions problem'' \cite{heywood1996artificial}
or, in the specific case of medical applications, 
``image-legacy problem''. Medical doctors were used to keep track of images but not of possible hemodynamic conditions to be used as boundary data! This means that we generally have patient-specific geometrical data but with a substantial lack, if not absence, of patient-specific boundary conditions. To leverage 
the abundance of geometrical data, many different approaches have been considered in the literature to fill this gap. Standard procedures consider (a) literature-based data for flow rates at the inflow, with an empirical selection of velocity profiles fitting these data; and (b) special boundary conditions at the outflow, based on a lumped parameter modeling of the peripheral circulation. We briefly detail these conditions hereafter.
\begin{enumerate}

\item {\it Inflow conditions}. The available data may generally refer to the flow rate or to the velocity (as a function of time) at one point of the inflow. While this 
is clearly not enough as a boundary condition, a popular and practical approach consists of assuming a velocity profile constructed around the available data. For instance, one can assume a parabolic profile (corresponding to the well-known Poiseuille solution) or a Womersley profile (corresponding to the time-dependent Womersley solution) \cite{womersley1955method} that fit the available flow rate or pointwise velocity.
This clearly introduces a bias in the solution since the choice of the profile
-- even if educated -- is arbitrary. To mitigate this aspect, an artificial extension of the inflow tract is added to the computational domain, called {\it flow extension}. 
An analysis of the error introduced by this approach can be found in \cite{veneziani2007approximate}. More sophisticated and mathematically sound approaches were introduced with the idea of identifying the best profile through an optimization approach (see \cite{formaggia2002numerical,veneziani2005flow,veneziani2007approximate,formaggia2008new,formaggia2010flow}),
yet they require an additional computational cost (and non-standard solvers)).
In the simulation considered in our work on AoD \cite{xu2018coupled, xu2020backflow}, we assumed flow rate data provided by the literature, with flow extensions at the inlet (the ascending aorta)
and the selection of a constant velocity profile. An analysis of the impact of idealized velocity profiles in AoD simulations can be found in \cite{armour2021influence}.

\item {\it Outflow Conditions}. 
For outflow conditions on the portion of the domain denoted by $\Gamma_{out}$, a common approach is to combine the 3D model based on the incompressible Navier-Stokes equations \eqref{eq:NS1}-\eqref{eq:NS2} with surrogate -- dimensionally reduced -- models representing the downstream circulations, in what has been called the {\it geometrically multiscale approach} \cite{formaggia1999multiscale,vignon2006outflow,formaggia2009multiscale,quarteroni2016geometric}. It is important to stress that, in this particular case of AoD, we may have some outflow sections referring to collateral vessels like the renal or the mesenteric arteries where not only do we lack patient-specific data, but it is also hard to find data in the literature. Yet, the inclusion of the renal flow is critical for a reliable assessment of the hemodynamics in an AoD. A popular approach, in this case, is to resort to the introduction of a lumped parameter model called 3-Element Windkessel\footnote{Windkessel was a device used by firemen to pump water from a reservoir, converting a periodic action into a quasi-steady flow, that is exactly what happens in the peripheral circulation, where the pulsatile aortic flow is eventually converted into a steady flow in the capillaries \cite{peiro2009reduced}.}, representing the downstream circulation at each outflow boundary (see, e.g., \cite{peiro2009reduced}). 
This approach leads to the prescription of a traction condition in the form
\begin{equation}\label{eq:3wk}
p \bn - \mu (\nabla \bu + \nabla^T \bu) \cdot \bn =
P_{wk} \bn, \qquad {\rm on} \quad \Gamma_{out},
\end{equation}
where $P_{wk}$ is a function of time defined as
$$
P_{wk}(t) = R_1 Q(t) +
e^{-\frac{t}{R_2 C}} \left(P_{wk} (0) - R_1 Q(0)  \right) +
\int_{0}^{t} \dfrac{e^{\frac{\tau -t}{R_2C}}}{C} Q(\tau) d \tau.
$$
Here, 
$Q(t) = \int_{{\Gamma}_{out}} \bu \cdot \bn \, d \gamma (t)$, and $R_1$, $R_2$, and $C$ are parameters representing the proximal and distal flow resistances and the compliance, respectively, of the downstream vascular district.
The condition \eqref{eq:3wk} turns out to be a sort of ``Robin'' condition (see Remark \ref{eq:BC_R}) for the NSE, mixing values of velocity and pressure, as the result of the multiscale coupling between the NSE and the Windkessel model. Each outflow section features its own set of parameters $R_1$, $R_2$, and $C$. The calibration of these parameters is usually performed with a combination of patient-specific/literature data with some empirical criteria (see, e.g., 
\cite{romarowski2018patient, pirola20194}). Although critical, this step is not central to the topics of this paper and was mentioned only for completeness.
\end{enumerate}

It is worth pointing out that, in aortic simulations, a numerical instability may occur on the boundary $\Gamma_{out}$ where traction conditions, like \eqref{eq:3wk}, are prescribed. This circumstance goes under the name of ``backflow instability,'' and it is related to a lack of energy control in the solution when a significant amount of fluid enters through $\Gamma_{out}$.
In detail, when computing the energy of the fluid, the boundedness of the Navier-Stokes solution 
can be proved under the condition that 
\begin{equation}\label{eq:backflow}
\int_{{\Gamma}_{out}} \rho |\bu|^2 \bu \cdot \bn d \gamma \geq 0.
\end{equation}
Unfortunately, the occurrence of incoming flow (that at outflow sections we can call {\it backflows}) such that $\bu \cdot \bn < 0$ may impair the control on the boundedness of the data, eventually leading to the so-called {\it backflow instabilities} \cite{bertoglio2018benchmark}.
On the other hand, backflows are physiological in the aortic circulation.
The numerical treatment of this problem has been discussed in several contributions, see, e.g., \cite{bruneau1994effective,bruneau1996new,armou2016modified,bertoglio2014tangential,bertoglio2016stokes,bertoglio2018benchmark}. 
A natural, simple approach in the specific case of Windkessel conditions \eqref{eq:3wk} consists of artificially increasing the proximal resistance $R_1$, as this results in damping the impact of the backflow on the energy balance. As we will see below, when using LES-EFR models, this turned out to be unnecessary. 

\paragraph{EFR in action.}
The occurrence of highly disturbed flow in the human ascending aorta is the consequence of the high inlet velocity at the aortic valve \cite{nichols2022mcdonald}. However, the alternation of fast and slow transients associated with the different phases of the heartbeat (``systole'' when the aortic valve is open, ``diastole'' when it is closed) generally prevents the transition to turbulence in physiological cases. 
Nonetheless, in the presence of small tears where the blood is strongly accelerated, we may have the transition to turbulence. Some studies on AoD report 
peak Reynolds numbers exceeding 10,000 \cite{chen2013patient}, 
and consider $k -\omega$ modeling for the turbulence. 
Other studies 
\cite{fatma2022numerical} find peak Reynolds numbers below 5,000, 
and consider laminar flow appropriate. This is a consequence of the fact that in computational hemodynamics, the patient-specific diversity and the high level of uncertainty may lead to different findings.
As we mentioned earlier, CACTs require  simulations for a reasonably large number of patients. Hence, computational efficiency is a criterion of paramount relevance, as is the capability to capture transitional and turbulent regimes.
The use of LES modeling, and in particular, EFR, is not only appropriate for the expected Reynolds numbers 
but is also allows for a significant reduction of the computational costs because it enables the use of coarse meshes 
that do not necessarily solve all the significant scales.

In \cite{xu2018coupled},
we compared a DNS for a patient-specific case with a mesh with
$1,063,000$ elements vs. a EFR 
simulation with a mesh with $654,000$ elements. 
Space discretization was performed with classical Taylor-Hood inf-sup compatible finite elements 
within the C++ library LifeV \cite{bertagna2017lifev}. 
The results obtained were similar and equivalent from the clinical point of view.
The reduction of the computational time, however, was significant (around $80 \%$).

For the Dirichlet conditions on the wall and at the inflow, we used the 
conditions for $\bv$ and $\overline{\bv}$ as specified in Sect. \ref{sec:EFR_NS}.
For the Windkessel conditions \eqref{eq:3wk}, after the time discretization, we have two options:
\begin{enumerate}
\item {\it implicit}, 
i.e., we actually evaluate
$Q^{n+1} = \int_{{\Gamma}_{out}} \bu^{n+1} \cdot \bn d \gamma$ so that 
we resort to Robin boundary conditions similar to the ones discussed in Remark \ref{eq:BC_R};

\item {\it explicit}, i.e., we use a time extrapolation of the velocity consistent with the time-discretization accuracy. 
For a first-order time advancing, for instance, one has
$Q^{n+1} \approx \int_{{\Gamma}_{out}} \bu^{n} \cdot \bn d \gamma$,
leading to classical traction (Neumann) conditions. 
\end{enumerate}
In the former case, for the filtered velocity, we follow the strategy indicated in Remark \ref{eq:BC_R},
and in the latter one, we adopt the homogeneous Neumann conditions specified in \eqref{eq:filter_BC_N}.

\paragraph{Filter radius and relaxation parameter.} 
The selection of the parameters $\alpha$ (radius) and $\chi$ (relaxation) is obviously critical and still an open problem.
Ideally, one would like to have an automatic selection of these parameters. This calls for the identification of general criteria to be applied to the specific case of interest. In the case of AoD, we have identified many different geometrical and fluid dynamic conditions that make the application of general criteria quite problematic. With a pretty long experience in different cases, we have identified some basic principles and heuristic criteria (see \cite{bertagna2016deconvolution,xu2020thesis,xu2020backflow}).

The radius $\alpha$ is calibrated to be related to the mesh size $h$.
However, we found that setting $\alpha = h_{max}$, i.e., the maximum size of the elements of the finite element mesh, may generate unnecessary dissipation. We decided to set $\alpha = h_{min}$ (smallest edges of elements in the mesh) to confine the effects of the diffusion introduced by the EFR method.

When selecting $\alpha$ proportional to the mesh size, a consistency argument for the
numerical EFR solution with the
original equations suggests to take $\chi \propto \Delta t$ \cite{bertagna2016deconvolution}. In particular, it has been observed that if we set
\begin{equation}\label{eq:chi}
\chi = \dfrac{\mu \Delta t}{c_0 \rho \eta \alpha^2} \max(h-\eta,0)
\end{equation}
we obtain that the viscous term of the EFR model is comparable to the viscous term in a standard NS solver
when the mesh size is the Kolmogorov scale $\eta$. 

While this approach to set $\chi$ is theoretically intriguing, in the case of AoD we need to consider the additional problem of backflow instabilities. The following result was proven \cite{xu2020backflow}:

\begin{theorem}
If the data of the problem are regular enough, assuming homogeneous boundary data,
for $\chi \rightarrow 1^-$ and a suitable choice of the radius $\alpha$, then
$$
\| \bu^n \|_{{L}^2} \leq \| \bu_0 \|_{{L}^2}.
$$
\end{theorem}

This theorem has an important consequence: the energy of the system, even in the presence of backflows, 
decreases since the diffusive term introduced by the EFR scheme controls the term \eqref{eq:backflow} 
responsible for the instability. 
In order to prevent backflow instabilities, the calibration of $\chi$ 
based on \eqref{eq:chi}
needs to be adjusted. 
 This is illustrated in Table
\ref{tab:1}, referring to a geometry designed to trigger backflows (a domain represented by a portion of a torus with a decreasing radius).

\begin{table}[h]
\centering
\begin{tabular}{|ccc|}
\hline
\multicolumn{3}{|c|}{$\alpha=h_{min}$}                                                                                          \\ \hline
\multicolumn{1}{|c|}{Mesh}                               & \multicolumn{1}{c|}{C}                      & F                      \\ \hline
\multicolumn{1}{|c|}{$\Delta t = \left\{ \begin{array}{c} {0.01}\\{0.005} \end{array} \right.$} & \multicolumn{1}{c|}{$ \begin{array}{c} {0.07}\\{0.04} \end{array} $} & $\begin{array}{c} {0.08}\\{0.06} \end{array}$ \\ \hline
\end{tabular} \quad
\begin{tabular}{|ccc|}
\hline
\multicolumn{3}{|c|}{$\alpha=0.9$}                                                                                          \\ \hline
\multicolumn{1}{|c|}{Mesh}                               & \multicolumn{1}{c|}{C}                      & F                      \\ \hline
\multicolumn{1}{|c|}{$\Delta t = \left\{ \begin{array}{c} {0.01}\\{0.005} \end{array} \right.$} & \multicolumn{1}{c|}{$ \begin{array}{c} {0.06}\\{0.04} \end{array} $} & $\begin{array}{c} {0.06}\\{0.05} \end{array}$ \\ \hline
\end{tabular}
\caption{Minimal value of the relaxation parameter that prevents backflows for different values of the radius and the time-step
for two meshes with two different refinements, coarse (C) and fine (F). See \cite{xu2020backflow} for more details.
}\label{tab:1}
\end{table}

In general, a larger $\alpha$ allows for a smaller $\chi$ and vice-versa.
However, the stabilization of backflows is obtained for values of the relaxation parameter independent of $\alpha$ and with a sublinear dependence on the time step.

A specific sensitivity analysis to elucidate the impact of the value of $\alpha$ on the simulation is reported below.

\begin{figure}
    \centering
    \includegraphics[width=0.75\linewidth]{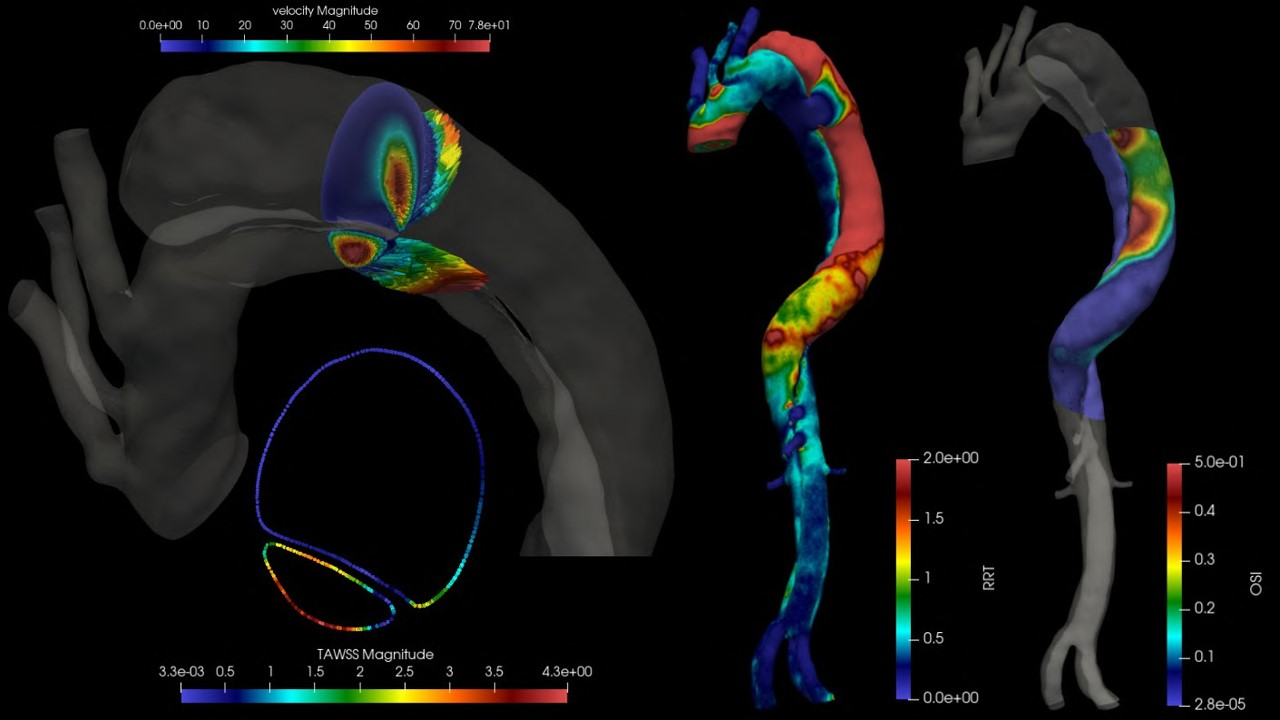}
    \caption{EFR simulation of the hemodynamics in a patient-pecific AoD: TAWSS in different regions of the false lumen.}
    \label{fig:AoDResults}
\end{figure}

\paragraph{Results.} With the method described above, we have simulated a number of patients in search of possible hemodynamics predictors that could improve the clinical guidelines to diagnose the evolution of the dissection. 
This work, still in progress \cite{yang2024predicting}, is in collaboration with the Department of Surgery at Emory University (Dr. B. Leshnower). While
the biomedical implications of our results are not the focus of this paper, we summarize here some major findings as evidence of the practical success of the EFR methodology. 
In \cite{xu2018coupled}, we considered a longitudinal study of a patient presented at Emory University in 2006 and followed for several years. Imaging in 2010 evidenced a significant growth of the false lumen, even though the PET was distal.
We co-registered the images in 2006 and 2010 to quantify the false lumen growth and correlate this growth with the hemodynamics computed with our method in the 2006 geometry. The results pointed out a potential role of the Time-Averaged WSS (TAWSS) defined as 
$$
\text{TAWSS} \equiv \dfrac{1}{T} \left\| \int_0^T \boldsymbol{\tau} (t)  dt \right\|,
$$
where $T$ is the heartbeat duration.
The growth is proportional to the TAWSS, but only below a certain threshold. Other hemodynamic indexes were found to be correlated to the possible thrombotic activity leading to a local shrinking of the false lumen (see \cite{xu2020thesis} for more details).
After analyzing several more patients with both growth and no-growth, we are now focusing on the difference between the TAWSS in the false and true lumen as a potential predictor of pathology evolution. See Fig.~\ref{fig:AoDResults}.

\paragraph{Sensitivity analysis for the filtering radius.}

\newcommand{\buo}{\overline{\bu}}

Sensitivity analysis can
be carried out in different ways. 
A possible approach is an experimental, yet educated, sampling of the solution for different values of the
parameters. In general, this requires several samplings, which, in our case, means
running the simulation for several values of $\alpha$.

{\it Local} sensitivity analysis consists of quantifying the 
dependence of the solution on a parameter by computing the corresponding (Gateaux) derivatives. These are the so called {\it sensitivities}. 
The numerical solution of these equations for a particular value of the parameter
allows us to determine in what parts of the computational domain the solution is more sensitive to such parameter.
This is a local analysis since it applies to a particular value of the parameter.
A complete derivation of the sensitivity equations to the radius $\alpha$
for Leray linear and nonlinear spatial filtering problems
with AD indicator functions \eqref{eq:a_lin} and
$a_{\cD_0}$ can be found in \cite{bertagna2018sensitivity}. 

The results in \cite{bertagna2018sensitivity} show that the velocity sensitivity can be
a reliable predictor of the regions in the domain with high sensitivity to
a specific parameter value.
In simulating the flow past an obstacle, for instance, the overdiffusive effects immediately past the obstacle
due to a excessively large value of $\alpha$ are correctly identified by the velocity sensitivity.
This leads to a possible ``adaptive'' algorithm for the calibration of the radius. Denoting by 
$\bs_u \equiv \Deriv{}{\bu}{\alpha}$ and by $s_p \equiv \Deriv{}{p}{\alpha}$
the sensitivities of velocity and pressure, one could solve the LES model and the 
sensitivity equations with a ``large'' value $\alpha_0$ of the radius  to attain stability, and then refine the solution
with a Taylor expansion
\begin{equation}\label{eq:leo}
\bu (\alpha) \approx \bu (\alpha_0) + \bs_u (\alpha_0) (\alpha - \alpha_0), \quad
p (\alpha) \approx p (\alpha_0) + s_p (\alpha_0) (\alpha - \alpha_0),
\end{equation}
with $\alpha < \alpha_0$. 
In the same paper, the sensitivity of the solution to a scheme with \eqref{eq:a_lin}
is demonstrated to be significantly higher than with the non-linear AD filter, pointing out the superiority of the latter. 

For specific cardiovascular applications, a {\it global} sensitivity analysis was performed too 
\cite{xu2021global}. The purpose of this analysis is to
understand what inputs affect the solution 
the most.
The inputs are regarded
as stochastic processes. Quantities of interest obtained from the numerical solution of the hemodynamics through our EFR method are consequently regarded as stochastic processes too. 
The sensitivity to the 
different inputs is quantified in terms of the variance of the outputs as a function of the input variances.
In \cite{xu2018coupled}, three inputs are considered: the geometry of the domain, the inflow rate, and the radius $\alpha$. For each of these variables, a probabilistic distribution is postulated.
For the radius, a Gaussian distribution with average $h_{min}$ and variance 
$0.25 h_{min}^2$ was deemed to be a correct description of the range of interest for $\alpha$.
The uncertainty in the geometry comes from the noise in the images and the consequent errors in the 
patient-specific reconstruction. Two cases were considered: (i) an idealized aortic arch, where the 
uncertainty relies on the geometrical parameters of the ideal geometry; (ii) a patient-specific case,
where the geometry was reconstructed in a time range of 6 years. In case (ii), after an image registration procedure, the
geometry was parametrized by one parameter specifying the morphing instant from the initial to the final geometry.
For this parameter, we postulated a uniform probabilistic distribution. To quantify the uncertainty, we used \textit{Polynomial Chaos Expansion}
(PCE).
In PCE, a variable of interest function of random variables is regarded as a polynomial expansion on a basis incorporating the stochastic features of the inputs. This basis is orthogonal with respect to a suitable probability measure. The coefficients of this (truncated) expansion can be computed with a pseudo-spectral approach
through appropriate quadrature rules. This yields to compute the so-called Sobol' indexes, which are variance-based
indicators to characterize the dependence of the output variance on
the stochastic inputs \cite{sobol1993sensitivity,sobol2001global}
and rank the influence of the inputs, especially for nonlinear models \cite{sudret2008global,crestaux2009polynomial}.
In short, the Sobol' index for a specific input $i$ on a quantity of interest $f$ is defined as
$$
S_i^f \equiv \dfrac{{\rm Var}(f^{PC}_i)}{{\rm Var}(f^{PC})}, 
$$
where  ${\rm Var}(f^{PC})$ denotes the variance of the stochastic process $f$ represented by the Polynomial Chaos expansion and due to all the stochastic inputs, while ${\rm Var}(f^{PC}_i)$ refers only to the variance induced by the input $i$.

\begin{figure}[hbt]
    \centering
    \includegraphics[width=\linewidth]{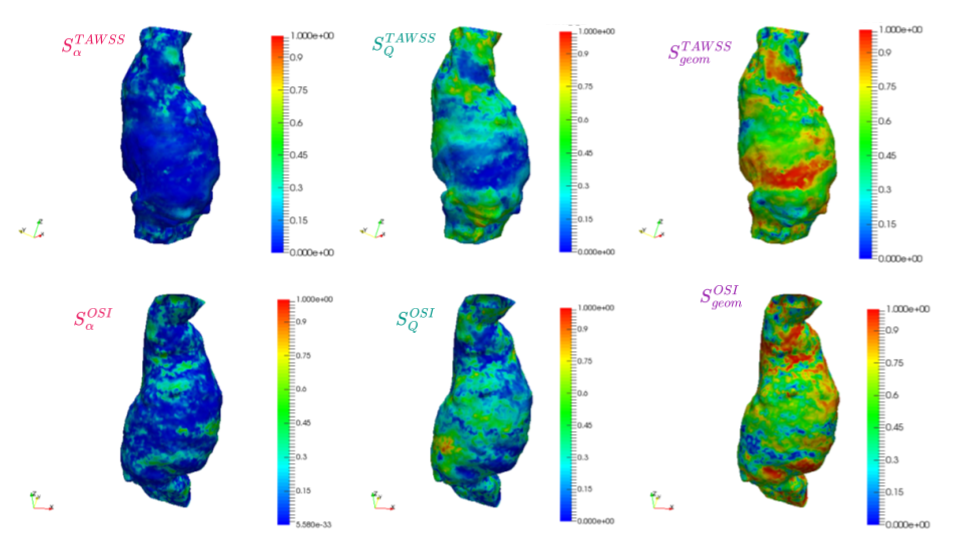}
    \caption{Sobol' indexes in a patient-specific geometry for the sensitivity of the TAWSS and the OSI to the radius $\alpha$ (left), the inflow rate $Q$ (center), and the geometry (right). 
    Blue regions identify parts of the domain weakly affected by variations of the input
    in comparison with the other uncertainties.}
    \label{fig:GSA}
\end{figure}

In \cite{xu2021global}, the outputs of interest are the TKE, the TAWSS, and the Oscillatory Shear Index (OSI). The OSI is a hemodynamic 
index ranging between 0 and 1/2, which quantifies for how long during the heartbeat the  
WSS is opposite to the direction of the flow. It is defined as
$$
\text{OSI} \equiv \dfrac{1}{2}\left(1 - \dfrac{\|\int_0^T \boldsymbol{\tau}(t) dt\| }{\int_0^T \|\boldsymbol{\tau} (t)\| dt} \right).
$$
Through the application of PCE, we learned a couple of important points:
\begin{enumerate}
\item The geometry is by far the most important factor in the results: an accurate geometrical reconstruction of the region of interest is critical for any biomedical analysis. See Fig. \ref{fig:GSA}, rightmost panels.

\item The impact of the radius on the TAWSS and the OSI is minimal. See Fig. \ref{fig:GSA}, leftmost and center panels. This means that the selection of the radius with the empirical rules used in our simulations is not expected to have a major impact on the clinical conclusions of the computational analysis.
\end{enumerate}    

This corroborates the confidence in our EFR modeling for AoD.

\subsubsection{Open problems}\label{sec:incom_op}

The EFR method gave excellent results in the simulation of AoD, providing the trade-off between the computational efficiency and the accuracy needed to process a significant number of patients. Yet, there are some aspects of the method that still need to be elucidated.

A first important point is the optimal selection of the parameters $\alpha$ and $\chi$.
As mentioned earlier, the selection is largely based on empirical and physical criteria.
An automatic, somehow optimal, selection would be certainly desirable. As noted,
a possible approach for the radius comes from the sensitivity local analysis and the
approximate expansion \eqref{eq:leo}. 
An initial conservative choice of the radius can be refined at each time step so long as some indicators does not suggest the presence of instabilities.
This approach has clearly the drawback of requiring the solution
of the sensitivity equations, which may be quite expensive in some applications, like  AoD hemodynamics. Yet, it can provide the backbone for the definition of an adaptive scheme once 
the sensitivity computations can be properly surrogated.
Likewise, the initial setting of $\chi$ can be based on linear dependence on $\Delta t$ and physical arguments, but the occurrence of backflows requires successive refinement. All these aspects need to be further investigated.

Another important problem in hemodynamics applications is related to 
``data assimilation'' procedures that may improve the overall reliability of numerical simulations. With more accurate measurement tools, like 4D MRI, and more storage space available, more hemodynamics data are progressively more available, in particular velocity data in some points of the region of interest. In general, these data are not on the boundary, so they do not provide boundary conditions, yet they give important information about the blood dynamics. 
However, these data can barely replace the wealth of information provided by a numerical simulation, particularly for hemodynamic indexes like the TAWSS or the OSI.
In a cooperative view, where data and models are not alternative but complementary, we need to find a solution both fitting the data (possibly loosely in the presence of high noise) and fulfilling the equations.
The problem has been well-known for years (see, e.g., \cite{d2012applications,d2012variational,d2013uncertainty}),
yet a ``standard'' solution may require even higher - unfeasible - computational costs. 
In this respect, the combination of Physically Informed Neural Networks (PINN) \cite{cai2021physics}
and our EFR scheme may provide a potential breakthrough that will be investigated in the years to come.


\subsection{Compressible flows}\label{sec:LESFOM_comp}

As mentioned in Sec.~\ref{sec:EFR_Euler}, 
despite many encouraging results for incompressible flows, EFR 
algorithms have been applied to simulate
compressible flows only very recently in \cite{clinco2023filter}. Therein, algorithm \eqref{eq:mom_td}-\eqref{eq:end-of-s-p} has been introduced
and tested on 
classical benchmarks for atmospheric flow. Here, we report 
the main results to support the following findings: i) for a given mesh, 
the EFR algorithm with the AD indicator function is less dissipative
than a Smagorinsky-like method (EFR with indicator function $a_S$ in \eqref{eq:a_smago}),
ii) the reason for point i) is the fact 
that $a_{\cD}$ is more selective in identifying the regions
of the domain where artificial diffusion is needed, and iii)
the additional cost for using $a_{\cD}$ instead of $a_S$
is marginal. 
For all the results reported here, we set $\chi = \xi = 1$ so that the artificial diffusion introduced by the EFR algorithm can easily be calculated (see \cite{clinco2023filter}).
See Sec.~\ref{sec:com_op} below for more on this. 

\begin{figure}[htb!]
     \centering
     \vspace*{0.2cm}
         \begin{overpic}[width=0.235\textwidth]{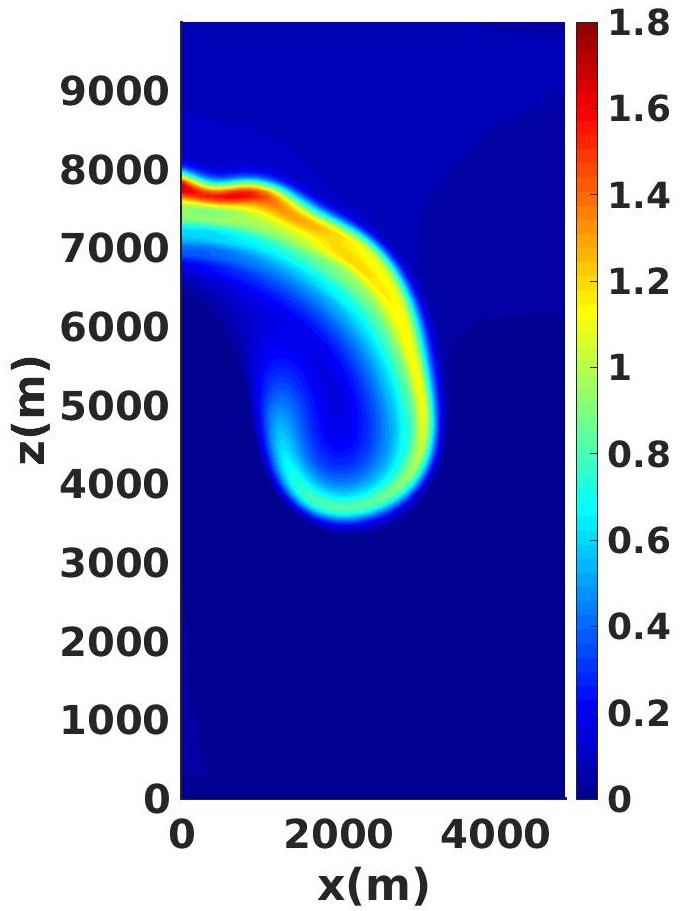}
    \put(50,102){{coarser mesh ($h=62.5$ m)}}
     \put(26,90){\textcolor{white}{\footnotesize{$a_S$, $\alpha=6$ m}}}
    \end{overpic}
    \begin{overpic}[width=0.24\textwidth]{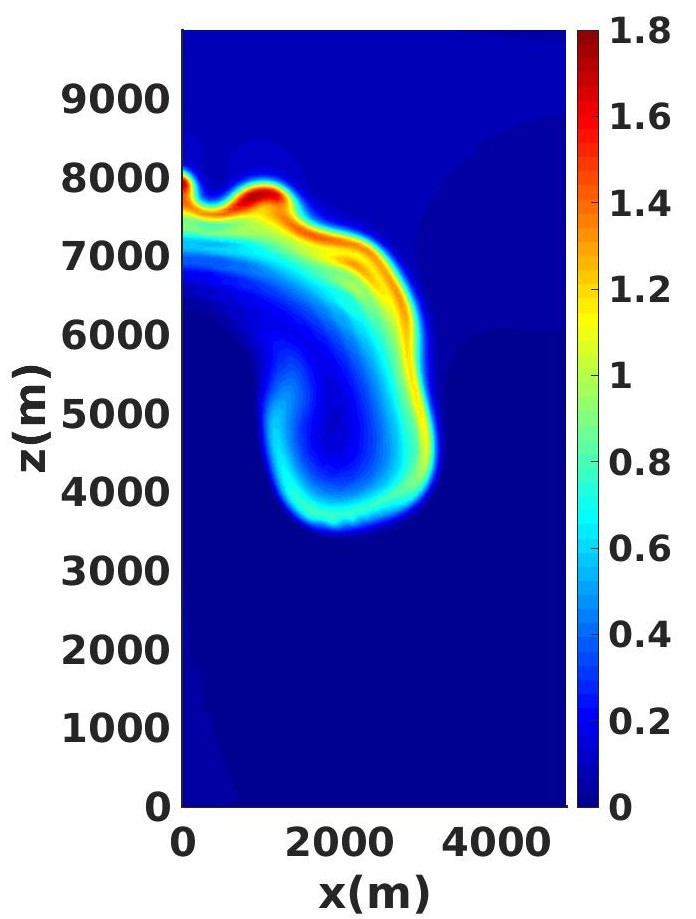}
     \put(26,89){\textcolor{white}{\footnotesize{$a_D$, $\alpha=6$ m}}}
    \end{overpic} 
     \begin{overpic}[width=0.245\textwidth,grid=false]
     {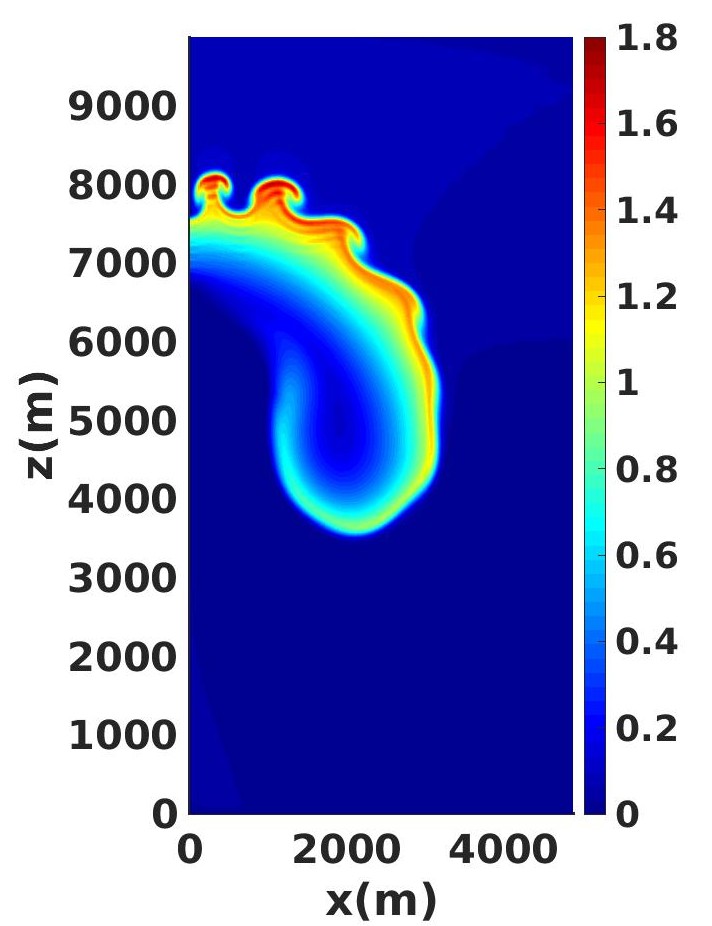}
     \put(50,100){{finer mesh ($h=31.25$ m)}}
     \put(27,88){\textcolor{white}{\footnotesize{$a_S$, $\alpha=3$ m}}}
    \end{overpic}
         \begin{overpic}[width=0.24\textwidth]
    {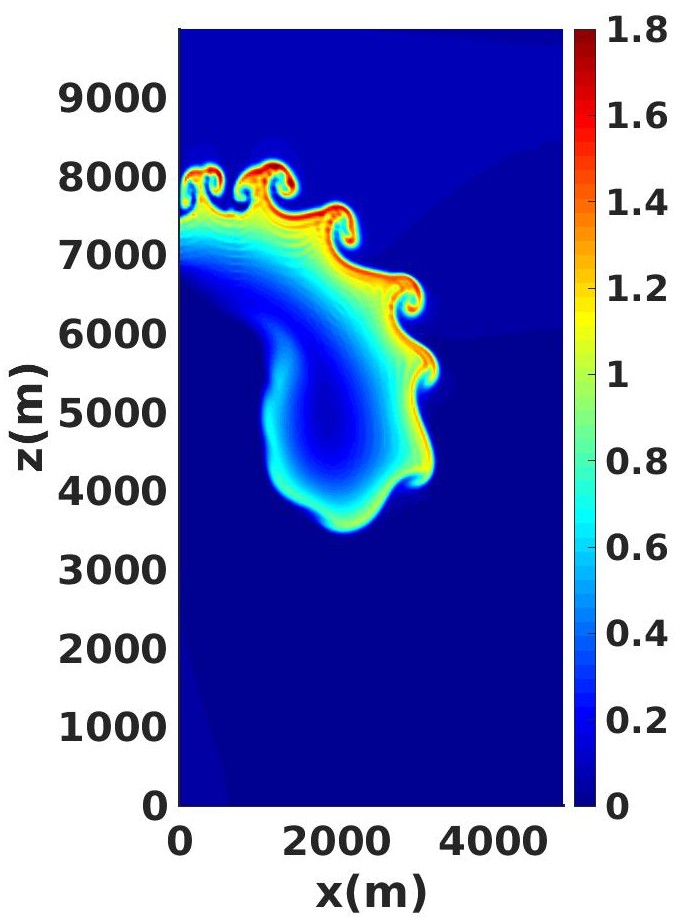}
     \put(26,88){\textcolor{white}{\footnotesize{$a_D$, $\alpha=3$ m}}}
    \end{overpic}
    \caption{Rising thermal bubble: perturbation of potential temperature $\theta'$ at $t = 1020$ s computed with the EFR and $a_S$ and $a_D$ with the coarser mesh (first two panels) and the finer mesh (last two panels).}
    \label{fig::TB_SM_DB}
\end{figure}

\begin{figure}[htb!]
    \centering
    %
    \vspace*{0.2cm}
    \begin{overpic}[width=0.48\textwidth]{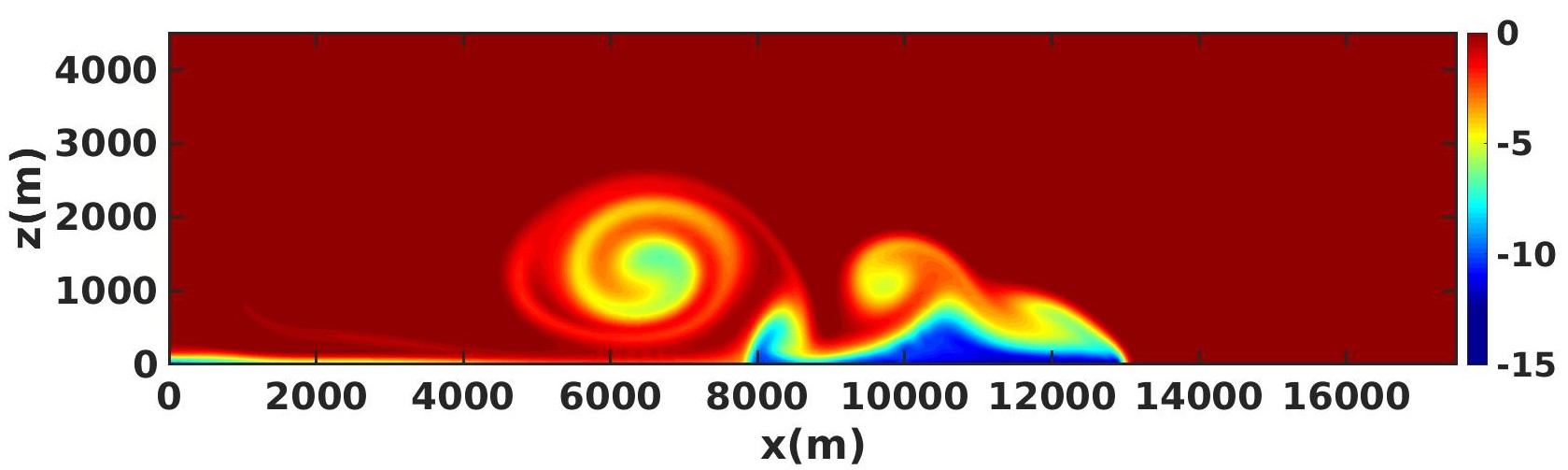}
    \label{fig:Smago_50_750}
    \put(85,32){{EFR with $a_S$, $\alpha = 11$}}
    \put(70,23){\textcolor{white}{\footnotesize{$\theta'$ at $t = 750$ s}}}
    \end{overpic}
        \begin{overpic}[width=0.48\textwidth]{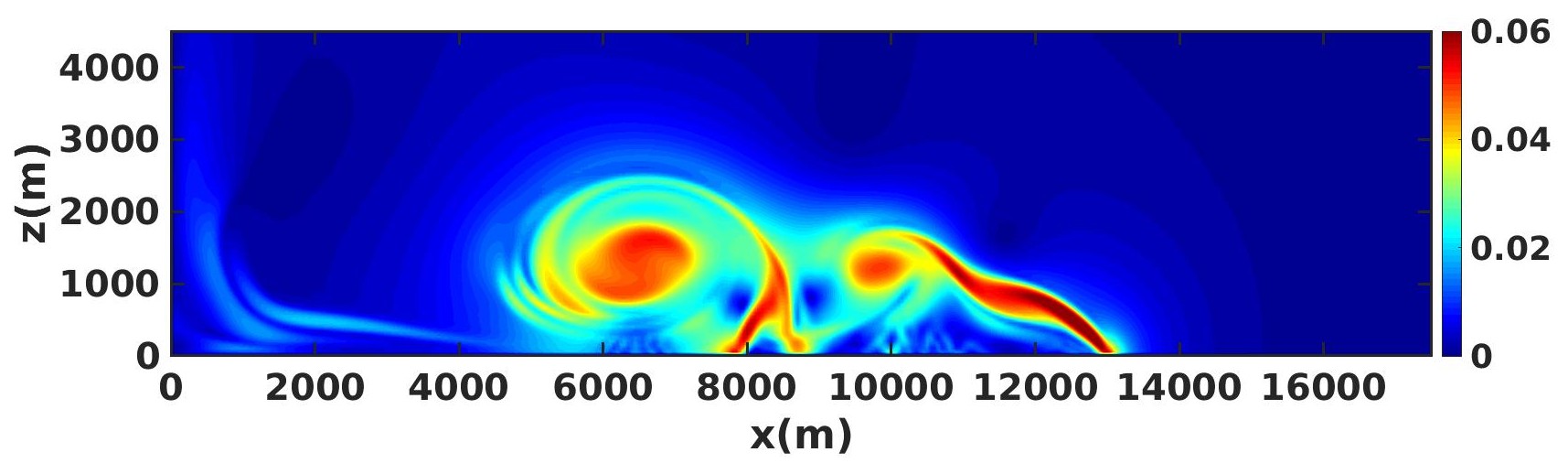}
    \put(67,23){\textcolor{white}{\footnotesize{$a_S$ at $t = 750$ s}}}
    \end{overpic}
    \\
        \begin{overpic}[width=0.48\textwidth]{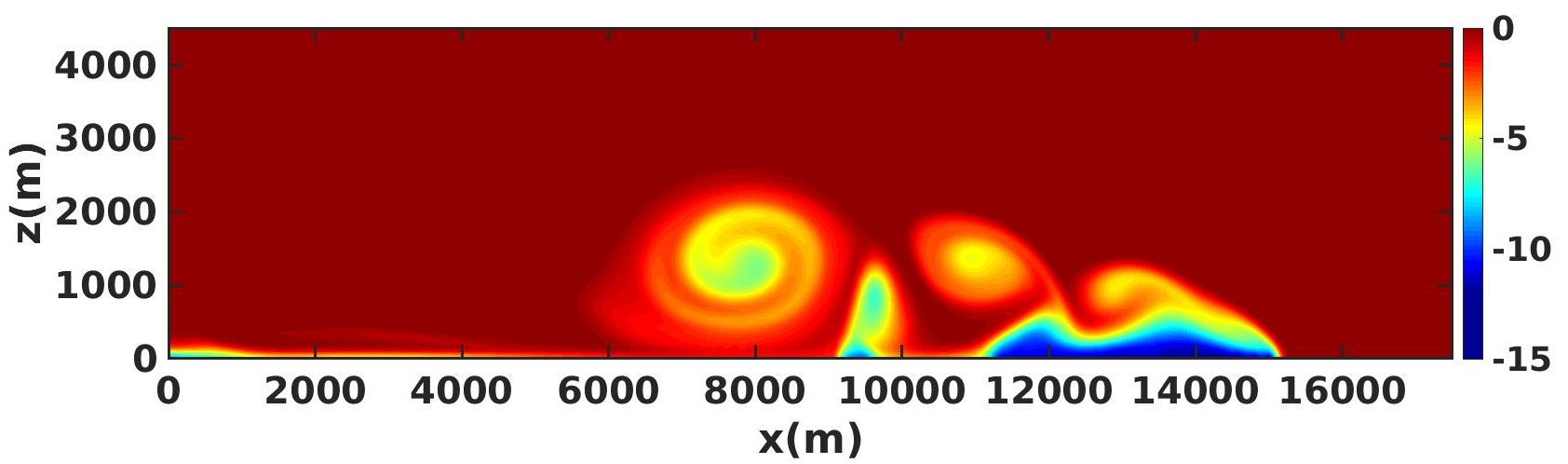}
    \label{fig:Smago_50_900}
    \put(70,23){\textcolor{white}{\footnotesize{$\theta'$ at $t = 900$ s}}}
    \end{overpic} 
        \begin{overpic}[width=0.48\textwidth]{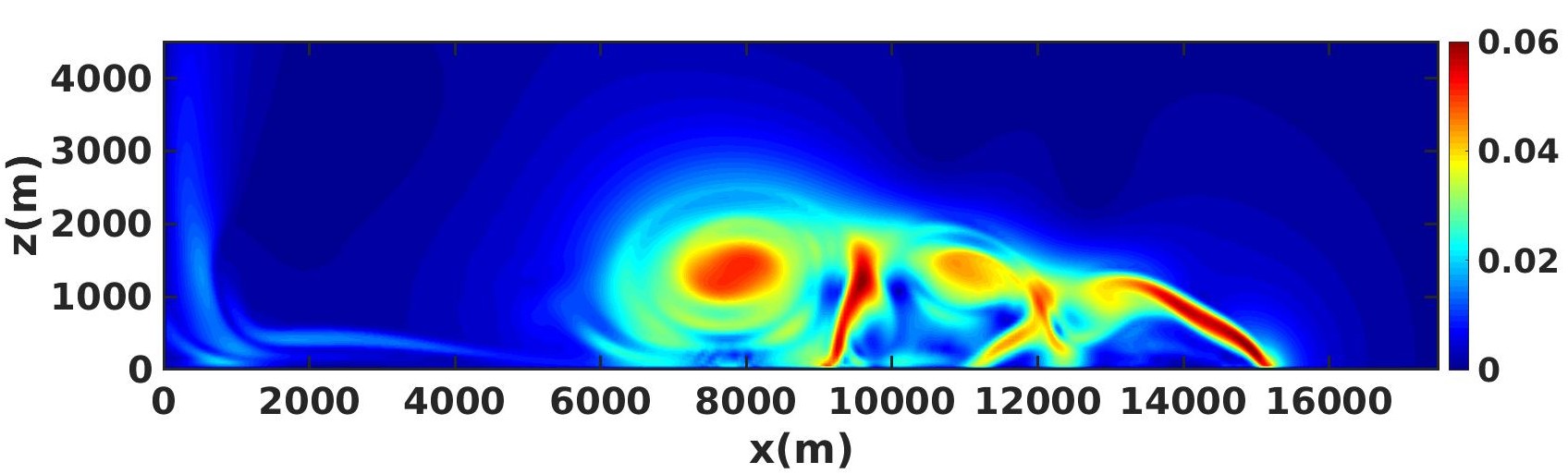}
    \put(67,23){\textcolor{white}{\footnotesize{$a_S$ at $t = 900$ s}}}
    \end{overpic} 
    \\
    \vskip .5cm
    \begin{overpic}[width=0.48\textwidth]{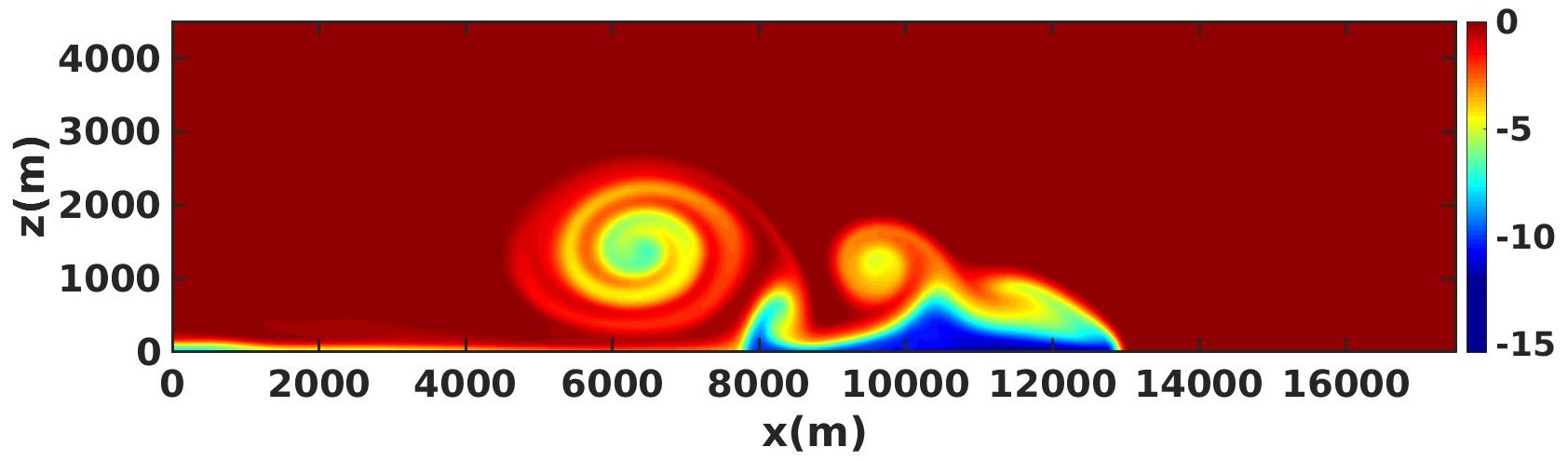}
    \label{fig:decon_125_750}
    \put(85,32){{EFR with $a_\cD$, $\alpha = 12$}}
    \put(70,23){\textcolor{white}{\footnotesize{$\theta'$ at $t = 750$ s}}}
    \end{overpic}
        \begin{overpic}[width=0.48\textwidth]{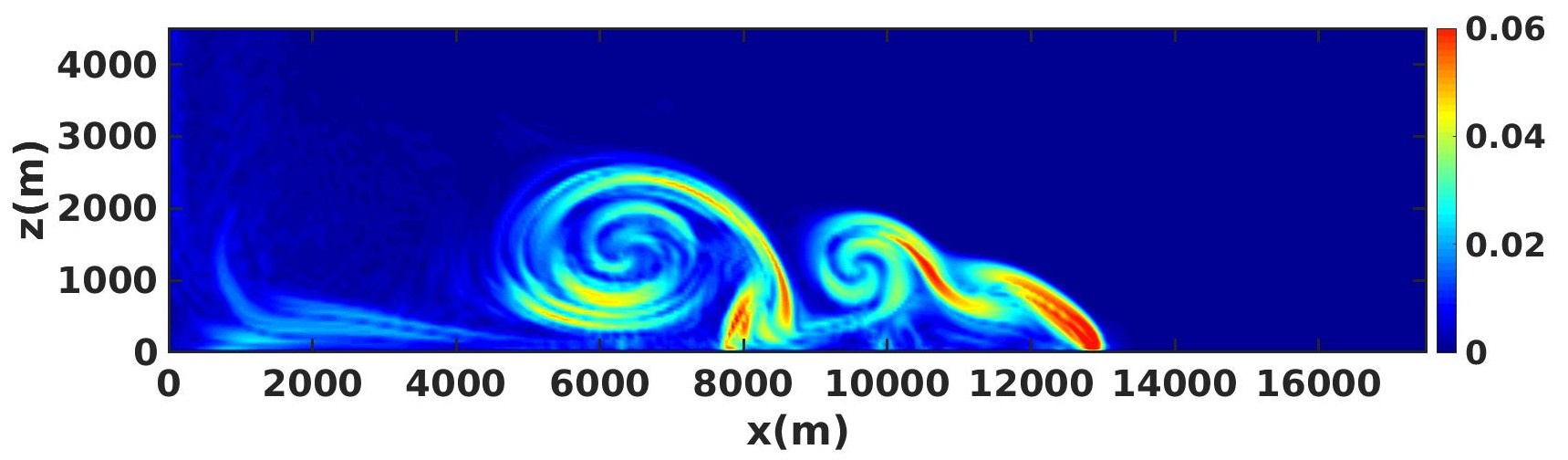}
    \put(67,23){\textcolor{white}{\footnotesize{$a_D$ at $t = 750$ s}}}
    \end{overpic}
    \\
    \begin{overpic}[width=0.48\textwidth]{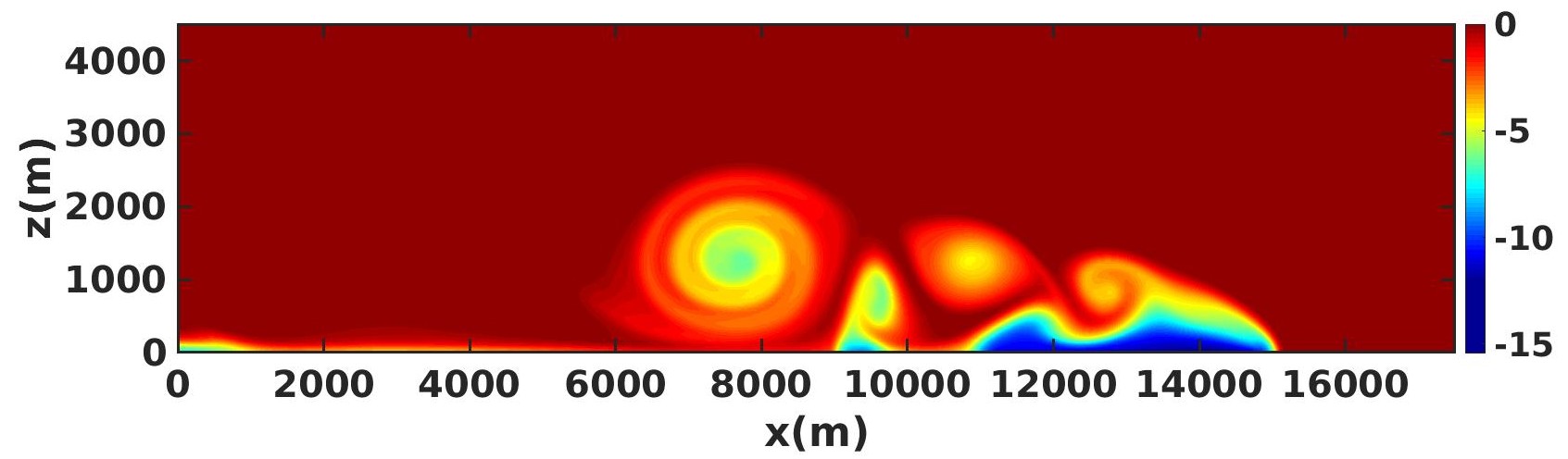}
    \label{fig:decon_125_900}
    \put(70,23){\textcolor{white}{\footnotesize{$\theta'$ at $t = 900$ s}}}
    \end{overpic}
        \begin{overpic}[width=0.48\textwidth]{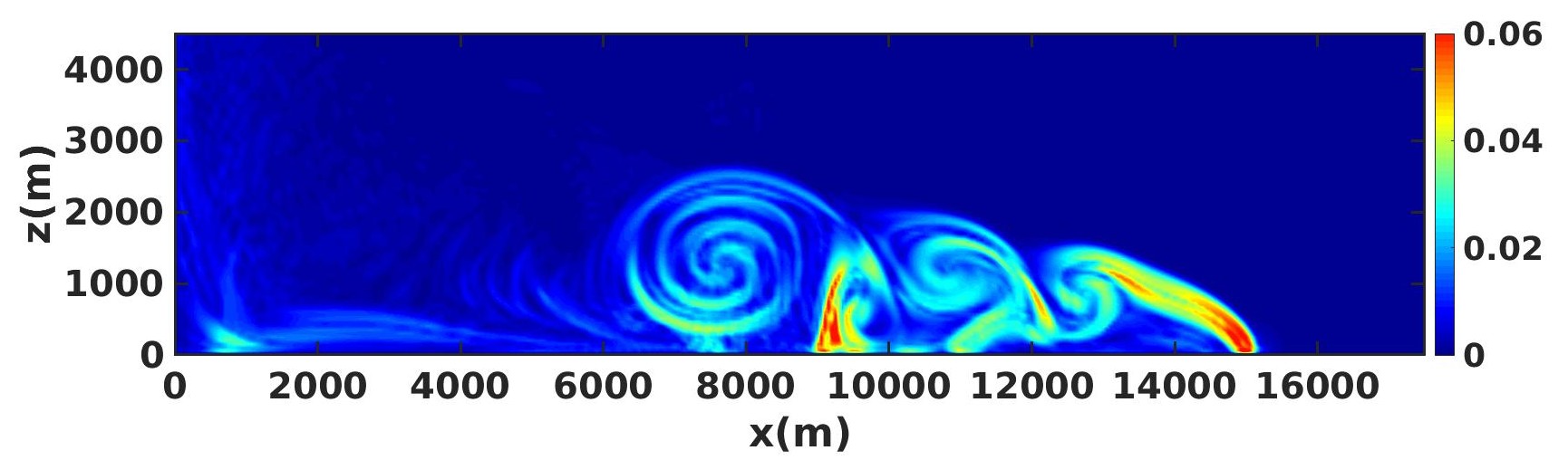}
    \put(67,23){\textcolor{white}{\footnotesize{$a_D$ at $t = 900$ s}}}
    \end{overpic} 
    \caption{Density current: potential temperature fluctuation $\theta'$ (left) and indicator function (right) for the EFR method with $a_S$ (top) and $a_\cD$ (bottom). The mesh size is $h = 50$ m.}
        \label{fig::SmagoRayModel50}
\end{figure}

To illustrate the above point i), we will first use results obtained for the
well-known rising thermal bubble benchmark. This is a 2D benchmark that
entails perturbing with a bubble of warm air a neutrally stratified atmosphere with uniform background potential temperature. 
In the time interval $(0, 1020]$ s,  
the air warmer than the ambient rises due to buoyancy and deforms due 
to shearing motion into a mushroom shape.
For the set-up, we refer to, e.g., 
\cite{ahmadLindeman2007,Feng2021,clinco2023filter}. The computational domain, 
which is $5 \times 10$ Km$^2$, is discretized with two meshes
with uniform resolution: a coarser one with $h = 62.5$ m  and a finer one with $h = 31.25$~m. 
The time step is set to $\Delta t = 0.1$ s for all the simulations. 
Fig.~\ref{fig::TB_SM_DB} shows the 
spatial 
distribution of $\theta'$ at $t = 1020$ s computed 
with the EFR algorithm, indicator functions $a_S$ or $a_{\cD}$, and the two meshes.
On a given mesh, the Rayleigh-Taylor instability at the edge of the bubble is more developed when 
using $a_{\cD}$ instead of $a_S$, which indicates that $a_{\cD}$ 
introduces less artificial viscosity than $a_S$.
This is corroborated by the results for a more complex benchmark called density current \cite{carpenterDroegemeier1990,strakaWilhelmson1993} {(see also~\cite{OIFSD07,OIF09,ozgokmen2009reynolds} for LES of density currents in incompressible flows)}, in which a neutrally stratified atmosphere 
with uniform background $\theta$ is perturbed with a bubble of cold air. In the time interval $(0,900]$ s, 
the cold air descends due to negative buoyancy and, when the cold air reaches the ground, it rolls up and forms a front. 
By time $t = 900$ s, a three-rotor structure emerges
as a result of the shear generated at the top boundary
of the cold front. The results obtained with the EFR method
and indicator functions $a_S$ and $a_{\cD}$ are shown in Fig.~\ref{fig::SmagoRayModel50}. Note that $\theta'$
computed with $a_{\cD}$ is sharper, less smeared than 
$\theta'$
computed with $a_{S}$. 
Additionally, the plots of the indicator functions show that 
$a_S$ at $t = 900$ s takes larger values over wider regions than $a_{\cD}$, illustrating the above point ii).

The computational time to run the simulations in Fig.~\ref{fig::SmagoRayModel50} on a common laptop
is reported in Table \ref{tab:6}.
Obviously, the total computational cost increases when switching
from $a_S$ to $a_\cD$ because, while $a_S$ requires a simple post-processing of the velocity field, $a_{\cD_0}$ in \eqref{eq:a_D0_a_D1} requires one application
of the linear Helmholtz filter.
However, note that that increase amounts to 
only 0.006 s per time step and the filter step with $a_\cD$ takes about half of the time needed for the evolve step. 
In total, the simulation takes only about 13 minutes. 

\begin{table}[htb!]
\begin{center}
\begin{tabular}{ | c | c | c | c | c | c | }
\hline
Model  & Evolve (s) & Filter (s) & Total (s)\\
 \hline
EFR, $a_S$  & 0.06 &  0.023  & 772\\
\hline
EFR, $a_D$ & 0.06 &  0.029  & 790\\
\hline 
\end{tabular}
\caption{Density current:
computational time taken by the evolve step and filter step per time step and total simulation time for the EFR algorithm with indicator functions $a_S$ and $a_D$ for mesh $h=50$ m.}\label{tab:6}
\end{center}
\end{table}

We remark that a DNS for this benchmark would require over 5000 billion DOFs. The results 
presented here
show that 
the {EFR method with a deconvolution-based indicator function is much less dissipative
than Smagorinsky-like LES methods while featuring comparable computational efficiency}. These findings suggest that EFR is a promising approach for compressible flows, as the studies on incompressible flows suggested.

\subsubsection{Open problems}\label{sec:com_op}

To obtain the results shown in Fig.~\ref{fig::TB_SM_DB} and \ref{fig::SmagoRayModel50}, we set  $\chi = \xi = 1$ (i.e., no relaxation), $N = 0$,
and performed a costly tuning of $\alpha$ to obtain 
results that compare well with 
the literature \cite{ahmadLindeman2007,marrasEtAl2013a,marrasNazarovGiraldo2015,ahmad2018,GQR_OF_clima}.
A clear understanding 
of the complex interplay of these parameters
and their effect on the solution accuracy is missing for compressible flows. 
In Sec.~\ref{sec:AoD}, we have seen that even for incompressible flows, for which much more work has been done, this understanding is still partial. 
A full understanding is obviously
central to the success of the EFR algorithm with the AD indicator function and the definition of 
automatic tuning procedures. Another open question is the identification of the most 
physically accurate and computationally efficient filter $\cF$ for each filtered variable. For this purpose, one could use various filters available in the literature (see, e.g, \cite{BIL05}). 

Just like in the case of complex hemodyanmics simulations, care must be taken in setting the boundary conditions for compressible flow problems. Indeed, for compressible flows, the presence of artificial boundaries may trigger unphysical reflections that need to be suppressed by ``non-reflecting'' conditions. These conditions mimic the evolution of the solution outside the domain along the characteristics 
in order to suppress artificial inflow components.
When non-reflecting boundary conditions are imposed, we need to devise conditions that minimize any artifact in the physical propagation. 
As the filter adds diffusivity, an initial idea could be to explore non-invasive conditions to minimize the impact of the filter. If this is not enough, one could explore different filter formulations. 

Finally, we would like to stress the importance of data assimilation for weather/climate prediction too. In the atmospheric sciences, there is an abundance of data, which are collected either in situ or through remote sensing. Such data (or observations) are routinely incorporated in numerical weather prediction through data assimilation techniques, which seek to optimally combine the weather model with the observations.
As mentioned in Sec.~\ref{sec:incom_op}, it would be very interesting to to study the computational efficiency and accuracy allowed by a combination of PINN and EFR.

\vspace*{-0.3cm}
\section{
        LES for Reduced Order Models}\label{sec:LES_ROM}










For over a couple of decades, ROMs (see, e.g., \cite{benner2020modelvol1,benner2020modelvol2,benner2020modelvol3,hesthaven2016certified, malik2017reduced,rozza2008reduced}
for reviews) have emerged as the 
methodology 
of choice to reduce the computational burden when traditional 
flow simulations
(also called Full Order Models - FOMs) have to be carried out for several parameter values, 
as in the case of uncertainty quantification, optimal control, and inverse problems. 
ROMs replace the FOM 
with a lower-dimensional approximation that captures the essential behavior of the flow. 
The reduction in computational time is achieved through a two-step procedure.
In the first step, called {\it offline
phase}, one constructs a database of several FOM solutions associated to given times and/or
physical parameter values. 
The FOM database 
is used to generate a reduced basis, which is (hopefully much) 
smaller than the high-dimensional FOM basis but still preserves the essential features of the flow. 
One of the most 
popular reduced bases is proper orthogonal decomposition (POD), which extracts the dominant modes from a FOM database. 
We remark, however, that the 
strategies we 
discuss are not limited to POD 
and 
work equally well with other methods (e.g., the reduced basis method~\cite{hesthaven2016certified,quarteroni2015reduced}).
In the second step, called {\it online phase}, one uses this reduced basis to quickly compute the 
solution for newly specified times and/or parameter values. Note that, while the offline phase is 
performed only once, 
the online phase is performed as many times as needed.

ROMs are very efficient surrogate models when 
the number of reduced basis functions is small, 
as is typically the case for diffusion-dominated flows. Unfortunately, though, a large number of basis functions is needed to capture the essential features of convection-dominated flows. If one 
retains such a large number of modes in order for the ROM to be accurate, then
the computational efficiency suffers. If the number of modes is otherwise kept low, a severe loss of information
hinders the accurate reconstruction of the solution. 
In fact, Galerkin projection-based ROMs of turbulent flows are affected by energy stability problems
related to the fact that, e.g., 
{POD} retains the modes biased toward large,
high-energy scales, while the 
TKE is dissipated at the level of
the small scales.  A possible way to tackle this challenging problem 
is to introduce dissipation via a closure model \cite{wang2012proper, AHLS88, ahmed2021closures} or stabilization~\cite{carlberg2011efficient,grimberg2020stability,baiges2012explicit,reyes2020projection, parish2020adjoint}. 

We note, however, that the 
current ROM closures and stabilizations are typically {heuristic}, and generally lack mathematical support.
Traditionally, researchers propose ROM closures and stabilizations that make sense intuitively, test them in limited computational settings, and declare success once they yield more accurate solutions than the corresponding non-stabilized ROM.
The main drawback of these ad hoc ROM closures and stabilizations is that, while they work well for the specific flow configuration on which they were trained, they can fail 
in different flow configurations and need to be retrained (which is a costly, tedious process) in order to perform adequately.

Recently, there have been several new research developments aimed at constructing novel ROMs for realistic turbulent flows.
The following are three significant research avenues in this direction:
LES-ROMs (Sec.~\ref{sec:les-rom}) aim at ``bridging'' two distinct research fields, LES and ROMs, to construct efficient and accurate ROMs for turbulent flows.
FOM-ROM coupling (Sec.~\ref{sec:les-rom-consistency}) aims at ``bridging'' the numerical methods used in the FOM and ROM in order to improve the stability and accuracy of the latter.
Finally, the numerical analysis of the FOM-ROM framework (Sec.~\ref{sec:les-rom-na}) leverages rigorous numerical analysis results to improve the ROM performance.

\subsection{LES-ROMs}
    \label{sec:les-rom}

For the past several years, our groups and collaborators have proposed, analyzed, and investigated a new strategy for constructing ROMs for turbulent flows.
This new strategy is based on {\it spatial filtering}, which is a mathematically and physically sound principle that is central in LES (see Sec.~\ref{sec:les-fom}). 
This new strategy aims squarely at bridging two research fields, LES and ROM, that have been treated separately until now.
We also emphasize that this new strategy is fundamentally different from current ROM closures and stabilizations, which 
are often disconnected from the numerical methods used at the FOM level.

These new LES-inspired ROMs, which we call {\it LES-ROMs}, are capable to operate over a range of flow problems with minimal user intervention.
Again, this is in stark contrast with current ROM closures and stabilizations, which need to be retrained for each new computational setting.

Just as in the FOM setting, the new LES-ROMs are constructed by leveraging a spatial filter $\cF$ to smooth out different terms in the 
governing
equations and eliminate spurious numerical oscillations that generally occur in the convection-dominated, under-resolved regime.
In Sec.~\ref{sec:rom-filters-ad}, we outline the main types of ROM filters and approximate deconvolution used to construct LES-ROMs.
Then, we outline 
several LES-ROMs: the EFR-ROM (Sec.~\ref{sec:efr-rom}), which is arguably the most popular LES-ROM, Leray ROM (Sec.~\ref{sec:leray-rom}), approximate deconvolution Leray ROM (Sec.~\ref{sec:leray-rom}), and time-relaxation ROM (Sec.~\ref{sec:leray-rom}).
Other types of  LES-ROMs are mentioned in Sec.~\ref{sec:other-les-rom}.

\subsubsection{ROM filters and approximate deconvolution}
    \label{sec:rom-filters-ad}


This section defines the {\it ROM spatial filters} and {\it ROM approximate deconvolution} used to construct the LES-ROMs presented in Secs.~\ref{sec:efr-rom}--\ref{sec:other-les-rom}.
As discussed in Sec.~\ref{sec:les-fom},
the {\it spatial filter} $\cF$ is crucial in the development of LES for FOMs~\cite{BIL05,sagaut2006large}. 
The spatial filter is also increasingly used to construct LES-ROMs for turbulent flows~\cite{wells2017evolve,gunzburger2019evolve} because the idea is straightforward: 
$\cF$ is used to eliminate the spurious numerical oscillations in stabilization strategies. 
Next, we outline the three spatial filters and the approximate deconvolution operators that are currently used for ROMs.
In what follows, let $\bu$ be a generic solution and
$\bu_r = \sum^r_{j=1} u_{r,j} \bphi_j(\bx)$ its ROM approximation, with 
$\bphi_j$ for $j=1,2,\ldots,r$ the ROM basis functions.
We denote by $\underline{u}_{r} = (u_{r,1}, \ldots, u_{r,r})^{\top}$ the vector of ROM coefficients of $\bu_r$.

\paragraph{ROM differential filter.}
The {\it ROM differential filter (DF)}~\cite{sabetghadam2012alpha,wells2017evolve,wells2015phd} is probably the most popular ROM spatial filter used to build LES-ROMs.
The ROM differential filter is a modular, simple, linear filter, which is a natural extension of the differential filter used to construct the EFR-FOM in Sec.~\ref{sec:EFR}.
It is defined as follows:
find $\obu_r(\bx) = \sum_{j=1}^r \ou_{r,j} \bphi_j(\bx)$
such that
\begin{eqnarray} 
    \biggl( 
        \obu_r - \alpha^2 \Delta 
        \obu_r , \bphi_i 
        \biggr) 
        = \biggl(\bu_r, \bphi_i \biggr)
    \qquad \forall \, i=1, \ldots r,
    \label{equation:df-weak}
\end{eqnarray}
where $\alpha$ is the filter radius.  
The DF weak form~\eqref{equation:df-weak}
yields the following linear system:
\begin{eqnarray} 
    \left( \mathbbm{I} + \alpha^2 
    A \right) \underline{\ou}_{r}
    = \underline{u}_{r},
    \label{equation:df-linear-system}
\end{eqnarray}
where $\underline{\ou}_{r}$ is the vector of ROM coefficients of $\obu_r$
and $\mathbbm{I}$ and $A$ are the identity and ROM stiffness matrices, respectively.  
We note that $\mathbbm{I}$ in \eqref{equation:df-linear-system} 
is a consequence of the POD basis functions being orthonormal in the $L^2$ norm. Thus, the ROM mass matrix is the identity matrix. 

We emphasize that~\eqref{equation:df-linear-system} is a low-dimensional, $r
\times r$ 
linear system, whose computational overhead is negligible (since $r$ is relatively small).  
Thus, the ROM differential filter will be used to construct LES-ROMs that increase the ROM accuracy without significantly increasing the computational cost.  

The differential filter 
was used in LES of turbulent flows with classical numerical
discretizations~\cite{germano1986differential,layton2012approximate, BIL05}.  
In reduced order modeling, the ROM DF was used to develop LES-ROMs for the Kuramoto-Sivashinsky equation~\cite{sabetghadam2012alpha}, the
NSE~\cite{wells2017evolve,strazzullo2022consistency}, and the quasi-geostrophic equations~\cite{girfoglio2023linear}.

\paragraph{ROM higher-order algebraic filter.}

The second filter we review is the {\it ROM higher-order algebraic filter (HOAF)}: 
find $\obu_r(\bx)
= \sum_{j=1}^r \ou_{r,j} \bphi_j(\bx)$
such that
\begin{eqnarray} 
    \left( \mathbbm{I} + \alpha^{2m} A^{m} \right) \underline{\ou}_{r}
    = \uu_{r}, 	
    \label{eqn:hodf}
\end{eqnarray}
where 
$m$ is a positive integer.  
As explained in~\cite{mullen1999filtering} in the Fourier setting, the exponent $m$ controls the percentage of filtering at different wavenumbers: 
as $m$ increases, the amount of filtering increases for the high wavenumber components
and decreases for the low wavenumber components.  
Just as the DF~\eqref{equation:df-linear-system}, the HOAF~\eqref{eqn:hodf} is a relatively low-dimensional, $r \times r$ 
linear system, that can be efficiently solved.  
Thus, the HOAF also leads to the development of accurate and efficient LES-ROMs.
We also note that, for $m=1$, the linear systems~\eqref{eqn:hodf} and
\eqref{equation:df-linear-system} are identical.  Thus, DF can be considered
a particular case of HOAF with $m=1$.  

The HOAF~\eqref{eqn:hodf} was proposed in~\cite{gunzburger2019evolve} and
was based on the HOAF introduced by Fischer and
Mullen~\cite{mullen1999filtering} in a spectral element method (SEM) setting.  
In~\cite{gunzburger2019evolve}, the HOAF~\eqref{eqn:hodf} was called the
higher-order differential filter, to be consistent with the SEM nomenclature.  
In \cite{tsai2023time}, we showed that the HOAF is related to, but slightly different from, the spatial discretization of a higher-order differential operator  
(they are the same in the periodic case). 
For clarity, in this paper we call the operator in~\eqref{eqn:hodf} the high-order
algebraic filter. 
In \cite{tsai2023time}
a theoretical and numerical investigation of HOAF is presented.


\paragraph{ROM projection.}
The third ROM spatial filter used to construct LES-ROMs is the {\it ROM projection filter}, which is defined as follows:
For a fixed $r_1 < r$ and a given $\bu_r \in \bX^r = \text{span} \{ \bphi_1, \ldots, \bphi_r \}$, the ROM projection (Proj) seeks $\obu_r \in \bX^{r_1} = \text{span} \{ \bphi_1, \ldots, \bphi_{r_1} \}$ such that
\begin{eqnarray}
    ( \obu_r , \bphi_j) 
    = ( \bu_r , \bphi_j)
    \qquad
    \forall j = 1, \ldots, r_1.
    \label{eqn:rom-projection}
\end{eqnarray}

To our knowledge, the first time the Proj has been used as an explicit ROM spatial filter was in~\cite{wang2012proper}, where it was leveraged to construct the ROM dynamic subgrid scale
(SGS) model. 
We note that the dynamic SGS is generally considered to be the state-of-the-art closure model in LES~\cite{sagaut2006large,BIL05}.
Next, the ROM projection was used in~\cite{wells2015phd, wells2017evolve} to construct the Leray ROM (see Sec.~\ref{sec:leray-rom}) and EFR-ROM (see Sec.~\ref{sec:efr-rom}). 

\paragraph{ROM filter radius.}

The filter radius $\alpha$ is fundamental in the definitions of both the ROM differential filter~\eqref{equation:df-weak} and the ROM higher-order algebraic  filter~\eqref{eqn:hodf}:
the radius of the ROM spatial filters represents the size of the scales that are captured by the filtered variables.
Furthermore, the filter radius is also essential in the approximate deconvolution process, which depends on the particular filter used.
Thus, a natural question is ``How do we choose the ROM filter radius?"
For example, the following practical question could be asked: 
\begin{eqnarray}
    \boxed{
        \text{
            Given $10$ ROM basis functions, what is the resolved lengthscale?
        }
    }
    \label{eqn:rom-lengthscale-question}
\end{eqnarray}

The most straightforward answer to question~\eqref{eqn:rom-lengthscale-question} is to choose the ROM filter radius the same way we choose the FOM filter radius.
As mentioned in Sec.~\ref{sec:les-fom}, 
the first choice is a {\it computational scale}, i.e., a scale based on the underlying numerical discretization.
We also mention that other $\alpha(h)$ scalings are also possible~\cite{BIL05,layton2012approximate}.
Another choice for the FOM differential filter radius is a {\it physical scale}, i.e., based on the physics of the underlying flow.
For example, a popular choice in this class is $\alpha \sim \eta$, where we recall $\eta$ is the Kolmogorov scale.

One limitation of using the computational scale or the physical scale to define the ROM filter radius is that these scales lack adaptivity, i.e., do not change when we vary the ROM dimension, $r$.
To address this limitation, in~\cite{mou2023energy} we put forth a new, {\it energy-based ROM lengthscale}, which was defined by using a judicious partition of the energy.
We showed that the new energy-based ROM lengthscale displays the correct asymptotic behavior: it converges to the smallest lengthscale in the system when the ROM dimension increases, and it converges to the largest lengthscale in the system when the ROM dimension decreases.
Furthermore, in the numerical simulation of the turbulent channel flow at Reynolds number $Re_{\tau}=395$, we compared the EFR-ROM equipped with the new energy-based ROM lengthscale and a classical ROM lengthscale based on dimensional arguments~\cite{HLB96}.
The numerical results showed that the former is more robust with respect to parameter changes than the latter.
Finally, we note that the new energy-based ROM lengthscale could also be useful in the preprocessing strategy advocated in~\cite{aradag2011filtered,farcas2022filtering} as a means to filter out the noise in the input data.

\paragraph{Approximate deconvolution.}
Although the AD strategy is central in image processing and inverse problems~\cite{bertero1998introduction,hansen2010discrete}, and it has been used in the LES development (see Sec.~\ref{sec:ind_f} and the research monographs~\cite{sagaut2006large,BIL05,layton2012approximate,john2004large,rebollo2014mathematical}), it is a relatively new concept in reduced order modeling.
The AD 
strategy can be formulated as follows:
Given an approximation of the filtered input 
variable, $\obu \doteq \cF \bu$, where $\cF$ is an invertible spatial filter, find an approximation of the unfiltered input signal, $\bu$.
Of course, the exact deconvolution, $\cF^{-1} \obu$, would seem a natural choice.
Computationally, however, the exact deconvolution is a very bad idea.
Indeed, as shown in~\cite{bertero1998introduction,hansen2010discrete}, 
the noise in the high wavenumber components of $\obu$ is amplified by the inverse filter, $\cF^{-1}$ in the exact deconvolution.
Thus, in practice, AD strategies are used instead of the exact deconvolution~\cite{bertero1998introduction,hansen2010discrete}. 
In LES of turbulent flows, the AD models were pioneered by Adams and Stolz for classical numerical discretizations~\cite{Stolz1999}.
To our knowledge, in reduced order modeling, the first AD model was proposed in~\cite{xie2017approximate}, where AD was used to develop a ROM closure model.
Recently, the AD strategy was also used to develop the AD Leray ROM~\cite{sanfilippo2023approximate}, where three different AD strategies were  investigated: 
the van Cittert AD approach~\cite[Section 3.2]{sanfilippo2023approximate}, the Tikhonov AD approach ~\cite[Section 3.3]{sanfilippo2023approximate}, and the Lavrentiev AD approach~\cite[Section 3.4]{sanfilippo2023approximate}.
We also note that the Lavrentiev AD strategy was used in~\cite{xie2017approximate} to develop a ROM closure model, and the Tikhonov approach was used in~\cite{cordier2010calibration,wang20162d,weller2009robust} for ROM regularizations.
These three AD methods were investigated numerically in~\cite[Section 5.2]{sanfilippo2023approximate}.

\subsubsection{EFR-ROM}
    \label{sec:efr-rom}
The most popular LES-ROM exports the EFR strategy~\eqref{eq:EFR1}--\eqref{eq:EFR3} to the reduced order level, in which the FOM variables $\bU$ and $\bV$ are replaced by the ROM variables $\bu_r$ and $\bv_r$, respectively, and the filter and deconvolution operators are replaced by their ROM counterparts.
This yields the {\it evolve-filter-relax ROM (EFR-ROM)}: 
\begin{itemize}
\item[-] \emph{Evolve}: find $\bv^{n+1}$ such that
\begin{align}
    \frac{\bv_r^{n+1} - \bu_r^{n}}{\Delta t} + \cM(\bv_r^{n+1}) = \bf{0}. \label{eqn:efr-rom-1}
\end{align}
\item[-] \emph{Filter}:  find filtered variable $\overline{\bv}_r^{n+1}$ such that 
\begin{align}
\overline{\bv_r}^{n+1} = \cF \bv_r^{n+1}.    \label{eqn:efr-rom-2}
\end{align}
\item[-] \emph{Relax}: set 
\begin{align}
\bu_r^{n+1} = (1-\chi){\bv_r}^{n+1} + \chi \overline{\bv_r}^{n+1},\label{eqn:efr-rom-3}
\end{align}
with relaxation parameter $\chi \in [0, 1]$.
\end{itemize}

In the evolve step of the EFR-ROM algorithm, one time step of the standard Galerkin projection ROM (G-ROM) time discretization is leveraged to advance the EFR-ROM approximation at the current time step, $\bu_{r}^{n}$, to an intermediate EFR-ROM approximation, $\bv_{r}^{n+1}$.
In the filter step of the EFR-ROM algorithm, one of the ROM spatial filters presented in Sec.~\ref{sec:rom-filters-ad} is used to filter the intermediate EFR-ROM approximation from the evolve step and obtain a smoother ROM approximation, without spurious numerical oscillations. Finally, in the relax step of the EFR-ROM algorithm, the EFR-ROM approximation at the next time step is defined as a convex combination of the unfiltered intermediate
EFR-ROM approximation obtained in the evolve step, $\bv_{r}^{n+1}$, and its filtered counterpart, $\overline{\bv}_{r}^{n+1}$.  
Just like in the case of EFR used as FOM (see Sec.~\ref{sec:EFR}), the goal of the relax step is to adjust the amount of dissipation introduced in the filter step by using a relaxation parameter, $0 \leq \chi \leq 1$. 
By varying the relaxation parameter $\chi$, one can produce a range of filter strengths, from no filtering at all ($\chi=0$) to  maximum filtering ($\chi=1$).  As pointed out in Sec. \ref{sec:AoD} for the EFR used at the FOM level, the choice of $\chi$ may be critical in the presence of backflows.
At the ROM level, the numerical investigation in~\cite{strazzullo2022consistency} has shown that EFR-ROM is sensitive with respect to $\chi$.

The evolve-filter ROM was introduced in~\cite{wells2017evolve} and EFR-ROM in~\cite{gunzburger2019evolve}.  
Since then, EFR-ROM has been developed in several directions, for example, numerical simulation of turbulent flows (e.g., the turbulent channel flow)~\cite{mou2023energy}
and ocean dynamics \cite{girfoglio2023linear}, the FOM-ROM  consistency~\cite{strazzullo2022consistency}, and feedback control~\cite{strazzullo2023new}.

\subsubsection{Leray ROM, approximate deconvolution Leray ROM, and 
time-relaxation ROM}
    \label{sec:leray-rom}

In this section, we present several LES-ROMs constructed by using ROM spatial filtering and approximate deconvolution.

\paragraph{Leray ROM.}

The {\it Leray ROM (L-ROM)} ~\cite{wells2017evolve,xie2018numerical,gunzburger2020leray} is an LES-ROM in which the spatial filter, $\cF$, is used to smooth out only the 
convective velocity  of the nonlinear term in the NSE, i.e.:
\begin{eqnarray}
			\left(
				\frac{\partial \bu_r}{\partial t} , \bphi_{i} 
			\right)
			+ Re^{-1} \, 
			\left( 
				\nabla \bu_r , 
				\nabla \bphi_{i} 
			\right)
			+ \biggl( 
				(\obu_r \cdot \nabla) \bu_r ,
				\bphi_{i} 
			\biggr)
			= 0,  \quad \forall i=1, \ldots, r,
   \label{eqn:l-rom}
\end{eqnarray}
where $\obu_r$ is the ROM velocity filtered with one of the ROM spatial filters, i.e., 
DF~\eqref{equation:df-linear-system},  HOAF~\eqref{eqn:hodf}, or Proj~\eqref{eqn:rom-projection}. By introducing the spatial filtering of
the convective term, the Leray ROM smooths out the spurious numerical oscillations that affect the standard G-ROM in the convection-dominated, under-resolved regime.

Leray regularization (see Sec.~\ref{sec:EFR}) was first used in the context of reduced order models in~\cite{sabetghadam2012alpha}  for the Kuramoto-Sivashinsky equations.  
For fluid flows, L-ROM was first used in~\cite{wells2017evolve} for the 3D flow past a circular cylinder at $Re=1000$.  
Since then, L-ROM has been successfully used as a stabilization technique for various under-resolved
flows: the NSE~\cite{girfoglio2021pod,girfoglio2023hybrid}, the stochastic
NSE~\cite{gunzburger2019evolve,gunzburger2020leray}, 
the Boussinesq equations~\cite{kaneko2020towards,tsai2022parametric},  the quasigeostrophic equations~\cite{girfoglio2023linear,girfoglio2023novel}, and the turbulent channel flow~\cite{tsai2023time}.
We also note that the ROM projection was leveraged in~\cite{kaneko2020towards,tsai2022parametric, tsai2023parametric} to build the Leray ROM for challenging applications.

\paragraph{Approximate deconvolution Leray ROM.}

The {\it approximate deconvolution Leray ROM (ADL-ROM)}~\cite{sanfilippo2023approximate} is an extension of the Leray ROM that leverages  
the AD operator $\cD$. The idea of the ADL-ROM is used to limit the amount of dissipation introduced in the Leray ROM when the filter radius is too large:
Given $\boldsymbol{u}_{r}^{n}$, 
find $\boldsymbol{u}_{r}^{n+1}$ such that 

\begin{equation}\label{ADLROMeq}
	\left(\frac{\boldsymbol{u}_{r}^{n+1}-\boldsymbol{u}_{r}^{n}}{\Delta t}, \boldsymbol{\varphi}_{i}\right)+ {Re^{-1}}  \left(\nabla\boldsymbol{u}_{r}^{n+1}, \nabla \boldsymbol{\varphi}_{i}\right)+\left(\left(\cD(\overline{\boldsymbol{u}}_{r}^{n+1}) \cdot \nabla\right) \boldsymbol{u}_{r}^{n+1}, \boldsymbol{\varphi}_{i}\right)=0, \quad \forall i=1, \ldots, r.
\end{equation}
The numerical investigation in~\cite{sanfilippo2023approximate} for convection-dominated systems shows that, when the filter radius is relatively large, the ADL-ROM is more accurate than the standard Leray ROM and less sensitive to model parameters. In addition, it shows
that the increased accuracy allowed by the ADL-ROM~\eqref{ADLROMeq} is not at the expenses of
numerical stability. 

We emphasize that this is yet another instance in which ideas from completely different fields are synthesized.
Indeed, AD is central in the image processing and inverse problems communities and has been successfully used to develop LES models~\cite{Stolz1999,Stolz2001}, as mentioned also in Sec.~\ref{sec:ind_f}.




\paragraph{Time-relaxation ROM.}

The {\it time-relaxation ROM}~\cite{tsai2023time} is an LES-ROM recently proposed with the aim to increase the ROM numerical stability by adding an extra term to the incompressible NSE. Specifically, the time-relaxation ROM reads as follows:
Find $\bu_r$ such that
\begin{eqnarray}
    \left( \frac{\partial \bu_r}{\partial t} , \bphi_{i} \right)
    + Re^{-1} \left( \nabla \bu_r , \nabla \bphi_{i} \right)
    + \biggl( (\bu_r \cdot \nabla) \bu_r , \bphi_{i}  \biggr)
    + \biggl( \chi_t (\bu_r - \obu_r) , \bphi_{i}  \biggr)
    = 0 , \quad \forall \, i=1, \ldots r,
    \label{eqn:tr-rom}    
\end{eqnarray}
where $\chi_t$ is the time-relaxation parameter, and $\obu_r$ is the ROM
velocity filtered with one of the ROM spatial filters introduced in
Sec.~\ref{sec:rom-filters-ad}. 
The role of this extra term is to add numerical stabilization only at the marginally resolved scales, without unnecessarily filtering the resolved scales.

In \cite{tsai2023time}, it was shown that 
the numerical simulation of the turbulent channel flow at Reynolds numbers $Re_{\tau}=180$ and $Re_{\tau}=395$ performed with the new time-relaxation ROM equipped with the ROM differential filter and the ROM higher-order algebraic filter yiels accurate results.

\subsubsection{Other LES-ROMs}
    \label{sec:other-les-rom}


There are numerous LES-inspired ROMs, and the list is continuously growing. For a review on ROM closures, see~\cite{ahmed2021closures,sanderse2024scientific}).
We note that many of the ROM closures in \cite{ahmed2021closures,sanderse2024scientific}
do not use ROM filtering or approximate deconvolution.
Examples include the eddy viscosity ROMs and machine learning ROM closures~\cite{ahmed2021closures,sanderse2024scientific}).
Since the main goal of this paper is to discuss LES-inspired ROMs constructed by using ROM spatial filtering and AD, we do not discuss these alternative ROM closures.
We outline, however, two types of LES-ROMs that are different from those we presented in Secs.~\ref{sec:efr-rom}--\ref{sec:leray-rom}:

The {\it approximate deconvolution ROM (AD-ROM)} was introduced in~\cite{xie2017approximate}.
To our knowledge, this is the first AD model used in reduced order modeling.
The AD-ROM is constructed by using AD (i.e., a Lavrentiev regularization) to construct an accurate ROM closure model.
In~\cite{xie2017approximate}, the AD-ROM was successfully tested in the numerical simulation of the three-dimensional flow past a circular cylinder at a Reynolds number $Re = 1000$.

The {\it variational multiscale (VMS)} methods 
\cite{hughes1995multiscale,hughes2001multiscale} 
are constructed by using the principle of locality of energy transfer, which states that energy is transferred mainly between neighboring scales or modes. 
Since ROMs use hierarchical bases in which the large and small structures are clearly displayed, the VMS framework naturally lends itself to be extended to the ROM setting.
Although VMS-ROMs do not explicitly use a ROM spatial filter with an explicit filter radius, they display a connection with the LES-ROMs presented in this paper:
In the VMS-ROMs, the decomposition of the flow variables into large scales and small scales can be regarded as the result of using the ROM projection operator~\cite{reyes2020projection,CODINA2021103599,mou2021data}. 
For more discussion on VMS-ROMs, we refer the reader to \cite[Section IV.A.5]{ahmed2021closures}.

\subsection{LES-ROM Consistency}
    \label{sec:les-rom-consistency}

As explained in~\cite{strazzullo2022consistency,ingimarson2022full}, ROMs are of two types:
\begin{itemize}
    \item {\it FOM-ROM consistent}, i.e., the ROM uses the same computational model and the same numerical discretization as the FOM (see~\cite[Definition 1.1]{ingimarson2022full}).
    \item {\it FOM-ROM inconsistent}, i.e., the ROM uses a  computational and/or a numerical discretization that are different from those used by the FOM.
\end{itemize}

We note that most ROMs (and data-driven models in general) are FOM-ROM inconsistent.
The reason is simple:
The FOM-ROM inconsistent ROMs (which are often called ``nonintrusive'') are much easier to use in practice, e.g., because they allow for the use of legacy codes at the FOM level. 
We emphasize, however, that recent numerical investigations~\cite{strazzullo2022consistency,ingimarson2022full} have shown that the FOM-ROM consistent ROMs yield more accurate results than the FOM-ROM inconsistent ROMs.
For example, in~\cite{strazzullo2022consistency} 
we generated the FOM data by using the EFR-FOM
and we considered two types of ROMs:
A FOM-ROM consistent ROM that uses the EFR strategy at the ROM level, and a FOM-ROM inconsistent ROM that does not use the EFR strategy at the ROM level.
Our numerical investigation showed that the FOM-ROM consistent ROM was more accurate than the FOM-ROM inconsistent ROM.
Other examples of FOM-ROM consistent ROMs based on the EFR strategy are presented in \cite{girfoglio2021pod,girfoglio2023hybrid} (see  \cite{giere2015supg,zoccolan2023stabilized,zoccolan2024streamline} for FOM-ROM consistent ROMs based on different stabilizations).

The LES-ROMs that we presented in this section are FOM-ROM consistent since they use an LES strategy both at the FOM and the ROM levels.
Thus, based on the preliminary studies in~\cite{strazzullo2022consistency} for the EFR method, we believe that LES-ROMs will yield more accurate results than their inconsistent counterparts.
We also note that, by ensuring the modeling consistency between the FOM and the ROM, we increase the ROM robustness with respect to changes in the FOM parameters.
For example, when the FOM resolution needs to be adapted by changing the FOM filter radius, the ROM filter radius is adapted automatically and the ROM does not have to be retrained.
We emphasize that this is a departure from most of the current ROM closures and stabilization, which require re-training (a costly process) every time an LES model parameter is changed at the FOM level.

Of course, as pointed out in~\cite[Section 5]{strazzullo2022consistency}, there are still many open questions:
For example, 
one could investigate the FOM-ROM consistency when different LES models are used at the FOM and ROM levels, e.g., the Leray model is used at the FOM level and the EFR model is used at the ROM level. 
One could also investigate the {\it parameter FOM-ROM consistency}, which is complementary to the model FOM-ROM consistency. 
To investigate the parameter FOM-ROM consistency, one could consider the same LES model at the FOM and ROM levels, but use different parameters (e.g., different $\alpha$ values) in these 
LES models. 
For example, one could invetigate whether using different $\alpha$ or $\chi$ values at the FOM and ROM levels (i.e., $\alpha^{FOM} \neq \alpha^{ROM}$ or $\chi^{FOM} \neq \chi^{ROM}$) could yield more accurate ROM solutions.

We also emphasize that the FOM-ROM consistency has been investigated mainly numerically, without providing theoretical support.  
For notable 
exceptions, see the numerical analysis performed in~\cite{giere2015supg,pacciarini2014stabilized} for FOM-ROM consistency of the streamline upwind Petrov-Galerkin (SUPG) stabilization, and~\cite{ingimarson2022full} for FOM-ROM consistency with respect to the discretization of the nonlinearity of the Navier–Stokes equations. 
We note, however, that the numerical analysis of the FOM-ROM consistency 
for LES-ROMs is still an open question.

\subsection{LES-ROM Numerical Analysis}
    \label{sec:les-rom-na}


At the FOM level, extensive mathematical support exists for LES closures and stabilizations with classical numerical discretizations (e.g., the finite element method). For example, the monographs~\cite{BIL05,john2004large,rebollo2014mathematical} present the mathematical analysis for many LES models, as well as the numerical analysis of their discretization.
Similarly, the monograph~\cite{roos2008robust} presents the mathematical and numerical analysis of classical stabilization strategies.
This extensive literature  provides answers not only to fundamental numerical analysis questions (e.g., stability and convergence) but also to {\it critical practical challenges} (e.g., parameter scalings for critical parameters like the filtering radius). 
In stark contrast, fundamental questions in the numerical analysis are still wide open for most of the ROM closure models:
Is the proposed ROM stable?
Does the ROM converge?
If so, what does it converge to?
What is the ROM rate of convergence?

Only the first steps in the numerical  analysis  of ROM closures and stabilizations have been taken:
Numerical analysis results (e.g., stability and convergence) of classic ROM closures was performed in~
\cite{borggaard2011artificial,iliescu2013variational,iliescu2014variational,ballarin2020certified,rebollo2017certified}. 
Numerical analysis was carried out in~\cite{iliescu2013variational,iliescu2014variational} for the eddy viscosity variational multiscale ROMs and in~\cite{ballarin2020certified,rebollo2017certified} for the Smagorinsky model in a reduced basis method setting.
The first numerical analysis of LES-ROMs was performed in
~\cite{gunzburger2020leray,xie2018numerical} for the  Leray ROM~\eqref{eqn:l-rom} (see also~\cite{azaiez2021cure} for related work), and in~\cite{strazzullo2023new} for the EFR-ROM.
We have also taken the first steps in proving FOM-ROM  {\it parameter scalings} for ROM stabilizations and closures~\cite{giere2015supg, iliescu2013variational,iliescu2014variational,john2022error}. 
For example, in~\cite{giere2015supg}, for the SUPG-ROM, we proved scalings between the FOM mesh size and the SUPG-ROM stabilization parameter.

Finally, we note that the most active research area in ROM closure modeling is in the development of data-driven ROM closures in which available data is utilized to build the ROM closure model.  
An example of data-driven ROM closure is the data-driven variational multiscale ROM (d2-VMS-ROM) proposed in~\cite{mou2021data,xie2018data,ivagnes2023pressure}.
The d2-VMS-ROM and its developments have been investigated numerically in~\cite{ivagnes2023pressure,ivagnes2023hybrid,koc2019commutation,mohebujjaman2019physically,mou2020data,mou2021data,xie2018data,xie2020closure,ahmed2023physics}.
However, providing mathematical support for the d2-VMS-ROM, and data-driven ROM closures in general, is an open problem.
Recently, in~\cite{koc2022verifiability}, we laid the mathematical foundations of data-driven ROM closures.
Specifically, in~\cite[Theorem 2]{koc2022verifiability}, we proved that the d2-VMS-ROM is \emph{verifiable}, i.e., the d2-VMS-ROM solution is accurate since the d2-VMS-ROM closure model is accurate.

Despite these first steps that our groups and our collaborators have taken, the numerical analysis of ROM closures and stabilizations is nowhere close to the numerical analysis for FOM closures and stabilizations.
Laying the mathematical foundations for LES-ROMs and, more importantly, providing numerical analysis guidance to practical LES-ROM choices (e.g., parameter scalings) are still open questions.

\section{Applications of LES for ROM}\label{sec:app_ROM}


In this section, we present some results of LES with ROM on test cases with an academic flavor and some real applications.
We also draw a road map of other potential applications for more real-world problems that motivated us in the development of the methodologies described above. We anticipate these applications to provide an excellent benchmark for the assessment and improvement of such methodologies.

\subsection{Incompressible flows}\label{sec:inc_ROM}

\paragraph{Flow 
past a cylinder.} 

Although simple (or maybe precisely because of that), the 
2D \cite{turek1996,John2004} and 3D \cite{turek1996,John2006,Bayraktar2012}
flow past a cylinder benchmarks 
have been widely used to assess numerical 
strategies, even those designed for higher 
Reynolds number flows. 
This is due to the fact
that, to obtain accurate results in the case of time-dependent Reynolds number $0 \leq Re(t) \leq 100$,
one needs a mesh with about 200K elements for the 2D benchmark and roughly 3 million elements for the 3D benchmark. Thus, one can use these benchmarks to demonstrate that accurate results can be obtained
also with much coarser meshes if suitable LES models are used. Indeed, for example, it has been shown
that with the EFR algorithm, one can get accurate results with meshes as coarse as roughly 15K elements in both 2D and 3D \cite{layton2012modular,bertagna2016deconvolution,abigail_CMAME,GIRFOGLIO201927,girfoglio2021pod, Girfoglio_ROM_Fluids,strazzullo2022consistency}.
Below, we present selected 2D results from \cite{girfoglio2023hybrid}, which touch upon several topics discussed in Sec.~\ref{sec:les-rom}.

The work in \cite{girfoglio2021pod,girfoglio2023hybrid} stemmed from the need for FOM-ROM consistency described in Sec.~\ref{sec:les-rom-consistency}. We proposed to use the EFR algorithm at both the full and reduced order level. 
In particular, we 
generate the snapshots with under-refined meshes. 
This is unlike other works, e.g., \cite{Xie2018_2,wells2017evolve,gunzburger2019evolve}, where 
the snapshots are obtained by DNS. 
By using the EFR method as FOM and ROM, we are adopting
the same mathematical framework during both the \emph{offline} and \emph{online} stage and thus
have a ROM that is fully consistent with the FOM.
In \cite{girfoglio2023hybrid}, we proposed to use the POD basis related to the evolve velocity to approximate the filtered velocity and compute the reduced pressure field with a Poisson Pressure Equation method 
\cite{Stabile2018,akhtar2009stability}. The main difference between \cite{girfoglio2023hybrid} and  \cite{girfoglio2021pod, Girfoglio_ROM_Fluids,strazzullo2022consistency}
lies in the indicator function: it is linear in 
\cite{girfoglio2021pod, Girfoglio_ROM_Fluids,strazzullo2022consistency} 
and AD-based \eqref{eq:a_D0_a_D1} in \cite{girfoglio2023hybrid}. The ROM in \cite{girfoglio2023hybrid} is called hybrid projection/data-driven because it
exploits a traditional projection method (G-ROM)
for the computation of the intermediate reduced velocity and pressure fields, and uses a
data-driven technique to compute the reduced coefficients of the indicator function field.
The data-driven approach leverages interpolation with Radial Basis Functions \cite{Lazzaro2002}.
Alternatives to the data-driven approach could be
using the same set of reduced coefficients for velocity, pressure, and indicator function
or an EIM/DEIM technique \cite{barrault04:_empir_inter_method,Chaturantabut2010}.
The former was considered 
for RANS in \cite{Lorenzi2016} and was shown in \cite{Hijazi2020} to provide less accurate results than a hybrid procedure. 

\begin{figure}[htb]
\centering
       \begin{overpic}[width=0.7\textwidth]{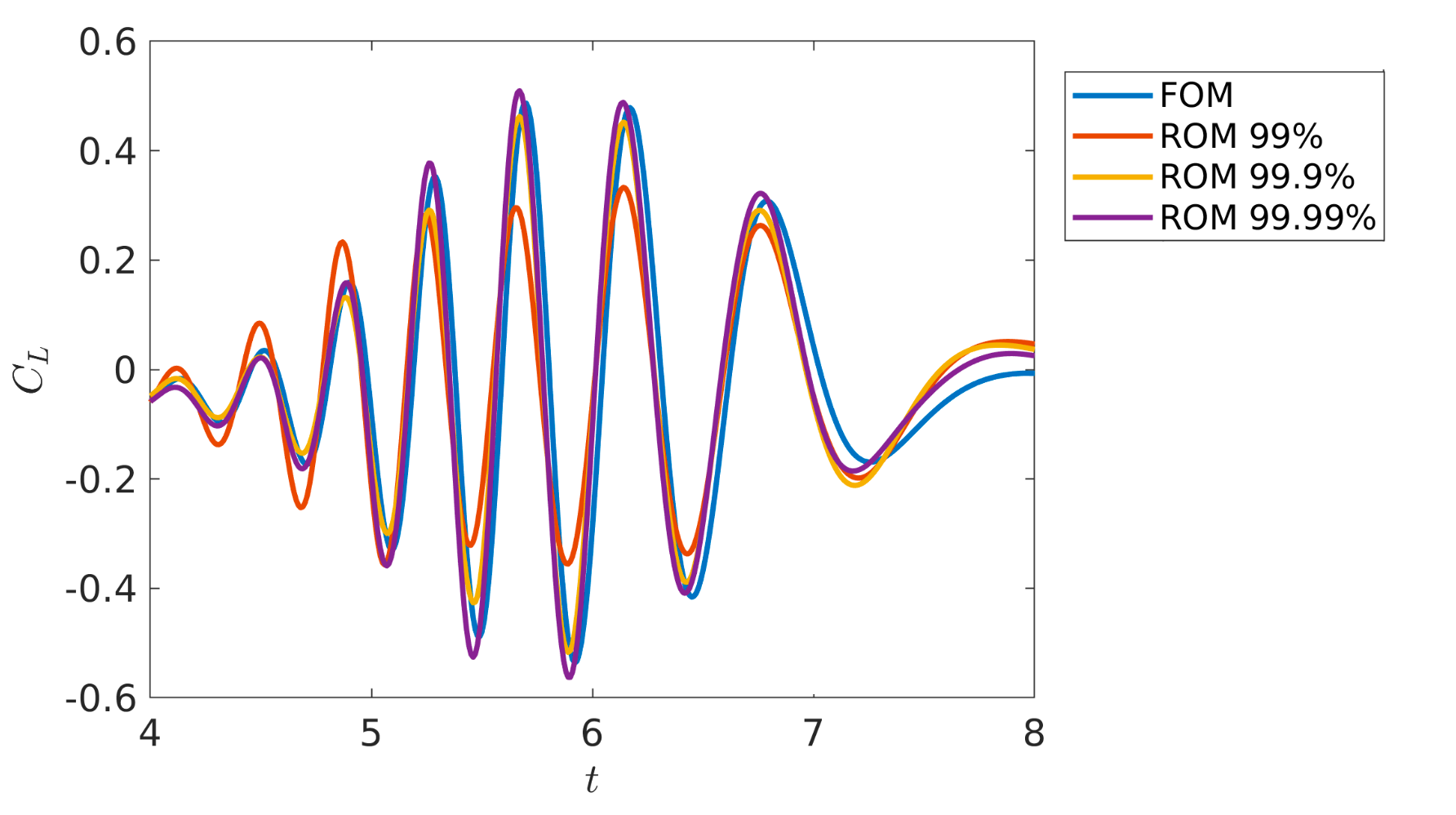}
      \end{overpic}\\
\caption{ 
Lift coefficient $C_L$ computed by the FOM and the projection/data-driven ROM from \cite{girfoglio2023hybrid} for different thresholds of cumulative energy.
}
\label{fig:coeff_t_nested}
\end{figure}

The 2D flow past a cylinder for $0 \leq Re(t) \leq 100$ is challenging because the flow is laminar for the first 4 s and then the well-known vortex shedding appears. 
In \cite{girfoglio2023hybrid}, the focus was on the time interval $[4, 8]$ s, for which 200 high-fidelity snapshots were collected. 
The first 50 most energetic POD modes were considered and a convergence test was performed based on three different energy thresholds: 
99\% (11 modes for the velocity, 5 modes for the pressure, and 20 modes for the indicator function), 
99.9\% (26 modes for the velocity, 12 modes for the pressure, and 44 modes for the indicator function), 
and 99.99\% (42 modes for the velocity, 24 modes for the pressure, and 50 modes for the indicator function).
Figure~\ref{fig:coeff_t_nested} compares the lift coefficient given by the FOM with the one given 
by the ROM. We observe that the evolution of 
lift coefficient computed
by the ROM is very accurate when 
99.9\% or 99.99\% of the eigenvalue energy is retained.
Similar accuracy was found for the 3D cylinder test. 
See \cite{girfoglio2023hybrid} for more details.

\begin{figure}[htb]
\centering
 \begin{overpic}[width=0.44\textwidth,grid=falso]{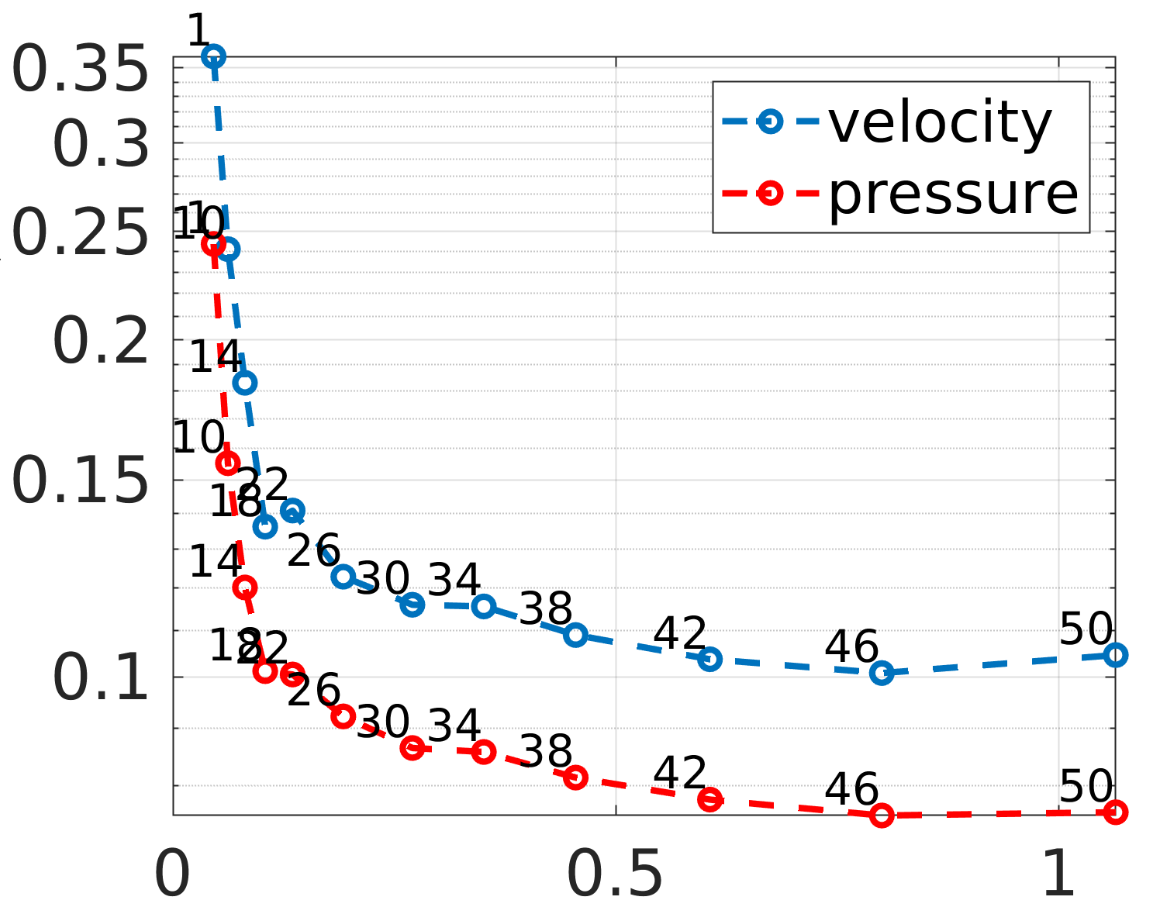}
        \put(40,-2){relative wall time}
        \put(-7,43){errors}
      \end{overpic}~
       \begin{overpic}[width=0.45\textwidth,grid=falso]{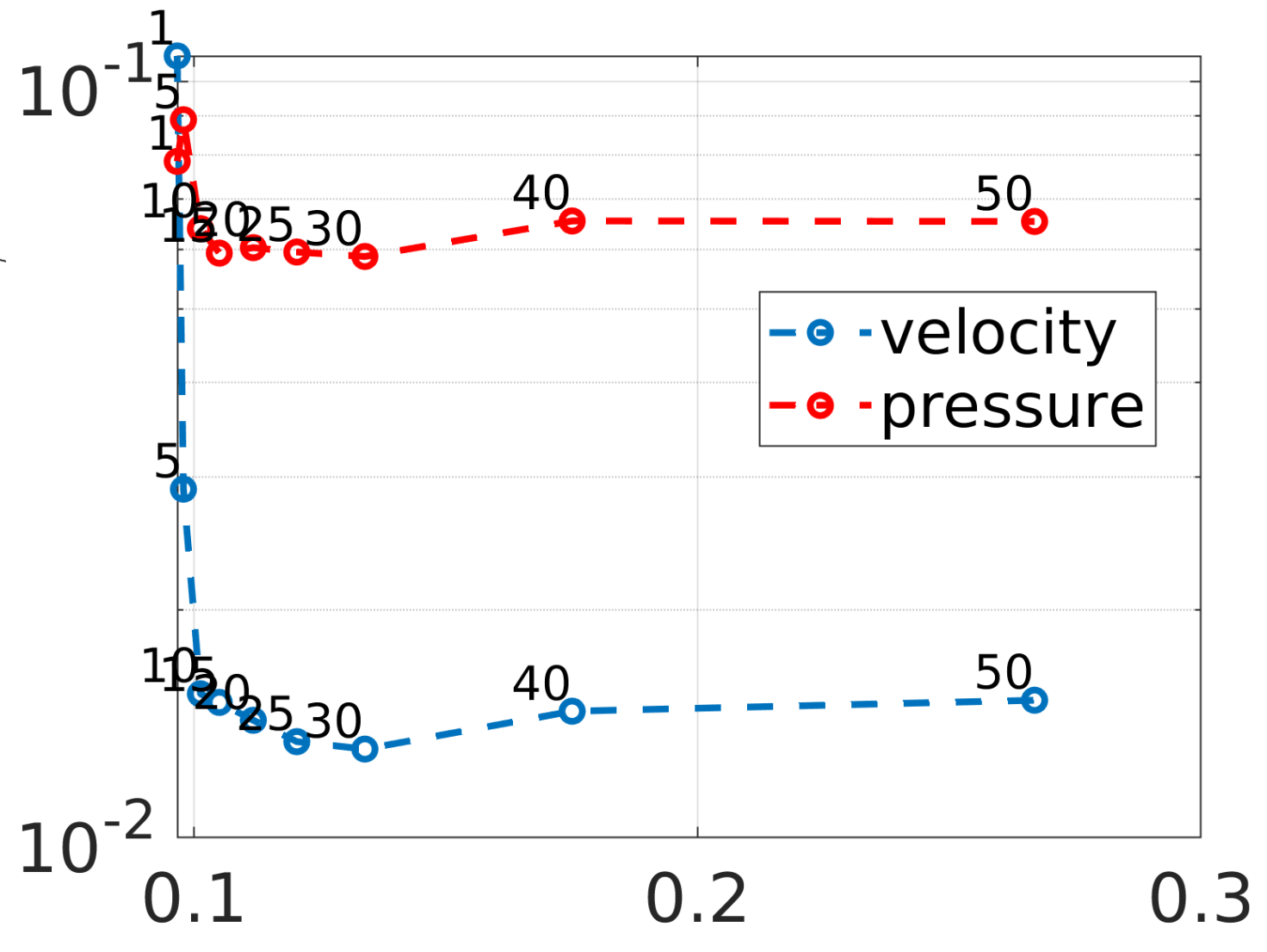}
        \put(40,-2){relative wall time}
        \put(-1,43){errors}
      \end{overpic}
      \caption{Pareto plots for the velocity and pressure: time-averaged relative $L^2$ error versus relative wall time when the number of basis
      functions for the velocity is varied for the 2D (left) and 3D (right) cylinder tests.}
\label{fig:SU_2D}
\end{figure}

To complete the picture, we need to connect the accuracy with the required computational time. 
Figure \ref{fig:SU_2D} (left) shows the time-averaged relative $L^2$ error for velocity and pressure versus the corresponding wall time when the number of basis
functions for the velocity is varied. The number
of basis functions for the pressure is fixed at 24, 
while the number
of basis functions for the indicator function is 50.
The relative wall time is given by the ratio between the CPU time needed by the ROM and 
the CPU time required by a FOM simulation (i.e., 398 s). As one would expected, the errors initially decrease and the relative wall time increases
as the number of basis functions increases.
The errors reach their minimum value for 46
basis functions for the velocity.
When the number of basis functions is increased to 50, two undesirable things happen: the relative wall time is larger than 1, meaning the ROM is more costly than the FOM (an absurdity), and the errors increase. 
Since 42 velocity modes are enough to retain 99.99\% of the eigenvalue energy, we suspect the errors increase
because of the noise introduced by the highest modes.
We note that the ROM is more computationally efficient in the 3D test.
See Figure \ref{fig:SU_2D} (right) and \cite{girfoglio2023hybrid} for details.

\paragraph{T-junction.}

\begin{figure}[h!]
    \centering
    \includegraphics[width=0.6\linewidth]{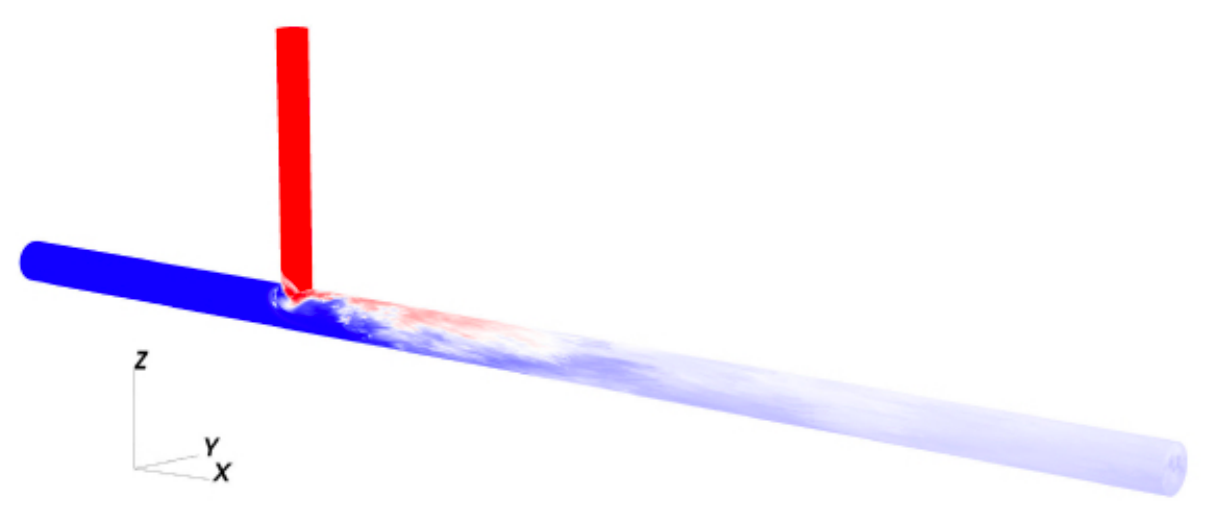}    
    \caption{
        T-junction test case:  Instantaneous temperature field at $Re = 10,000$
        for an investigation on thermal striping, which is critical in nuclear engineering~\cite{tsai2023parametric}.
    }
    \label{fig:t-junction}
\end{figure}

Here, we summarize the results of the numerical investigation performed in~\cite{tsai2023parametric}.
We investigate three of the LES-ROMs discussed in Sec.~\ref{sec:les-rom} (i.e., the EFR-ROM, Leray ROM, and time-relaxation ROM) in the numerical simulation of a T-junction problem. 
When streams of rapidly moving flow merge in a T-junction, large oscillations can occur at the scale of the diameter, $D$, with a period scaling as $\mathcal{O}(D/U)$, where $U$ is the characteristic flow speed. 
If the streams are at different temperatures, the oscillations result in fluctuations (see Fig.~\ref{fig:t-junction}). This 
phenomenon, known as thermal striping, can accelerate thermal-mechanical fatigue at the pipe wall in the outlet branch and ultimately cause pipe failure. Since thermal striping
is of critical importance in nuclear engineering, the nuclear energy modeling and simulation community established a T-junction benchmark~\cite{obabko2011cfd} to test the ability of CFD codes to predict thermal striping.

\begin{figure}[h!]
    \centering    
    \includegraphics[width=\linewidth]{
    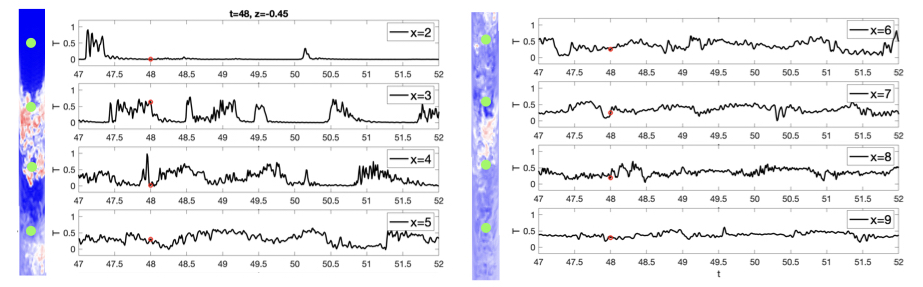}    
    \caption{
        Near-wall temperature history at $y = 0$, $z = 0.45$ for several $x$ locations in the outlet branch. 
    \label{fig:t-junction-temperature}
    }
\end{figure}

We consider the T-junction problem at the Reynolds number $Re = 10,000$. 
The FOM is based on a spectral element discretization with about $21$ million mesh points.
Further details about the computational setting and FOM are given in~\cite{tsai2023parametric}.
We test the standard G-ROM and the three LES-ROMs in the predictive regime, that is, we consider a time interval that is larger than the interval where snapshots are collected. 
We consider $r = 250$ for the G-ROM and $r = 100$ for the LES-ROMs. 
To assess the ROMs’ performance, 
we compare
the near-wall temperature at several locations ($(x,\pm0.5, 0), (x,\pm0.45, 0), (x, 0,\pm0.5)$, and $(x, 0,\pm0.45)$ 
for $x = 2, 3, . . . , 9$) with the FOM data 
(see Fig.~\ref{fig:t-junction-temperature}).

\begin{figure}[h!]
    \centering    
    \includegraphics[width=0.9\linewidth]{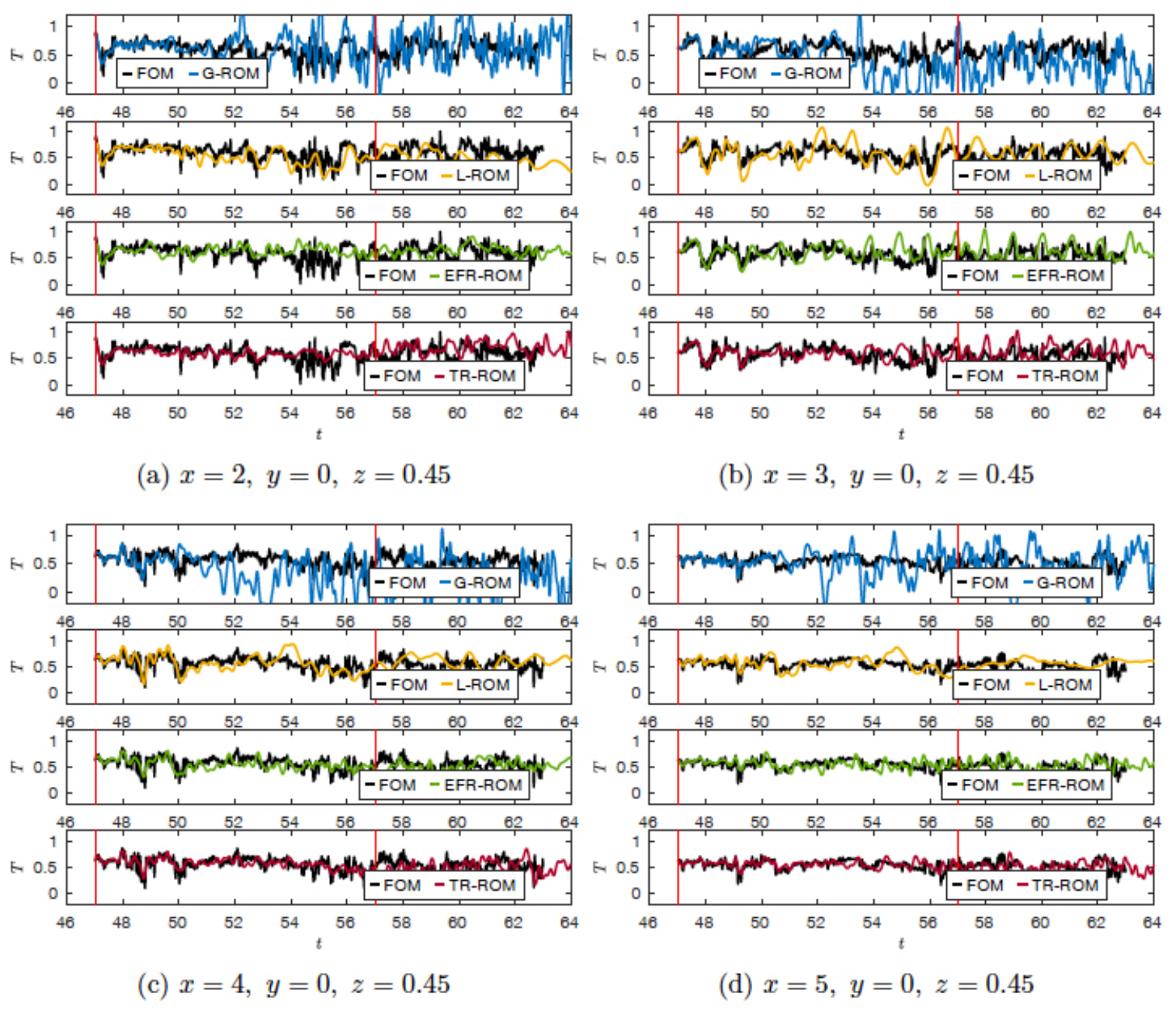}    
    \caption{
        T-junction at $Re = 10,000$: 
        Comparison of the near-wall temperature history at $z = 0.45$ between the FOM, the G-ROM, and the LES-ROMs for $x = 2, 3, 4, 5$ (outlet branch).
    }
    \label{fig:t-junction-les-rom-1}
\end{figure}

\begin{figure}[h!]
    \centering    
    \includegraphics[width=0.9\linewidth]{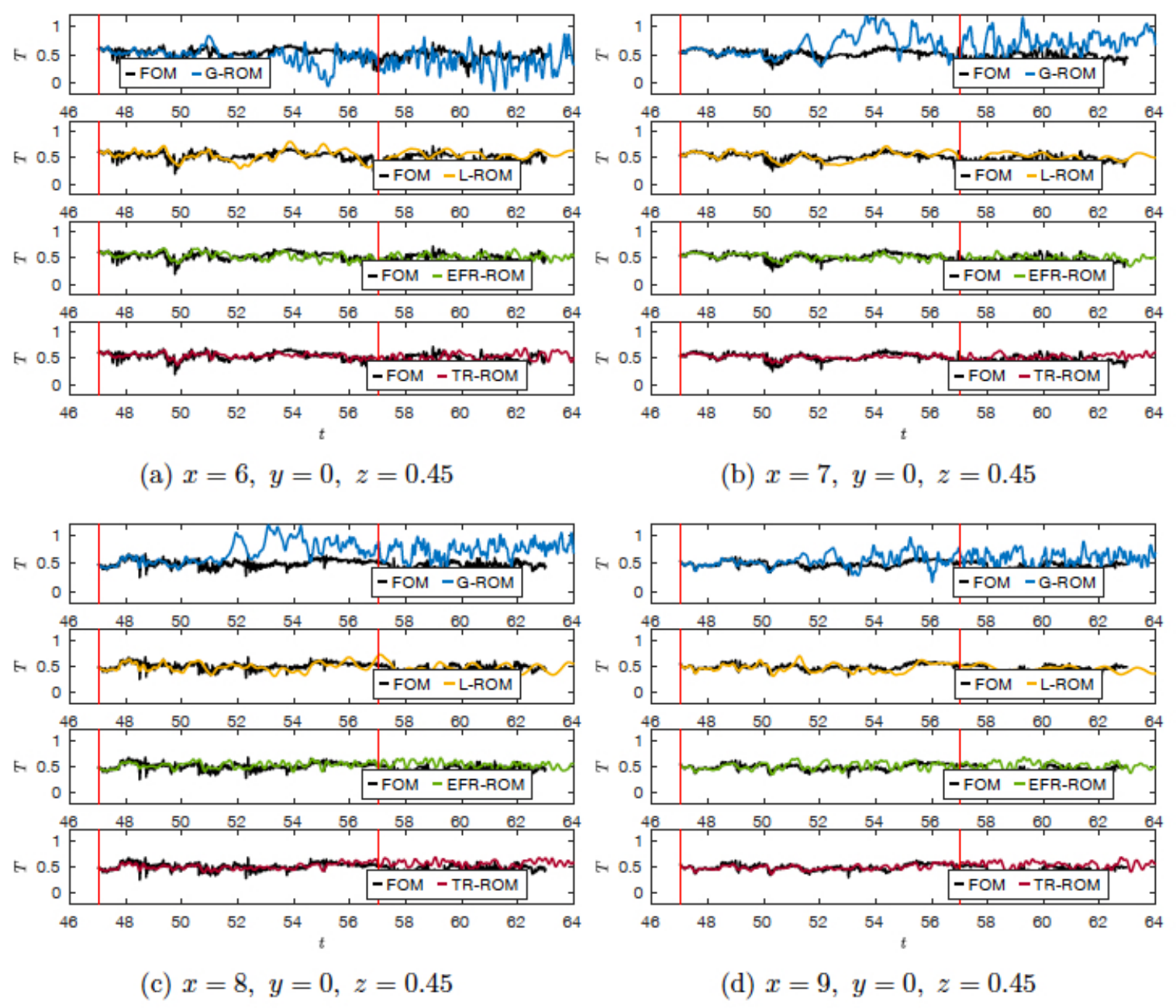}    
    \caption{
        T-junction at $Re = 10,000$: 
        Comparison of the near-wall temperature history at $z = 0.45$ between the FOM, the G-ROM, and the LES-ROMs for $x = 6, 7, 8, 9$ (outlet branch).
    }
    \label{fig:t-junction-les-rom-2}
\end{figure}

Figures~\ref{fig:t-junction-les-rom-1}--\ref{fig:t-junction-les-rom-2} show the near-wall (i.e., $z = 0.45$)
temperature of the down-stream pipe at several $x$ locations for the G-ROM, the Leray ROM, the
EFR-ROM, and the time-relaxation ROM. 
The two vertical red lines denote the time interval in which the snapshots were collected.
The results indicate that the near-wall temperature of the G-ROM initially agrees with the
FOM but becomes unstable in a relatively short time ($\approx 6$ 
convective time units after the initial condition) even with $r = 250$. This indicates that the G-ROM 
is not accurate even in the reconstructive regime. 
Furthermore, larger fluctuations are observed for small $x$ values. 
In contrast, the LES-ROM temperature stays on the overall FOM trajectory, and
in the predictive regime (i.e., outside the region between the two vertical red lines) 
all three LES-ROMs are significantly more accurate than the G-ROM.

\paragraph{Hemodynamics applications.}

As mentioned in Sec.~\ref{sec:inc_FOM}, blood flow is generally in a laminar regime, and it does not require specific modeling for turbulence.
AoDs are an exception where the combination of the specific flow features induced by the tiny entry tears and the need for an efficient solver to retrieve information from many patients of a (computer-assisted) clinical trial
call for specific modeling solutions like EFR.

\begin{figure}[hbtp]
    \centering
    \includegraphics[width=0.75\linewidth]{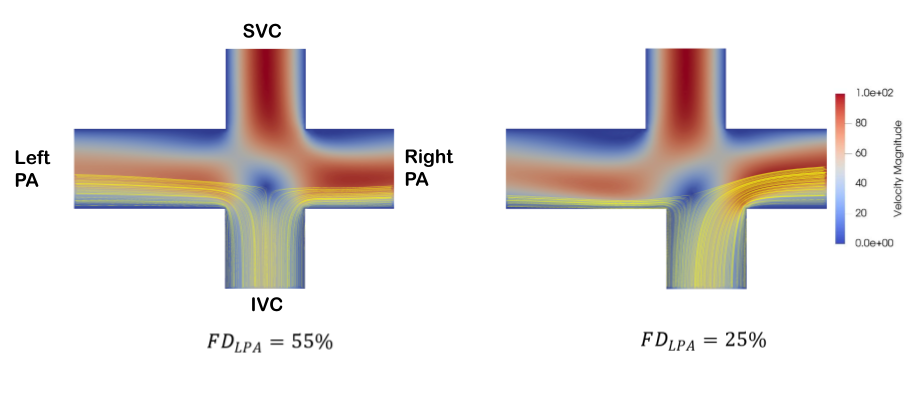}
    \caption{Simplified geometry of a TCPC. The vertical vessel is the vena cava (VC: superior at the top - SVC, inferior at the bottom - IVC). The pulmonary artery (PA) is the horizontal vessel. The inflow sections are at the SVC and at the IVC. This generates colliding fronts. The picture reports the results corresponding to two different surgical options. The difference is in the flow distribution from the IVC (the so-called Hepatic Flow Distribution): $FD_{LPA}$ is the fraction of hepatic flow directed to the Left PA. An even flow distribution (i.e., $FD_{LPA}\approx 50\%$) is desirable. Notice that the different $FD_{LPA}$ are created by different offsets between the SVC and IVC.}
    \label{fig:tcpc}
\end{figure}

There are other problems in the field of computational hemodynamics where the need for extremely efficient 
solvers is critical. Typically, these are problems where the use of specific medical devices requires an optimization procedure. While ``optimization'' can be done by trial-and-error approaches, translational mathematicians (i.e., mathematicians engaged in bringing state-of-the-art mathematical and computational procedures to the forefront of medical practice) feel the need for introducing automatic or, at least, semi-automatic rigorous, mathematically sound procedures. The cost of several attempts to find the optimal solution, either as the result of empirically educated guesses or iterations of a mathematical optimization procedure, may be prohibitive.
In general, in this kind of application, efficiency is privileged with detrimental consequences for accuracy.
For this reason, RANS models are often preferred. 

An example of great interest in pediatric surgery is related to the so-called 
{\it Total Cavopulmonary Connection (TCPC)}. This is a palliative surgery for newborn babies with a severe heart malformation called Left Ventricle Hypoplasia Syndrome. This condition 
(called {\it univentricular circulation}) is not compatible with life and requires specific treatments. Years ago, 
the TCPC was introduced to buy time before a transplant
\cite{de1988total}. It consists of an artificial connection between the superior and inferior venae cavae  (SVC and IVC, respectively) with the pulmonary artery (PA) to route systemic circulation to the pulmonary circulation. 
An oversimplified representation of this shape is reported in Fig.~\ref{fig:tcpc}. 
The functioning ventricle pumps blood into the large circulation. The connection between the venae cavae and the PA has a peculiar cruciform shape, shown in Fig.~\ref{fig:tcpc},  with many interesting implications in terms of fluid dynamics. The biomedical engineering community has investigated these implications in a huge number of excellent contributions, see, e.g., \cite{sharma1996vitro,migliavacca1999computational,ensley1999toward,dubini2004ten,hsia2016multiscale,schiavazzi2015hemodynamic,pekkan2005total,tang2014geometric,khiabani2012effect,dasi2009fontan}.
From the computational point of view, the presence of the colliding fronts injected at the SVC and IVC
raises some challenges since even moderate Reynolds numbers may induce flow disturbances. In general,
the shape of the cruciform connection was speculated to have a major role in the long-term healthy conditions of the patient. For instance, an even splitting of the flow coming from the IVC (called {\it Hepatic Flow}) is critical and is determined by
the offset between the SVC and IVC left after the surgery (see Fig. \ref{fig:tcpc}).
However, a rigorous shape optimization based on mathematical procedures in the operating room is still a dream for many reasons. One is the identification of all the relevant criteria that define a reliable and robust ``optimal'' solution. Another one is the huge computational cost of numerical optimization.
On top of this, a more recent variant of the surgery requires the introduction of a pump in the cross-shaped junction to establish hemodynamic conditions closer to the physiology of bi-ventricular circulation \cite{rodefeld2003cavopulmonary}. The regime induced by the pump requires specific turbulence modeling \cite{delorme2013large}.
High computational costs currently led to the use of RANS models, even though the superiority of LES modeling for the flow regimes characteristic of TCPC is commonly recognized \cite{sarfare2023computational}. Our intention is to bring LES-ROM modeling into this field as a potential breakthrough that enables 
not only accurate LES modeling as the routinary approach,
but also rigorous shape optimization procedures as part of regular surgical planning.

\paragraph{Wind energy applications.}
LES-based wind modeling presents significant challenges, with key questions posed on turbulence, mostly focused on mean-flow kinetic energy (MKE) entrainment and dissipation in large wind farms \cite{meneveau2019big}. Understanding the upper limit of power production is linked to MKE entrainment. Recent studies, such as \cite{antonini2021spatial}, identified spatial constraints and the influence of the Coriolis parameter on power output through idealized atmospheric simulations run with the Weather Research and Forecasting model. An entrainment-based model for wind farm flow and power prediction was recently proposed in \citep{bempedelis2023turbulent}. Wind energy, projected to meet 35\% of global electricity needs by 2050, faces significant challenges, including optimizing wind farm layouts and minimizing wake interactions that reduce power output and affect turbine longevity. With this in mind, it is worth noting that enhancing energy production by just 1\% could yield 30 TWh annually, equivalent to adding 3,600 turbines and \$1 billion in revenue \cite{howland2022collective}. 

The filtering-based approaches reviewed in this paper can provide accurate modeling tools to predict wind farm wakes, addressing spatio-temporal variability, unsteady wake interactions, and atmospheric turbulence. Moreover, improved turbine wake prediction and cost-effective surrogate models can optimize turbine layout and support wind farm deployment \cite{pawar2022towards}. For further insights on LES of wind applications, we refer the reader to references \cite{bempedelis2023turbulent,antonini2021spatial,luzzatto2018entrainment,sedaghatizadeh2018modelling,mehta2014large,meneveau2012top,porte2011large}. 
Morover, a recent study \cite{stadtmann2023digital} offers a comprehensive overview of digital twin technology, with a focus on wind energy applications. The study consolidates digital twin definitions, identifies the current state-of-the-art in modeling and simulation techniques, and outlines research needs in the wind energy sector from an industrial perspective. It proposes solutions to these challenges from research institutes' viewpoints and provides recommendations for stakeholders to facilitate technology adoption, with the methodologies highlighted in our article seen as key enablers for this digital technology adaptation.

\subsection{Compressible flows}

ROMs combined with LES modeling has the potential to 
be a major player in reducing the computational cost associated with weather forecast. 
This section is meant to convey an ideal of such potential and show some limitations of current methodologies, as an incentive to do more work in this 
research area. Mainly, we report results from \cite{HAJISHARIFI2024104050} obtained with selected
data-driven ROMs for the same benchmarks 
for atmospheric flow described in Sec.~\ref{sec:LESFOM_comp}. However, let us start
with a brief literature review. 

While POD, which is also referred to as Empirical Orthogonal Function (EOF) analysis in the geophysical
fluid dynamics community,
has been used for a long time,
it is only recently that ROMs have been applied for the simulation of
atmospheric flows. 
EOF has been applied to identify spatio-temporal coherent 
meteorological patterns, e.g., the Madden-Julian Oscillation, the Quasi-Biennial Oscillation, 
and the El Ni\~{n}o-Southern Oscillation. See, e.g., \cite{lario2022neural,pawar2022equation,schmidt2019spectral,chen2014predicting,chen2018predicting}.
The EOF analysis for these phenomena uses data on the global scale and 
considers time (ranging from several months to many years) as the only parameter. 
In addition, it is mostly limited to data analysis as a mean to understand the weather system, i.e., EOF 
is not used for forecasts. 
Only very recently, a data-driven ROM based on EOF analysis 
has been used for pattern prediction, specifically to forecast the weekly average sea surface temperature \cite{pawar2022equation}.  
Other data-driven methods borrowed from 
machine learning have been applied to global weather forecasting. See, e.g., \cite{pathak2018model,rasp2021data,schultz2021can,weyn2019can,GraphCast,FourCastNet,Pangu-Weather}.

The work in \cite{HAJISHARIFI2024104050} is a first attempt to apply ROMs that generate a reduced order basis (unlike machine learning methods) to both reconstruct and forecast of regional atmospheric flows (unlike EOF, which is not used to forecast). Additionally, since 
the spatial scale is a few kilometers and the time scale is a few hours, there is an obvious difference in resolution from EOF analysis. The considerd data-driven ROMs are: Dynamic Mode Decomposition (DMD), Hankel Dynamic Mode Decomposition (HDMD), and Proper Orthogonal Decomposition with Interpolation (PODI). DMD was specifically designed to predict the future behavior of a system \cite{kutz2016dynamic,schmid2010dynamic, schmid2011applications, tu2013dynamic} and has been successfully applied to
several incompressible fluid dynamics
problems \cite{schmid2011application,duke2012experimental,seena2011dynamic}. HDMD enhances 
the DMD algorithm with time-delay embedding \cite{arbabi2017ergodic,  curtis2023machine, fujii2019data, jiang2015study,vasconcelos2019dynamic, yang2021synchronized} and is able to
predict more accurately and for longer periods of time systems exhibiting strong nonlinear dynamics \cite{frame2022space,vasconcelos2019dynamic}. Successful applications include
periodic cavity flow \cite{arbabi2017ergodic,Hess2023}, electromechanical systems  \cite{yang2021synchronized}, 
and biological systems \cite{fujii2019data}.
Unlike DMD and HDMD, 
PODI was not designed to forecast the system evolution, but rather to interpolate solutions in a parameter space, where time is one of possibly many parameters of interest. 
So far, PODI has been applied to perform parametric studies for problems in hemodynamics \cite{girfoglio2022non}, chemical \cite{hajisharifi2023non} and naval \cite{demo2018efficient, demo1} engineering, 
and aeronautics \cite{ripepi2018reduced}.

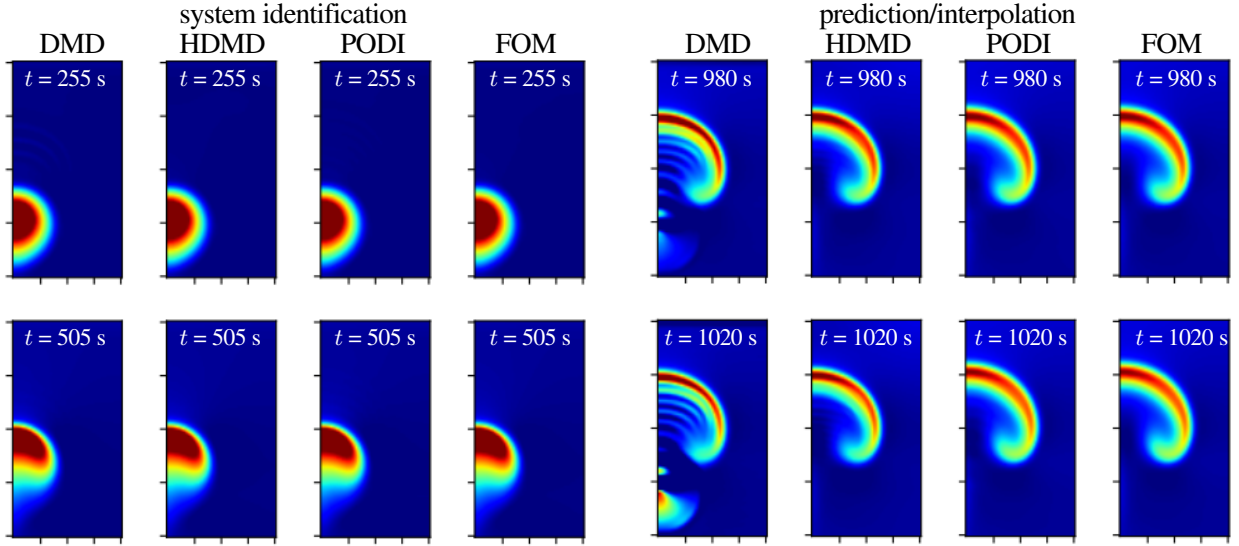
\begin{figure}[htb!]
     \centering
     \vspace*{0.4cm}
         \begin{overpic}[width=0.48\textwidth, grid=false]{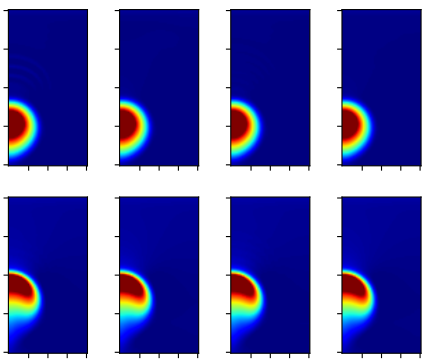}
    \put(30,88){{system identification}}
    \put(6.5,83){DMD}
    \put(30,83){HDMD}
    \put(57,83){PODI}
    \put(83,83){FOM}
    \put(4,77.5){\footnotesize{\textcolor{white}{$t$ = 255 s}}}
    \put(30,77.5){\footnotesize{\textcolor{white}{$t$ = 255 s}}}
    \put(56,77.5){\footnotesize{\textcolor{white}{$t$ = 255 s}}}
    \put(82,77.5){\footnotesize{\textcolor{white}{$t$ = 255 s}}}
    \put(4,34){\footnotesize{\textcolor{white}{$t$ = 505 s}}}
    \put(30,34){\footnotesize{\textcolor{white}{$t$ = 505 s}}}
    \put(56,34){\footnotesize{\textcolor{white}{$t$ = 505 s}}}
    \put(82,34){\footnotesize{\textcolor{white}{$t$ = 505 s}}}
    \end{overpic} \hfill
             \begin{overpic}[width=0.48\textwidth, grid=false]{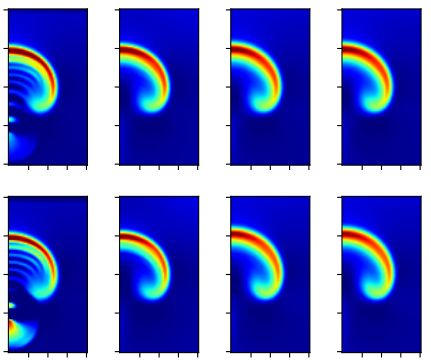}
    \put(29,88){{prediction/interpolation}}
    \put(6.5,83){DMD}
    \put(30,83){HDMD}
    \put(57,83){PODI}
    \put(83,83){FOM}
    \put(4,77.5){\footnotesize{\textcolor{white}{$t$ = 980 s}}}
    \put(30,77.5){\footnotesize{\textcolor{white}{$t$ = 980 s}}}
    \put(56,77.5){\footnotesize{\textcolor{white}{$t$ = 980 s}}}
    \put(82,77.5){\footnotesize{\textcolor{white}{$t$ = 980 s}}}
    \put(3,34){\footnotesize{\textcolor{white}{$t$ = 1020 s}}}
    \put(29,34){\footnotesize{\textcolor{white}{$t$ = 1020 s}}}
    \put(55,34){\footnotesize{\textcolor{white}{$t$ = 1020 s}}}
    \put(82,34){\footnotesize{\textcolor{white}{$t$ = 1020 s}}}
    \end{overpic}
    \caption{Rising thermal bubble: $\theta'$
   given by the ROMs and the FOM at time values within (left) and
   outside (right) the training dataset.}
    \label{fig:RTB}
\end{figure}

For the rising thermal bubble, we collect a database of potential temperature perturbations consisting of 204 snapshots, i.e., the computed $\theta^\prime$ every 5 seconds. 
For simplicity, we used a constant eddy viscosity $\nu_a = 15$ m$^2$/s, which is frequently used in the literature (see, e.g. 
\cite{ahmadLindeman2007,marrasNazarovGiraldo2015,GQR_OF_clima}).
Then, 
90\% of the database (i.e., 184 snapshots) is used for training. 
In the case of DMD and HDMD, these 184 solutions are the first 184 in the database (associated to the time interval $(0, 920]$ s). In the case of PODI, 
these 184 solutions are selected randomly over the entire time interval $[0, 1020]$ s. The remaining 20 solutions form the validation set. This difference in the training and validation sets reflects the different nature of PODI and DMD/HDMD algorithms. 
To retain 99\% of the eigenvalue energy in the database, one needs 17 POD modes for DMD and PODI and 46 modes for HDMD. 
Figure~\ref{fig:RTB} shows a qualitative comparison of the ROM solutions with the FOM solution. Among the four time instants chosen for the visualization, two of them  ($t = 255$ and $t = 505$ s)  correspond to solutions belonging to the training set and allow us to asses the ability of each ROM technique to identify the system dynamics. 
We see that all three ROMs perform well in system identification.
The remaining two times ($t = 980$ and $t = 1020$ s) are not associated with the training set and thus are used to check the accuracy of the ROM in predicting (for DMD and HDMD) or interpolating (for PODI) the system dynamics. For these times, 
DMD provides a poor approximation of $\theta'$, while HDMD and PODI reconstruct the solution well. The problem with the reconstruction provided by DMD appears to be related to assigning large weights to  basis functions
associated to previous times.  Indeed, in the DMD solution for $t = 1020$ s we can observe the time history of the system dynamics, i.e., the rising of the bubble.

All the simulations whose results are reported in Fig.~\ref{fig:RTB} were run on a common laptop. 
Table~\ref{tab:cost} reports the computational time needed to construct the reduced basis offline and to perform a simulation online for each of the ROM we consider. 
The computational times for DMD and PODI are comparable, which for the online phase is explained by the fact that the DMD and PODI reduced basis have the same size, while HDMD is computationally more expensive.

\begin{table}[ht!]
\centering
\begin{tabular}{|c|c|c|}
\hline 
Method & Basis const. & Online run \\
\hline
DMD & 0.085 s  & 0.02 s \\
\hline
HDMD & 2.865 s & 2.95 s \\
\hline
PODI & 0.1 s   & 0.018 s  \\
\hline
 
\end{tabular}
 \caption{Rising thermal bubble: 
 computational time needed to construct the reduced basis offline and to perform a simulation online for DMD, HDMD, and PODI.}
 \label{tab:cost}
\end{table}

\begin{figure}
    \centering
\vspace*{0.4cm}
         \begin{overpic}[width=\textwidth, grid=false]{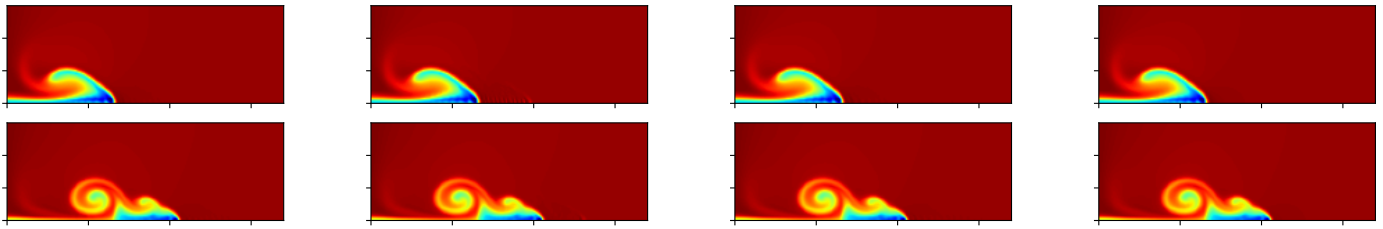}
    \put(40,20){{system identification}}
    \put(8,17){DMD}
    \put(33,17){HDMD}
    \put(60,17){PODI}
    \put(86,17){FOM}
    \put(12,14){\footnotesize{\textcolor{white}{$t$ = 400 s}}}
    \put(38,14){\footnotesize{\textcolor{white}{$t$ = 400 s}}}
    \put(65,14){\footnotesize{\textcolor{white}{$t$ = 400 s}}}
    \put(91,14){\footnotesize{\textcolor{white}{$t$ = 400 s}}}
    \put(12,5.5){\footnotesize{\textcolor{white}{$t$ = 600 s}}}
    \put(38,5.5){\footnotesize{\textcolor{white}{$t$ = 600 s}}}
    \put(65,5.5){\footnotesize{\textcolor{white}{$t$ = 600 s}}}
    \put(91,5.5){\footnotesize{\textcolor{white}{$t$ = 600 s}}}
    \end{overpic} \\
    \vspace*{1cm}
     \begin{overpic}[width=\textwidth, grid=false]{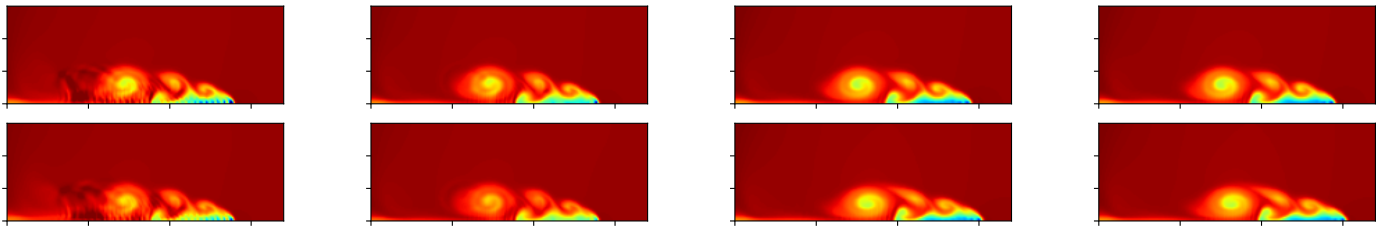} 
    \put(40,20){{prediction/interpolation}}
    \put(8,17){DMD}
    \put(33,17){HDMD}
    \put(60,17){PODI}
    \put(86,17){FOM}
    \put(12,14){\footnotesize{\textcolor{white}{$t$ = 852 s}}}
    \put(38,14){\footnotesize{\textcolor{white}{$t$ = 852 s}}}
    \put(65,14){\footnotesize{\textcolor{white}{$t$ = 852 s}}}
    \put(91,14){\footnotesize{\textcolor{white}{$t$ = 852 s}}}
    \put(12,5.5){\footnotesize{\textcolor{white}{$t$ = 900 s}}}
    \put(38,5.5){\footnotesize{\textcolor{white}{$t$ = 900 s}}}
    \put(65,5.5){\footnotesize{\textcolor{white}{$t$ = 900 s}}}
    \put(91,5.5){\footnotesize{\textcolor{white}{$t$ = 900 s}}}
      \end{overpic}
      \caption{Density current: $\theta'$
   given by the ROMs and the FOM at time values within (top) and
   outside (bottom) the training dataset
      }
        \label{fig:POD_DMD_HDMD_Alltime_DC}
\end{figure}
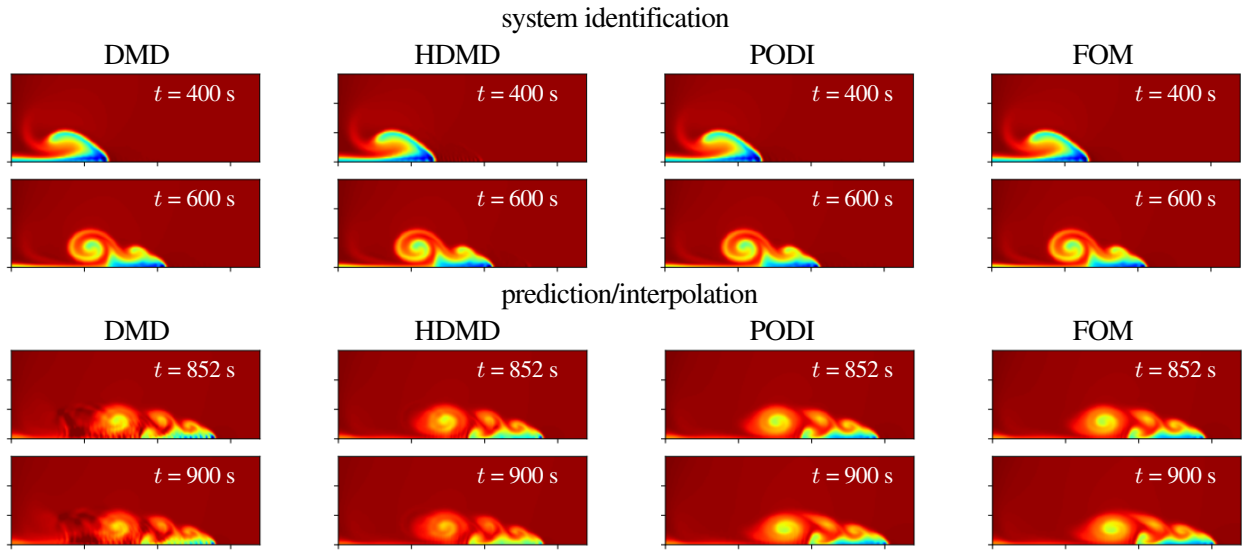

The density current benchmark proved to be more challenging for the fact that it features mainly vertical dynamics for part of the time interval (bubble descending) and mainly horizontal dynamics for the rest of the time (after the bubble hits the ground and a cold front propagates). 
To separate the two dynamics \cite{girfoglio2023hybrid, strazzullo2022consistency}, we collect snapshots associated only with the cold front propagation, i.e., the computed $\theta^\prime$ every 4 seconds for $t \geq 280$. Also in this case, we set the eddy viscosity to a constant value ($\mu_a = 65$
m$^2$/s) taken from the literature \cite{ahmadLindeman2007,strakaWilhelmson1993}.
Out of the snapshot database, $85\%$ is used for 
training and $t =[808,900]$ s is used as prediction window. 
We set again the eigenvalue energy threshold to 99\%, which leads to retaining 49 POD modes for DMD, 96 modes for HDMD, and 52 modes for PODI. Figure~\ref{fig:POD_DMD_HDMD_Alltime_DC}
compares the evolution of the potential temperature perturbation given by DMD, HDMD, and PODI with the evolution computed by the FOM. The top two rows in 
Fig.~\ref{fig:POD_DMD_HDMD_Alltime_DC} correspond to solutions belonging to the training set. Once again, 
we see that all three ROMs perform well in system identification.
The bottom two rows in 
Fig.~\ref{fig:POD_DMD_HDMD_Alltime_DC} correspond to solutions not within the training set. The DMD solutions are affected by the same problem observed in Fig.~\ref{fig:RTB}: we can see the evolution of $\theta^\prime$ in each panel because large weights are assigned to basis functions associated to previous times. The HDMD solutions are a clear improvement, but the the PODI solutions compare more
favorably with the FOM solutions.

In conclusion, although DMD and HDMD are intended for forecasts, the accuracy in the prediction of the system dynamics is low even when 99\% of the eigenvalue energy is retained and the snapshots in the database are tailored to the problem at hand. Thanks to the interpolatory approach, PODI maintains a good level of accuracy during the entire time interval of interest. In \cite{HAJISHARIFI2024104050}, it is shown that 
this is true also when a physical parameter (not just time) is varied 
within a parametric study and for 3D results. 
While still preliminary, these results show that a lot more work is needed if we want to use ROMs to cut the computational time required to forecast the weather for long prediction windows and in high-dimensional parameter space.
The results for the T-junction benchmark test summarized in Sec.~\ref{sec:inc_ROM} suggest that L-ROM, EFR-ROM, and TR-ROM could likely 
improve the results presented in this section in terms of accuracy.
Additional improvements in terms of efficiency could come from Machine Learning-based techniques that can better detect and reproduce the nonlinear behavior exhibited by the FOM, e.g., convolutional autoencoders and long-short time memory
 \cite{gonzalez2018deep,maulik2021reduced, mohan2019compressed, shi2015convolutional}. 

\section{Concluding Remarks and Outlook}\label{sec:concl}

In this paper, we connect two important research fields that have been treated separately until now: 
LES and 
ROMs by reviewing ROMs that are inspired from LES strategies, which we call LES-ROMs.
In our review, we tried to emphasize two essential features of LES-ROMs:
(i) the LES-ROM achievements in reduced order modeling of realistic, convection-dominated, under-resolved flows; and
(ii) their rigorous, mathematical foundations, that enable a clear physical interpretation of the results. 

Convection-dominated flows in the under-resolved regime are central in important engineering, scientific, and medical applications.
Standard ROMs generally fail to produce acceptable results in these challenging settings. 
Thus, alternatives are urgently needed.
In our review, we emphasized the significant achievements of LES-ROMs in the efficient and accurate numerical simulation of convection-dominated, under-resolved flows in important applications in engineering (e.g., turbulent flows in aerospace industry), science (e.g., weather/climate modeling), and medicine (e.g., cardiovascular flows).
In these critical applications, LES-ROMs have yielded accurate results, while their cost was orders of magnitude lower than the cost of traditional numerical methods (e.g., the finite element method), also called 
FOMs. 
We also emphasized that an important advantage of LES-ROMs is that they often are extremely easy to implement.
Indeed, one can start with a ``legacy'' ROM code (e.g., one of the current ROM software libraries~\cite{rbenicsx,libROM}) 
and, in a matter of {\it minutes}, implement an efficient and accurate LES-ROMs.
We illustrated the significant achievements of LES-ROM for both incompressible flows (e.g., the incompressible Navier-Stokes equations used in computational hemodynamics) and compressible flows (e.g., the weakly compressible Euler equations used in weather prediction).

Our review revolves around the mathematical tools that are used to construct LES-ROMs, which not only provide sound support for LES-ROMs, but also connect (``bridge'') two important research fields, LES and ROMs, that have been investigated separately.
Our presentation focused on {\it spatial filters}, both at a FOM level and at the ROM level, and {\it approximate deconvolution (AD)} operators, which provides efficient and stable approximations to the inverse of the spatial filter.
The FOM spatial filters we discussed are the differential filter and the nonlinear filter.
We then used these filters to build LES models.
The ROM spatial filters we presented are the ROM differential filter, the ROM higher-order algebraic filter, and the ROM projection.
We then illustrated how these ROM spatial filters are leveraged to construct LES-ROMs.
As for the AD, we presented several strategies (e.g., Lavrentiev, Tikhonov, and Van Cittert) and showed how they can be used to increase the accuracy of both LES models and LES-ROMs.

Equipped with these mathematical tools (i.e., spatial filtering and AD), we then showed how they can be used to construct LES models and LES-ROMs.
For clarity of presentation, we focused on one of the most popular and successful model, the {\it evolve-filter-relax (EFR)} strategy, and discussed it at both the FOM and the ROM levels.
At the FOM level, we showed how the EFR model is constructed for both the incompressible Navier-Stokes equations and the weakly compressible Euler equations.
Then, we illustrated some of the EFR's achievements in incompressible flows (computational hemodynamics) and compressible flows (the rising thermal bubble and the density current benchmarks).
At the ROM level, we constructed the EFR-ROM and several other LES-ROMs (the Leray ROM, AD Leray ROM, and time-relaxation ROM).
Then, we illustrated some LES-ROMs' achievements for both incompressible flows (the classical flow past a cylinder and the T-junction test case with thermal striping) 
and compressible flows (again the rising thermal bubble benchmark and density currents).

We are aware that this is not an exhaustive review of LES-ROMs. However, in writing this paper, 
our hope was to provide a brief and friendly introduction to this exciting research area, which we believe has a lot of potential in practical numerical simulation of convection-dominated flows in engineering, science, and medicine.
We hope we conveyed to the reader the reason for our excitement: LES-ROM are extremely easy to implement and use, highly efficient, and accurate in reproducing average flow quantities of interest, even in the case of  challenging, critical applications of convection-dominated flows.

Despite the LES-ROMs' success, there are still many research directions that need to be further explored.
This is natural since the LES-ROMs are much more recent than classical LES models.
Among the promising research directions in the LES-ROM development, we mention the construction, analysis, and investigation of other LES-ROMs, spatial filters, and approximate deconvolution operators.
Indeed, at the FOM level, there is a plethora of LES models that are constructed with various types of spatial filters and closure modeling strategies.
A few of those have been already used to build LES-ROMs, but there are many more that have not been investigated.
Another interesting research direction is the construction of new, data-driven spatial scales (lengthscales) at the ROM level.
Indeed, at the FOM level the spatial filter radius can be defined based on the mesh used in the numerical discretization.
At the ROM level, however, how should we define the filter radius?
In this paper, we outlined a few possibilities, but much more remains to be done.
Another important research direction is continuing to lay the mathematical foundations of LES-ROMs.
Only the first steps in this direction have been made, as we briefly discussed in this paper.
There is, however, much more that needs to be done to bring the mathematical support of LES-ROMs to the same level as the mathematical support of LES (see, e.g., the research monographs~\cite{BIL05,john2004large,layton2012approximate,rebollo2014mathematical} on the mathematical theory of LES).




\bibliographystyle{acm}
\bibliography{bibsync,ReferencesNSF,literatur,bibliography_completa,mybib,traian,alex,san}{}

\end{document}